\documentclass[11pt]{article}
\usepackage{amssymb}
\usepackage{amsmath}
\usepackage{amscd}
\usepackage{amsthm}

\numberwithin{equation}{section}

\begin{document}
\bibliographystyle{plain}


\vspace{2cm}

\title{\bfseries Aspects of Higher-Abelian Gauge Theories at zero and finite temperature: Topological Casimir effect, duality and Polyakov loops}

\author{Gerald Kelnhofer\footnote{email: gerald.kelnhofer@gmail.com} \\
Faculty of Physics \\
University of Vienna \\
Boltzmanngasse 5, A-1090 Vienna, \\ Austria}
\date{}
\maketitle
\begin{abstract}
Higher-abelian gauge theories associated with Cheeger-Simons differential characters are studied on compact manifolds without boundary. The paper consists of two parts: First
the functional integral formulation based on zeta function regularization is revisited and extended in order to provide a general framework for further applications. A field
theoretical model - called extended higher-abelian Maxwell theory - is introduced, which is a higher-abelian version of Maxwell theory of electromagnetism extended by a
particular topological action. This action is parametrized by two non-dynamical harmonic forms and generalizes the $\theta$-term in usual gauge theories. In the second part the
general framework is applied to study the topological Casimir effect in higher-abelian gauge theories at finite temperature at equilibrium. The extended higher-abelian Maxwell
theory is discussed in detail and an exact expression for the free energy is derived. A non-trivial topology of the background space-time modifies the spectrum of both the
zero-point fluctuations and the occupied states forming the thermal ensemble. The vacuum (Casimir) energy has two contributions: one related to the propagating modes and the
second one related to the topologically inequivalent configurations of higher-abelian gauge fields. In the high temperature limit the leading term is of Stefan-Boltzmann type
and the topological contributions are suppressed. With a particular choice of parameters extended higher-abelian Maxwell theories of different degrees are shown to be dual. On
the $n$-dimensional torus we provide explicit expressions for the thermodynamic functions in the low- and high temperature regimes, respectively. Finally, the impact of the
background topology on the two-point correlation function of a higher-abelian variant of the Polyakov loop operator is analyzed.
\\ \\ \\ \\ \\
\end{abstract}




\pagebreak

\tableofcontents

\section{Introduction and summary}

\subsection{Motivation and objective}


Since its discovery by Casimir in 1948 \cite{Casimir}, the Casimir effect at zero and nonzero temperature has been a subject of intense research both theoretically and
experimentally, ranging from string theory to condensed matter physics, nanotechnology and cosmology. For a general review see \cite{PMG, MT, K-A-Milton, BMM, BKMM, BD} and
further references therein.\par

In brief, the Casimir effect is caused by a modification of the vacuum energy of a quantum field subjected to boundary conditions as compared to the quantum field without any
constraints. At finite temperature not only the vacuum energy but also the energy of the thermal excitations is affected giving rise to a modified free energy. These constraints
on the quantum field may be imposed either by the presence of material boundaries, interfaces, domain walls or by a topologically non-trivial space-time background. The effect
implied by the latter is usually called \textit{topological Casimir effect} and indicates a deep relation between the global properties of underlying space-time and the quantum
phenomenon. The topological Casimir effect at zero and finite temperature has been considered for many years in different contexts (see e.g. \cite{Ford 1975, Ford 1976,
Dowker-Critchley 1977, Dowker-Kennedy 1978, Altaie-Dowker 1978, Rubin-Roth 1983, Dowker 1984, Accetta-Kolb 1986, Goncharov-Bytsenko-1987, BKM-1988,
 Bytsenko-Goncharov 1991, Cognola-Vanzo-Zerbini 1992, Zhuk-Kleinert 2006-1, Zhuk-Kleinert 2006-2, Bytsenko-Cognola-Vanzo-Zerbini 1996, Brevik-Milton-Odintsov 2002,
 Muller-Fagundes-Opher 2002, Elizalde
 2003, Dowker 2004, OS, Elizalde 2006, Lima-Muller 2007, Lim-Teo, Saharian-Setare 2008, GKM, BS-2010, Teo-2009, Saharian-Mkhitaryan 2010, Bezerra-Klimchitskaya-Mostepanenko-Romero 2011,
 Bezerra-Mostepanenko-Mota-Romero 2011, EOS, Saharian 2012, Asorey-Beneventano-Ascanio-Santangelo 2013, Bezerra-Mota-Muniz 2014, Mota-Bezerra 2015})\footnote{The
reviews \cite{PMG, MT, K-A-Milton, BMM, BKMM, BD} and the references therein cover also further aspects of the topological Casimir effect. Due to the vast amount of literature
we are well aware that our selection may by no means be complete.}. These areas of applications include for example the impact of the topological Casimir effect in models of a
compact universe with non-trivial topology, its role for the stabilization of moduli in multidimensional Kaluza-Klein type theories, its occurrence in brane-world cosmological
scenarios or its consequences for compactified condensed matter systems. The interest in the topological Casimir effect at finite temperature is motivated in a large part by the
standard cosmological model and inflationary models, where at very early stages the evolution of our universe undergoes a high temperature phase. In addition to the
modifications of the partition function and the thermodynamic functions, a non-trivial topology of the background space-time also affects the (thermal) vacuum expectation values
(VEVs) of physical observables.\par

Most of the research has been devoted to scalar fields and to the electromagnetic field, but to a lesser extent to antisymmetric tensor fields. In fact, these fields are
represented by differential forms $A\in\Omega^{p}(M)$ of a certain degree $p\leq n$ on a $n$-dimensional manifold $M$ and are subjected to gauge transformations $A\mapsto
A+d\xi$, where $\xi\in\Omega^{p-1}(M)$. The most prominent example is the so-called $p$-form Maxwell theory \cite{KR, HT}, which is governed by the classical action

\begin{equation}
S(A)=\frac{q_p}{2}\int_{M}dA\wedge\star\ dA,\label{p-form Maxwell}
\end{equation}
and whose dynamical (i.e. propagating) degrees of freedom are represented by co-exact $p$-forms on $M$ ($q_p$ denotes the coupling constant). During the last years several
aspects of the topological Casimir effect in $p$-form Maxwell theory were discussed on different manifolds at zero \cite{CT, ELV-1, ELV-2, BGM, Dowker-4, Dowker-5} and at finite
temperature \cite{D-2002, BMT, BSHS}.\par

However, antisymmetric tensor fields represent just a special class of higher-abelian gauge fields. Compared to the former which are topologically trivial, the latter possess
topologically inequivalent configurations giving rise to quantized charges and fluxes. These higher-abelian fields play an important role in string theory and supergravity and
nowadays gain importance also in other branches of physics. A suitable mathematical framework for their description is provided by differential cohomology, which, in brief, can
be viewed as a specific combination of ordinary (e.g. singular) cohomology with the algebra of differential forms on smooth manifolds. Following \cite{Szabo}, we may call
\textit{a higher-abelian (or generalized abelian) gauge theory} any field theory on a smooth manifold whose space of gauge inequivalent configurations (i.e. the gauge orbit
space) is modeled by a certain differential cohomology group. A review of differential cohomology and its application in physics can be found in \cite{Szabo, Freed, Moore}.\par

For sake of clarity and due to the fact that differential cohomology is uniquely determined up to equivalence \cite{Simons-Sullivan 2008, Baer-Becker 2014}, we focus on
higher-abelian gauge theories whose gauge orbit space is the abelian group of Cheeger-Simons differential characters $\widehat{H}^{p}(M)$ \cite{Cheeger-Simons} of some degree
$p$. Let us recall that differential characters are $U(1)$-valued group homomorphisms on the abelian group $Z_{p}(M)$ of smooth singular $p$-cycles in $M$ that satisfy a certain
smoothness condition. To each differential character $\hat{u}\in \widehat{H}^{p}(M)$ is associated a closed $p+1$ differential form $\delta_1(\hat{u})$ with integer periods,
called curvature (or field strength), and a so-called characteristic class $\delta_2(\hat{u})$ which belongs to the integer cohomology group $H^{p+1}(M;\mathbb{Z})$. As
$\hat{u}$ varies, the assigned characteristic classes label the different topological sectors or in physical terms the topologically inequivalent configurations of
higher-abelian gauge fields.\par

It is natural to raise the question which impact a non-trivial topology of the background space has on these topologically inequivalent configurations which are not present in
theories of ordinary antisymmetric tensor fields. In regard to the applications mentioned above, particularly the combined effect of topology and temperature seems to deserve
closer attention. However, up to now there does not exist exists an exhaustive analysis of the topological Casimir effect in higher-abelian gauge theories.\par

Our main motivation is to fill this gap and to present a detailed study of higher-abelian gauge theories at finite temperature at equilibrium on a $n$-dimensional compact and
closed spatial background. The aim is to shed some light on the impact of the topology of the background on both the vacuum (Casimir) energy and the free energy associated with
the occupied states forming the thermal ensemble of higher-abelian gauge fields at equilibrium. The present paper is a continuation of previous work on abelian gauge theories at
zero \cite{kelnhofer-3,kelnhofer-1} and non-zero temperature \cite{kelnhofer-2}.

\subsection{Outline and summary of results}

We perform our analysis in the functional integral formalism by using zeta function regularization. It is well known that especially in gauge theories at finite temperature
particular care has to be taken to obtain a correct functional integral measure including all relevant temperature and volume dependent factors\footnote{For a discussion on the
necessity to retain the ghost-for-ghost term in order to get the correct number for the degrees of freedom in pure QED at finite temperature we refer to the seminal paper
\cite{Bernard}. See also analogous statements regarding the need for the ghost-for-ghost factor in antisymmetric tensor field theories \cite{Townsend} or in supergravity
\cite{Nielsen}. For Maxwell theory on topologically non-trivial manifolds we refer again to \cite{kelnhofer-2} and references therein.}. This justifies to review the
corresponding construction principle before discussing the topological Casimir effect. \par

For this reason the present paper is split into two main parts. The first part which contains section 2 is devoted to the functional integration of higher-abelian gauge
theories. In order to make the presentation self-contained we begin in section 2.1 with a short introduction to Cheeger-Simons differential characters. Since higher-abelian
gauge theories are affected in general by Gribov ambiguities \cite{kelnhofer-1}, the conventional Faddeev-Popov method to fix the gauge needs to be modified. By revisiting in
section 2.2 the construction of the functional integral representation for the partition function \cite{kelnhofer-1}  to resolve the Gibov problem  we find, however, that
higher-abelian gauge theories exhibit a larger global gauge symmetry than discussed previously. After extracting the volume of the corresponding gauge group, the evaluation of
the functional integral along the gauge inequivalent configurations yields a field independent Jacobian which is formed by two factors. One part derives from the conventional -
due to the existence of zero modes suitable regularized - ghost-for-ghost contribution of the propagating modes. The other contribution to the Jacobian now is related to the
volumes of the subgroups of flat differential characters (i.e. differential characters with vanishing curvature). In the thermal case this topological factor depends on
temperature and volume and hence - like the conventional ghost-for-ghost contribution - has to be retained in the partition function. In summary, we obtain a general formula
from which the partition function can be derived once the higher-abelian gauge model is specified.\par

In this paper we consider an extended version of higher-abelian Maxwell theory which we introduce in section 2.3. In its original form, higher-abelian Maxwell theory
\cite{Szabo, Moore, kelnhofer-1, FMS1, FMS2} is governed by the classical action

\begin{equation}
S_{0}(\hat{u})=\frac{q_p}{2}\int_{M}\delta_1(\hat{u})\wedge\star\ \delta_1(\hat{u}),\qquad \hat{u}\in\widehat{H}^p(M)., \label{generalized Maxwell 0}
\end{equation}
and generalizes \eqref{p-form Maxwell} by replacing the trivial field strength $dA$ by $\delta_1(\hat{u})$. We will extend this action furthermore by adding a specific
topological action $S_{top}=S_{top}(\hat{u},\gamma ,\theta)$ to $S_0$, such that the classical equation of motion is unchanged. The topological action is parametrized by two
arbitrary harmonic forms $\gamma$ and $\theta$ on $M$ representing two de Rham cohomology classes of degree $p+1$ and $n-p$, respectively. They can be regarded as two external,
non-dynamical topological fields. Although the addition of $S_{top}$ does not change the dynamics classically, it contributes non-trivially to the quantum partition function.
This resembles the situation of the $\theta$-term in ordinary gauge theory and is related to the fact that the quantization is not unique on topologically non-trivial
backgrounds. For short we may call this model with total action $S_{tot}=S_0+S_{top}$ \textit{extended higher-abelian Maxwell theory}. We compute the corresponding partition
function and show that the ratio of the partition functions of degrees $p$ and $n-p-1$ becomes independent of the metric of $M$ if certain conditions on the coupling constants
and on the topological fields are met. Additionally, in odd dimensions and for topological fields with integer periods the corresponding partition functions are equal, meaning
that the two extended higher-abelian Maxwell theories are exactly dual to each other. In the special case of vanishing topological fields we recover the result previously
obtained in \cite{DMW}. This closes the first part of the present paper.\par

The second part which contains sections 3 and 4 is devoted to the application of the general framework of the first part to the thermal case. In section 3.1. we determine an
exact formula for the regularized free energy and derive a corresponding asymptotic expansion for the high-temperature regime. These results are applied in section 3.2 to
extended higher-abelian Maxwell theory. It is shown that the topologically inequivalent configurations of the higher-abelian gauge fields give rise to additional contributions
in the thermodynamic functions. As a consequence the total vacuum (Casimir) energy is the sum of the vacuum energy of the propagating modes and the vacuum energy generated by
the topological fields. The latter vanishes whenever the topological fields are absent or have integer periods. This generalizes our findings obtained previously in Maxwell
theory at finite temperature \cite{kelnhofer-2}, where the effect of these topological contributions was studied. This has initiated a lot of further investigations with
interesting applications \cite{CCZ,Z-1,Z-2,Z-3}. One of the main conclusions was that these topological contributions should be regarded as physically observable. They can be
traced back to tunneling events between different topological sectors. In analogy we would suppose that the same arguments apply to the topological Casimir effect in extended
higher-abelian Maxwell theory as well.\par

At low temperatures the free energy is governed by the vacuum (Casimir) energy. For the high-temperature limit an asymptotic expansion for the free energy is computed in terms
of the Seeley (heat kernel) coefficients. The leading contribution is the extensive Stefan-Boltzmann term in $n$ dimensions generated by the propagating degrees of freedom,
followed by subleading terms related to finite size and topological effects. Due to regularization ambiguities and a non-trivial topology of the spatial background the
invariance of the free energy under constant scale transformations is violated leading to a modified equation of state compared to an ideal Bose gas in flat Euclidean space. As
a consequence of the duality relation discussed in section 2 we infer that the thermodynamic functions of extended higher-abelian Maxwell theories of degrees $p$ and $n-p-1$ are
equal implying that the two theories are exactly dual to each other.\par

As an explicit example we consider extended higher-abelian Maxwell theory on the $n$-dimensional torus $\mathbb{T}^{n}$ in section 3.3. Exact expressions for the thermodynamic
functions are computed for both the low- and high temperature regime in terms of Epstein-zeta functions. If the topological field $\theta$ has half-integral periods, the entropy
converges in the zero temperature limit to a non-vanishing universal term indicating that the ground state is degenerate. In the high-temperature limit each thermodynamic
function is dominated by its respective Stefan-Boltzmann term followed by a subleading topological contribution. All other terms decrease exponentially with increasing
temperature. Since the topological contributions are suppressed in the high-temperature limit we find that the leading term in the entropy to energy ratio becomes linear in the
inverse temperature and depends only on the dimension of the $n$-torus.\par

Of particular importance for thermal gauge theories is the so-called Polyakov loop operator, a variant of the Wilson loop operator, which measures the holonomy of the gauge
fields along the periodic (thermal) time. In section 4 we investigate how the two-point correlation function of a higher-abelian version of the Polyakov loop operator is
affected by the topology of the spatial background. By analogy with point-like static charges in gauge theory this two-point correlation function is interpreted as change in the
free energy in the presence of a pair of oppositely charged, static and closed branes in a bath of thermal higher-abelian gauge fields. Apart from a temperature independent
interaction term of Coulomb type the topological sectors are shown to give rise to both a temperature independent and temperature dependent contribution.\par

Appendix A contains some information on Riemann Theta functions and Epstein zeta functions needed in this paper. We choose the conventions $c=\hbar =k_B=1$.

\section{Higher-abelian gauge theories}

In this section we want to introduce the geometrical setting for the formulation of higher-abelian gauge theories as well as the notation and conventions that will be used in
this paper. In the second step the configuration space is studied and the functional integral over the group of Cheeger-Simons differential characters is introduced.

\subsection{The configuration space - Cheeger-Simons differential characters}

Let $M$ be a $(n+1)$-dimensional compact, oriented and connected Riemannian manifold without boundary, $(\Omega ^{\bullet}(M);d^{M})$ the de-Rham complex of smooth differential
forms and $\Omega_{cl}^{\bullet}(M)=\ker d_{\bullet}^{M}$ the space of closed differential forms. Furthermore, let $(C_{\bullet}(M;\mathbb{Z});\partial )$ be the complex of
smooth singular chains in $M$ with boundary map $\partial_{\bullet}\colon C_{\bullet}(M;\mathbb{Z})\rightarrow C_{\bullet -1}(M;\mathbb{Z})$. For an abelian group $G$, let
$C_{\bullet}(M;G)=C_{\bullet}(M;\mathbb{Z})\otimes G$ denote the smooth singular complex with coefficients in $G$ and let $C^{\bullet}(M;G)=Hom(C_{\bullet}(M;\mathbb{Z}),G)$ be
the complex of abelian groups of $G$-valued cochains with coboundary $\delta_{\bullet}\colon C^{\bullet}(M;G)\rightarrow C^{\bullet +1}(M;G)$. Furthermore,
$Z_{\bullet}(M;G)=\ker(\partial_{\bullet}\otimes id)$ is the subcomplex of smooth singular $G$-valued cycles and $Z^{\bullet}(M;G)=\ker\delta_{\bullet}$ is the subcomplex of
smooth singular $G$-valued cocycles. The corresponding singular homology and cohomology groups with coefficients in $G$ are denoted by $H_{\bullet}(M;G)$ and $H^{\bullet}(M;G)$,
respectively.\par

Let $i_p\colon\Omega ^{p}(M)\rightarrow C^{p}(M;U(1))$ be the map $i_p(A)(\Sigma )=e^{2\pi i\int _{\Sigma}A}$, where $\Sigma\in C_{p}(M;\mathbb{Z})$. Moreover, let $\Omega
_{\mathbb{Z}}^p(M)$ denote the abelian group of closed smooth differential $p$-forms with integer periods, i.e. for $\omega\in\Omega _{\mathbb{Z}}^p(M)$ its integral
$\int_{\Sigma}\omega$ over any $p$-cycle $\Sigma\in Z_p(M;\mathbb{Z})$ gives an integer.

The abelian group of \textit{Cheeger-Simons differential characters} \cite{Cheeger-Simons} of degree $p$ is defined by

\begin{equation}
\widehat{H}^p(M)=\{ \hat{u}\in Hom\left( Z_p(M;\mathbb{Z}),U(1)\right) |\quad \hat{u}\circ\partial \in i_{p+1}(\Omega _{ \mathbb{Z}}^{p+1}(M))\},
\end{equation}
with group multiplication $(\hat{u}_1\cdot \hat{u}_2)(\Sigma )=\hat{u}_1(\Sigma )\cdot\hat{u}_2(\Sigma )$. By definition there exists a $(p+1)$-form
$\delta_{1}(\hat{u})\in\Omega _{\mathbb{Z}}^{p+1}(M)$ such that $\hat{u}(\partial\Sigma )=e^{2\pi i\int _{\Sigma}\delta_{1}(\hat{u})}$ for all $\Sigma\in C_{p+1}(M;\mathbb{Z})$.
This correspondence $\hat{u}\mapsto \delta_{1}(\hat{u})$ defines a homomorphism $\delta_{1}\colon\widehat{H}^p(M)\rightarrow\Omega _{\mathbb{Z}}^{p+1}(M)$, which is called the
curvature (or field strength) of $\hat{u}$. On the other hand, since $Z_{p}(M;\mathbb{Z})$ is a free $\mathbb{Z}$ module, there exists a homomorphism $\tilde{u}\colon
Z_{p}(M;\mathbb{R})\rightarrow \mathbb{R}$ such that $\hat{u}(\Sigma )=e^{2\pi i\tilde{u}(\Sigma )}$. But then the map $\breve{u}\colon
C_{p+1}(M;\mathbb{Z})\rightarrow\mathbb{R}$, defined by $\breve{u}(\Sigma )=\int_{\Sigma}\delta_{1}(\hat{u})-\tilde{u}(\partial\Sigma )$, belongs to $Z^{p+1}(M;\mathbb{Z})$ and
gives rise to the cohomology class $[\breve{u}]\in H^{p+1}(M;\mathbb{Z})$. This assignment $\hat{u}\mapsto [\breve{u}]$ generates a homomorphism $\delta
_2\colon\widehat{H}^p(M)\rightarrow H^{p+1}(M,\mathbb{Z})$ and $\delta _2(\hat{u})=[\breve{u}]$ is called the characteristic class of $\hat{u}$.\par

Let $R^{p+1}(M)=\{(\omega ,c)\in\Omega _{\mathbb{Z}}^{p+1}(M)\times H^{p+1}(M,\mathbb{Z})|[\omega]=j_{\ast}(c)\}$, where $j_{\ast}\colon H^{p+1}(M,\mathbb{Z})\rightarrow
H^{p+1}(M,\mathbb{R})$ is induced by the inclusion $\mathbb{Z}\hookrightarrow\mathbb{R}$. Each class $[v]\in H^p(M,U(1))$ defines a differential character
$j_1([v])\in\widehat{H}^p(M)$ by $j_1([v])(\Sigma):=v|_{Z_p(M;\mathbb{Z})}(\Sigma)=<[v],[\Sigma]>$ where $[\Sigma]$ denotes the homology class of the $p$-cycle $\Sigma\in
Z_p(M;\mathbb{Z})$. On the other hand each class $[A]\in\Omega ^p(M)/\Omega _{\mathbb{Z}}^p(M)$ defines a differential character $j_2([A])\in\widehat{H}^p(M)$ by setting
$j_2([A])(\Sigma ):=e^{2\pi i\int _{\Sigma}A}$. The homomorphisms $j_1$, $j_2$, $\delta _1$ and $\delta _2$ fit into the following exact sequences of abelian groups
\cite{Cheeger-Simons}

\begin{equation}
\begin{split}
& 0 \rightarrow H^p(M,U(1))\xrightarrow{j_1} \widehat{H}^p(M)\xrightarrow{\delta _1} \Omega _{\mathbb{Z}}^{p+1}(M) \rightarrow 0 \\
& 0 \rightarrow \Omega ^p(M)/\Omega _{\mathbb{Z}}^p(M) \xrightarrow{j_2}\widehat{H}^p(M)\xrightarrow{\delta _2}H^{p+1}(M,\mathbb{Z})\rightarrow 0\\
& 0\rightarrow H^{p}(M;\mathbb{R})/j_{\ast}(H^{p}(M;\mathbb{Z}))\rightarrow\widehat{H}^p(M)\xrightarrow{(\delta _1,\delta _2)}R^{p+1}(M)\rightarrow 0 .
\end{split}\label{cheeger-simons}
\end{equation}
Furthermore $\delta _1(j_2([A]))=dA$ and $\delta _2(j_1(v))=-\delta ^{\ast}([v])$, where $\delta ^{\ast}\colon H^p(M,U(1))\rightarrow H^p(M,\mathbb{Z})$ is the Bockstein
operator \cite{Bredon} induced from the short exact sequence $0\rightarrow\mathbb{Z}\rightarrow\mathbb{R}\rightarrow U(1)\rightarrow 0$. Since $\delta _2(j_2([A]))=0$, the
differential characters induced by $p$-forms are called \textit{topologically trivial}, whereas differential characters induced by classes in $H^p(M,U(1))$ are called
\textit{flat}, since $\delta _1(j_1([v]))=0$. The group of flat differential characters is a compact abelian group and fits into the short exact sequence

\begin{equation}
0 \rightarrow H^{p}(M;\mathbb{R})/j_{\ast}(H^{p}(M;\mathbb{Z}))\rightarrow H^p(M,U(1))\xrightarrow{\delta^{\ast}} H_{tor}^{p+1}(M,\mathbb{Z})\rightarrow 0,
\label{sequence-global}
\end{equation}
where $H_{tor}^{p+1}(M,\mathbb{Z})$ is the torsion subgroup in $H^{p+1}(M,\mathbb{Z})$.\par

In fact, $\widehat{H}^{\bullet}(.)$ is a graded functor from the category of smooth manifolds to the category of $\mathbb{Z}$-graded abelian groups, equipped with natural
transformations $j_1$, $j_2$, $\delta_1$, $\delta_2$. It provides a specific model for a differential cohomology theory\footnote{While there exist many different models in
addition, like Deligne cohomology \cite{Bry}, the de Rham-Federer model \cite{HLZ} or Hopkins-Singer differential cocycles \cite{HS}, they were all shown \cite{Simons-Sullivan
2008, Baer-Becker 2014} to be isomorphic to $\widehat{H}^{\bullet}(.)$.}. Moreover $\widehat{H}^p(M)$ can be given the structure of an infinite dimensional abelian Lie group,
whose connected components are labeled by $H^{p+1}(M,\mathbb{Z})$ (for a detailed account on its Frechet-Lie group structure we refer to \cite{BSS}). In physical terms
$\widehat{H}^p(M)$ is interpreted as gauge orbit space of higher-abelian gauge theories. In this context a differential character $\hat{u}\in \widehat{H}^p(M)$ represents an
equivalence class of a higher abelian gauge field\footnote{More generally, higher-abelian gauge fields can be naturally modeled as objects of an action groupoid whose morphisms
are the local gauge transformations and whose automorphisms represent the global gauge transformations. The set of isomorphism classes of objects gives then the corresponding
gauge orbit space. An explicit model is provided by the category of Hopkins-Singer differential cocycles \cite{HS}. This resembles the situation encountered in ordinary gauge
theory where one works with gauge fields rather than gauge equivalence classes.}. The equivalence class induced by an antisymmetric tensor field corresponds to a differential
character whose field strength is an exact form and whose characteristic class vanishes.\par

A $L^2$-inner product on $\Omega ^p(M)$ is defined by

\begin{equation}
<\upsilon _1,\upsilon _2>=\int _{M}\ \upsilon _1\wedge\star\ \upsilon _2\qquad\upsilon _1, \upsilon _2\in\Omega ^p(M),\label{inner-product}
\end{equation}
where $\star$ is the Hodge star operator on $M$ with respect to a fixed metric $g$. On $p$-forms the Hodge star operator satisfies $\star ^2=(-1)^{p(n+1-p)}$. Together with the
related co-differential operator $(d_p^{M})^{\dag}=(-1)^{(n+1)(p+1)+1}\star d_{n+1-p}^{M}\star \colon\Omega ^p(M)\rightarrow\Omega ^{p-1}(M)$ the Laplace operator on $p$-forms
is defined by $\Delta _p^{M} =(d_{p+1}^{M})^{\dag}d_p^{M}+d_{p-1}^{M}(d_{p}^{M})^{\dag}$.\par

Let $\mathcal{H}^p(M):=\ker\Delta _p^{M}$ denote the space of harmonic $p$-forms on $M$ and let $\mathcal{H}^{p}(M)^{\perp}$ be its orthogonal complement. The Green\rq s
operator is given by

\begin{equation}
G_p^{M}\colon\Omega ^p(M)\rightarrow \mathcal{H}^{p}(M)^{\perp},\quad G_p^{M}:= (\Delta _p^{M}\vert _{\mathcal{H}^{p}(M)^{\perp}})^{-1}\circ \pi ^{\mathcal{H}^{p\perp}},
\end{equation}
where $\pi ^{\mathcal{H}^{p\perp}}$ is the projection onto $\mathcal{H}^{p}(M)^{\perp}$. By construction $\Delta _p^{M}\circ G_p^{M}=G_p^{M}\circ\Delta _p^{M} =\pi
^{\mathcal{H}^{p\perp}}$. The dimension of $\mathcal{H}^p(M)$ is given by the $p$-th Betti number $b_p(M)$ of $M$. The abelian subgroup of harmonic $p$-forms with integer
periods is denoted by $\mathcal{H}_{\mathbb{Z}}^{p}(M):= \mathcal{H}^p(M)\cap\Omega _{\mathbb{Z}}^p(M)$.\par

Since the homology of $M$ is finitely generated we can choose a set of p-cycles $\Sigma_i^{(p)}\in Z_p(M)$, $i=1,\ldots ,b_p(M)$, whose corresponding homology classes
$[\Sigma_i^{(p)}]$ provide a Betti basis of $H_p(M;\mathbb{Z})$. Let $(\rho _M^{(n+1-p)}) _j\in\mathcal{H}_{\mathbb{Z}}^{n+1-p}(M)$ be a basis associated to the Betti basis
$[\Sigma_i^{(p)}]$ via Poincare duality, where $j=1,\ldots ,b_{n+1-p}(X)$. A dual basis $(\rho _M^{(p)})_i\in\mathcal{H}_{\mathbb{Z}}^{p}(M)$ can be adjusted such that $\int
_{\Sigma_j^{(p)}}(\rho _M^{(p)})_i=\int _{M}(\rho _M^{(p)})_i\wedge (\rho _M^{(n+1-p)})_j=\delta _{ij}$ holds, implying that $\int _{\Sigma_j^{(p)}}B =\int _{M}B\wedge (\rho
_M^{(n+1-p)})_j$ for any $[B]\in H^p(M;\mathbb{R})$. Then

\begin{equation}
(h_{M}^{(p)})_{ij}=<(\rho_{M}^{(p)})_i,(\rho_{M}^{(p)})_j>=\int _{M}(\rho_{M}^{(p)})_i\wedge\star\ (\rho_{M}^{(p)})_j\label{harmonic-basis}
\end{equation}
is the induced metric on $\mathcal{H}^p(M)$. With respect to this basis, the projector onto the harmonic forms reads

\begin{equation}
\pi ^{\mathcal{H}^{p}}(A)=\sum_{j,k=1}^{b_p(M)}(h_{M}^{(p)})_{jk}^{-1}< A,(\rho _M^{(p)})_j> (\rho _M^{(p)})_k=\sum_{j=1}^{b_p(M)}\left( \int_{M}A\wedge (\rho
_M^{(n+1-p)})_j\right)\ (\rho _M^{(p)})_j.
\end{equation}
Two such bases of harmonic forms with integer periods are connected by a unimodular transformation.\par

Since $im(d_{p-1}^{M})\cong im(d_{p}^{M})^{\dag}$, one has equivalently $\Omega _{\mathbb{Z}}^p(M)\cong im(d_{p}^{M})^{\dag}\times\mathcal{H}_{\mathbb{Z}}^{p}(M)$. In the
following we will utilize this isomorphic representation of $\Omega _{\mathbb{Z}}^p(M)$. Correspondingly, the action of $\Omega _{\mathbb{Z}}^p(M)$ on $\Omega^p(M)$ has the form
$A_{p}\cdot (\tau_{p-1},\eta_{p}):= A_{p}+d\tau_{p-1}+\eta_{p}$, where $(\tau_{p-1},\eta_{p})\in im(d_{p}^{M})^{\dag}\times\mathcal{H}_{\mathbb{Z}}^{p}(M)$.

\subsection{The partition function - general case}

In this section we revisit the functional integral quantization of higher-abelian gauge theories in terms of differential characters extending \cite{kelnhofer-1}. Our aim is to
provide a general framework for the functional integral formulation of these theories without referring to a particular action or observable.\par

Let $S=S(\hat{u}_{p})$, $\hat{u}_{p}\in\widehat{H}^p(M)$, be an Euclidean action for a higher-abelian gauge theory of degree $p$ on the manifold $M$. Additionally, the action
may depend on other external (i.e. non dynamical) fields. The vacuum expectation value (VEV) of an observable $\mathcal{O}=\mathcal{O}(\hat{u}_{p})$ is defined by

\begin{equation}
\langle\mathcal{O}\rangle :=\frac{\mathcal{Z}_{M}^{(p)}(\mathcal{O})}{\mathcal{Z}_{M}^{(p)}},\label{VEV-3}
\end{equation}
where

\begin{equation}
\mathcal{Z}_{M}^{(p)}(\mathcal{O})=\int_{\widehat{H}^p(M)}\mathcal{D}\hat{u}_{p}\ e^{-S(\hat{u}_{p})}\mathcal{O}(\hat{u}_{p}).\label{functional-integral-original}
\end{equation}
Furthermore $\mathcal{Z}_{M}^{(p)}:=\mathcal{Z}_{M}^{(p)}(1)$ denotes the partition function of the theory. The quite formal functional integral
\eqref{functional-integral-original} can be given a definite meaning as follows:\par

Since the abelian group $\Omega ^p(M)/\Omega _{\mathbb{Z}}^p(M)$ is divisible, the second exact sequence in \eqref{cheeger-simons} splits. Accordingly, we can assign to each
cohomology class $\frak c_{M}^{(p+1)}\in H^{p+1}(M;\mathbb{Z})$ a differential character, denoted by $\hat{\frak c}_M^{(p+1)}\in\widehat{H}^{p}(M)$, such that
$\delta_{2}(\hat{\frak c}_M^{(p+1)})=\frak c_{M}^{(p+1)}$. We call $\hat{\frak c}_M^{(p+1)}$ the \textit{background differential character} associated to $\frak c_{M}^{(p+1)}$.
Let $\widehat{\mathcal{H}}^{p}(M):=\{\hat{u}_{p}\in\widehat{H}^{p}(M)|\delta_{1}(\hat{u}_{p})\in\mathcal{H}_{\mathbb{Z}}^{p}(M)\}$ denote the subgroup of harmonic differential
characters \cite{GM}. By using the Hodge decomposition theorem one can always select a family of harmonic background differential characters $\hat{\frak
c}_M^{(p+1)}\in\widehat{\mathcal{H}}^{p}(M)$ associated to $\frak c_{M}^{(p+1)}$.\par

Since $H^{p+1}(M;\mathbb{Z})$ is finitely generated it admits a (non-canonical) splitting of the form $H^{p+1}(M;\mathbb{Z})\cong H_{free}^{p+1}(M;\mathbb{Z})\oplus
H_{tor}^{p+1}(M;\mathbb{Z})$ into its free and torsion subgroup. For the free part we choose a Betti basis $(\frak f_M^{(p+1)})_i$, where $i=1,\ldots ,b_{p+1}(M)$. The torsion
subgroup is generated by the basis $(\frak t_M^{(p+1)})_{j}$, where $j=1,\ldots ,r_{p+1}(M)$. Let $w_{p+1}^{1},\ldots , w_{p+1}^{r_{p+1}(M)}\in\mathbb{N}$ be such that
$w_{p+1}^j(\frak t_M^{(p+1)})_{j}=0$ for each $j=1,\ldots ,r_{p+1}(M)$, then the order of the torsion subgroup is given by
$|H_{tor}^{p+1}(M;\mathbb{Z})|=\prod_{j=1}^{r_{p+1}(M)}w_{p+1}^{j}$. With respect to these choices any class $\frak c_{M}^{(p+1)}\in H^{p+1}(M;\mathbb{Z})$ admits the following
non-canonical decomposition

\begin{equation}
\frak c_{M}^{(p+1)}=(\frak c_{M}^{(p+1)})_{f}+(\frak c_{M}^{(p+1)})_{t}= \sum _{i=1}^{b_{p+1}(M)}m_{p+1}^{i}(\frak f_M^{(p+1)})_i+\sum _{j=1}^{r_{p+1}(M)}\
\tilde{m}_{p+1}^{j}(\frak t_M^{(p+1)})_{j},\label{cohomology-class}
\end{equation}
where $m_{p+1}^{i}, \tilde{m}_{p+1}^{j}$ are integer coefficients. Let us choose $(\hat f_M^{(p+1)})_i\in \widehat{\mathcal{H}}^p(M)$, such that

\begin{alignat}{3}
\delta _1\left( (\hat{\frak f}_M^{(p+1)})_{i} \right) &=(\rho _M^{(p+1)})_i, & \qquad \delta _2\left( (\hat{\frak f}_M^{(p+1)})_{i} \right) &=(\frak f_M^{(p+1)})_i, &\qquad
i=1,\ldots ,b_{p+1}(M), \label{6-1}
\end{alignat}
and let us introduce $(\hat{\frak t}_M^{(p+1)})_{j}\in\widehat H^{p}(M)$ with $j=1,\ldots ,r_{p+1}(M)$, satisfying

\begin{equation}
\delta _1\left((\hat{\frak t}_M^{(p+1)})_{j}\right)=0,\qquad \delta _2\left((\hat{\frak t}_M^{(p+1)})_{j}\right)= (\frak t_M^{(p+1)})_{j},\qquad ((\hat{\frak
t}_M^{(p+1)})_{j})^{w_{p+1}^j}=1. \label{6-11}
\end{equation}
In summary, the family of harmonic background differential characters is given by

\begin{equation}
\hat{\frak c}_M^{(p+1)}=\prod _{i=1}^{b_{p+1}(M)}(\hat{\frak f}_M^{(p+1)})_{i}^{m_{p+1}^{i}}\prod _{j=1}^{r_{p+1}(M)}(\hat{\frak
t}_M^{(p+1)})_{j}^{\tilde{m}_{p+1}^{j}},\label{back}
\end{equation}
which satisfies

\begin{equation}
\delta _1(\hat{\frak c}_M^{(p+1)})=\sum _{i=1}^{b_{p+1}(M)}m_{p+1}^{i}(\rho _M^{(p+1)})_i,\qquad\delta _2(\hat{\frak c}_M^{(p+1)})=\frak c_{M}^{(p+1)},\qquad
d^{\ast}\delta _1(\hat{\frak c}_M^{(p+1)})=0. \label{cohomology}
\end{equation}
For each fixed cohomology class $\frak c_{M}^{(p+1)}\in H^{p+1}(M;\mathbb{Z})$, $\delta_{2}^{-1}(\frak c_{M}^{(p+1)})$ is a torsor for the abelian group $\Omega ^p(M)/\Omega
_{\mathbb{Z}}^p(M)$ of topologically trivial differential characters. Hence there exists a unique class $[A_{p}]\in\Omega ^p(M)/\Omega _{\mathbb{Z}}^p(M)$ for each
$\hat{u}_{p}\in \delta_{2}^{-1}(\frak c_{M}^{(p+1)})$ such that $\hat{u}_{p}=\hat{\frak c}_M^{(p+1)}\cdot j_2([A_{p}])$. However, the choice of a gauge field from the class
$[A_{p}]$ is not unique. From a field theoretical viewpoint this ambiguity can be interpreted as a gauge symmetry. Thus for fixed class $\frak c_{M}^{(p+1)}$ the higher-abelian
gauge theory can be interpreted as an "ordinary" abelian gauge theory with gauge orbit space $\Omega ^p(M)/\Omega _{\mathbb{Z}}^p(M)$, action $S^{\hat{\frak
c}_M^{(p+1)}}([A_{p}]):=S(\hat{\frak c}_M^{(p+1)}\cdot j_2([A_{p}]))$ and observable $\mathcal{O}^{\hat{\frak c}_M^{(p+1)}}([A_{p}])=\mathcal{O}(\hat{\frak c}_M^{(p+1)}\cdot
j_2([A_{p}]))$. This interpretation suggests to define the formal functional integral in \eqref{functional-integral-original} as functional integral over the gauge orbit space
$\Omega ^p(M)/\Omega _{\mathbb{Z}}^p(M)$ with respect to a yet to be defined measure $\mathcal{D}[A_{p}]$, followed by summing over all characteristic classes $\frak
c_{M}^{(p+1)}$, namely

\begin{equation}
\mathcal{Z}_{M}^{(p)}(\mathcal{O}):=\sum_{\frak c_{M}^{(p+1)}\in H^{p+1}(M;\mathbb{Z})} \int_{\Omega ^p(M)/\Omega _{\mathbb{Z}}^p(M)}\mathcal{D}[A_{p}]\ e^{-S^{\hat{\frak
c}_M^{(p+1)}}([A_{p}])}\ \mathcal{O}^{\hat{\frak c}_M^{(p+1)}}([A_{p}]).\label{functional-integral-1}
\end{equation}
To calculate the sum over the cohomology classes the recipe is to use the explicit expression \eqref{back} and to sum over the integers $m_{p+1}^{i}$ and $\tilde{m}_{p+1}^{j}$,
respectively.\par

Due to the principle of locality we want to replace the integration over the gauge orbit space by an integration over the whole configuration space $\Omega ^p(M)$ subjected to a
certain gauge fixing condition and followed by dividing by the volume of the gauge group in order to factor out redundant gauge degrees of freedom.\par

Two questions arise: First, what is the total gauge group and second is it possible to fix the gauge globally? An answer to the first question is motivated by the following
exact sequence which derives directly from \eqref{cheeger-simons}:

\begin{equation}
0 \rightarrow H^{p-1}(M,U(1))\xrightarrow{j_1} \widehat{H}^{p-1}(M)\xrightarrow{\delta _1} \Omega ^{p}(M)\xrightarrow{\tilde{j_2}}\widehat{H}^p(M) \xrightarrow{\delta
_2}H^{p+1}(M,\mathbb{Z})\rightarrow 0,\label{cheeger-simons-1}
\end{equation}
where $\tilde{j_2}(A_p):=j_2([A_p])$.\par For an interpretation of \eqref{cheeger-simons-1} let us first consider the case $p=1$: Recall that Maxwell theory of electromagnetism
can be equivalently described in terms of differential cohomology. In fact, the holonomy associated to a gauge field, which is represented geometrically by a connection on a
principal $U(1)$-bundle over $M$, defines a certain differential character whose field strength is the curvature of that connection and whose characteristic class is the first
Chern class of the underlying principal $U(1)$-bundle. This correspondence induces a bijection between the set of isomorphism classes of principal $U(1)$-bundles with
connections and $\widehat{H}^1(M)$. In physical terms, $\widehat{H}^1(M)$ represents the gauge orbit space of Maxwell theory. For $p=1$ the exact sequence
\eqref{cheeger-simons-1} reduces to

\begin{equation}
0 \rightarrow U(1)\xrightarrow{j_1} C^{\infty}(M;U(1)) \xrightarrow{\delta _1} \Omega ^{1}(M)\xrightarrow{\tilde{j_2}}\widehat{H}^1(M) \xrightarrow{\delta
_2}H^{2}(M,\mathbb{Z})\rightarrow 0.\label{cheeger-simons-2}
\end{equation}
where we have used that $H^{0}(M,U(1))\cong U(1)$ and $\widehat{H}^{0}(M)\cong C^{\infty}(M;U(1))$. Hence $H^{0}(M,U(1))$ and $\widehat{H}^{0}(M)$ can be identified with the
groups of global and local $U(1)$ gauge transformations, respectively. Therefore the sequence \eqref{cheeger-simons-2} can be interpreted as follows: Since
$\delta_{1}(\hat{u}_{0})=\frac{1}{2\pi i}\hat{u}_{0}^{-1}d\hat{u}_{0}$ holds for a gauge transformation $\hat{u}_{0}\in C^{\infty}(M;U(1))$, the homomorphism $\delta_1$
generates apparently the gauge transformation $A\mapsto A+\delta_{1}(\hat{u}_{0})$ on the space of topologically trivial gauge fields. Evidently, the kernel of $\delta_{1}$
consists of the global gauge transformations. On the other hand the homomorphism $\tilde{j_2}$ assigns to each $\eta\in \Omega^{1}(M)$ the trivial $U(1)$-bundle on $M$ with
connection $\eta$ modulo connection preserving isomorphisms. Finally, $\delta_{2}$ assigns to each equivalence class of principal $U(1)$-bundles the corresponding first Chern
class.\par

By analogy we interpret $\widehat{H}^{p-1}(M)$ as the total gauge group of a higher-abelian gauge theory of degree $p$. According to \eqref{cheeger-simons-1} the curvature
$\delta_{1}(\hat{u}_{p-1})$ of a gauge transformation $\hat{u}_{p-1}\in \widehat{H}^{p-1}(M)$ gives a morphism connecting two topologically trivial $p$-form gauge fields $A_{p}$
and $A_{p}^{\prime}$ by $A_{p}^{\prime}-A_{p}=\delta_{1}(\hat{u}_{p-1})$. The isotropy group of this action is the subgroup of flat differential characters of degree $(p-1)$,
namely $H^{p-1}(M,U(1))$, which represents the group of global gauge transformations. The quotient $\widehat{H}^{p-1}(M)/H^{p-1}(M,U(1))$ gives the reduced gauge group, which is
isomorphic to $\Omega _{\mathbb{Z}}^p(M)$. Finally, the subgroup of topologically trivial differential characters $\Omega ^p(M)/\Omega _{\mathbb{Z}}^p(M)$ is the corresponding
gauge orbit space. \par

Concerning the second question raised above the answer is negative. It was shown in \cite{kelnhofer-1} that in general the principal fiber bundle $\pi_{\Omega^{p}}\colon\Omega
^p(M)\rightarrow\Omega ^p(M)/\Omega _{\mathbb{Z}}^p(M)$ is not trivializable implying that the theory suffers from a Gribov ambiguity. Therefore the conventional Faddeev-Popov
method to fix the gauge does not longer apply.\par

To overcome this drawback we follow an approach, which was introduced and studied in detail in \cite{kelnhofer-3, kelnhofer-1, Huffel-kelnhofer}. The idea is to modify the gauge
invariant measure $\mathcal{D}A_{p}$ on $\Omega^{p}(M)$, which is induced by the metric \eqref{inner-product} in such a way that the resulting measure is damped along the gauge
degrees of freedom yielding a finite value for the functional integral of gauge invariant observables. Since this modification is possible only locally, these modified but local
measures are pasted together in such a way that the vacuum expectation values of gauge invariant observables become independent of the particular way the gluing is provided. The
construction is as follows: Let us introduce a contractible open cover $\{V_{a}|a\in J\}$ of the 1-torus $\mathbb{T}$ (where $J$ is an appropriate index set) and introduce local
sections $s_{a}\colon V_{a}\rightarrow\mathbb{R}$ of the universal cover $e^{2\pi i(.)}\colon\mathbb{R}\rightarrow\mathbb{T}$. A contractible open cover of $\mathbb{T}^{b_p(M)}$
is provided by

\begin{equation}
\mathcal{V}^{(p)}=\{V_a^{(p)}:=V_{a_1}\times\cdots\times V_{a_j}\times\cdots\times V_{a_{b_p(X)}}\vert\quad a:=(a_1,\ldots ,a_j,\ldots ,a_{b_p(M)}),\quad a_{j}\in J,\quad
\forall j\}.
\end{equation}
The family of local sections $s_{a}=(s_{a_1},\ldots ,s_{a_{b_p(M)}})\colon V_a^{(p)}\rightarrow\mathbb{R}^{b_{p}(M)}$ provides a bundle atlas for
$\mathbb{R}^{b_{p}(M)}\rightarrow\mathbb{T}^{b_p(M)}$. The projection $\pi^{\prime}\colon\Omega ^p(M)/\Omega _{\mathbb{Z}}^p(M)\rightarrow\mathbb{T}^{b_p(M)}$, given by

\begin{equation}
\pi^{\prime}([A_{p}])= \left(e^{2\pi i\int _M A_{p}\wedge (\rho _M^{(n+1-p)})_1},\ldots , e^{2\pi i\int _M A_{p}\wedge (\rho _M^{(n+1-p)})_{b_p(M)}}\right)\label{projection}
\end{equation}
induces an open cover $\{U^{(p)}_{a}:=\pi^{\prime -1}(V_{a}^{(p)})\}$ of the gauge orbit space. For the principal $\Omega _{\mathbb{Z}}^p(M)$-fiber bundle
$\pi_{\Omega^{p}}\colon\Omega ^p(M)\rightarrow\Omega ^p(M)/\Omega _{\mathbb{Z}}^p(M)$ a bundle atlas $\{U^{(p)}_{a},\varphi_{a}^{(p)}\}$ is provided by the family of maps
$\hat{\varphi}_{a}^{(p)}\colon\pi_{\Omega^{p}}^{-1}(U^{(p)}_{a})\rightarrow\Omega _{\mathbb{Z}}^p(M)$,
$\hat{\varphi}_{a}^{(p)}(A_{p})=\left((d_{p}^{M})^{\dag}G_{p}^{M}A_{p},\sum _{j=1}^{b_p(M)} \varepsilon_{a_j}(A_p)(\rho _M^{(p)})_j\right)$, where

\begin{equation}
\varepsilon_{a_j}(A_p):=\int_{M}A_{p}\wedge (\rho _M^{(n+1-p)})_j-s_{a_j}\left( e^{2\pi i\int_{M}A_{p}\wedge (\rho _M^{(n+1-p)})_j}\right)\in\mathbb{Z}\qquad j=1,\ldots,
b_p(M).\label{triv_0}
\end{equation}
Let us introduce a (smooth) real-valued function $R_{p}$ on $\Omega _{\mathbb{Z}}^p(M)$ in such a way that the regularized group volume

\begin{equation}
vol(\Omega _{\mathbb{Z}}^p(M)):=\sum_{\eta_{p}\in\mathcal{H}_{\mathbb{Z}}^{p}(M)}\int_{im(d_{p}^{M})^{\dag}}\mathcal{D}\tau_{p-1}\
e^{-R_{p}(\tau_{p-1},\eta_{p})},\label{regularized-group}
\end{equation}
becomes finite. $\mathcal{D}\tau_{p-1}$ is the measure on $im(d_{p}^{M})^{\dag}$ induced by the inner product \eqref{inner-product}. Now we replace $\mathcal{D}A_{p}$ by the
product $\mathcal{D}A_{p}\cdot\Xi(A_{p})$, where

\begin{equation}
\Xi(A_{p}):=(2\pi )^{-\frac{b_p(M)}{2}}\sum_{a}\wp_{a}^{(p)}([A_p])\ e^{-\hat{\varphi}_{a}^{(p)\ast}R_{p}(A_p)}.\label{partition-function-regularized}
\end{equation}
Here $\{\wp_{a}^{(p)}\}$ denotes a partition of unity subordinate to the open cover $\{U_a^{(p)}\}$ and the pre-factor is chosen for convenience. We will see, however, that the
final result is independent of the concrete choice for $R_p$\footnote{A natural choice for $R_p$ is
$R_{p}(\tau_{p-1},\eta_{p})=\frac{1}{2}\|\Delta_{p-1}^{M}|_{im(d_{p}^{M})^{\dag}}\ \tau_{p-1}\|^{2}+\frac{1}{2}\|\eta_{p}\|^{2}$, which on $\pi_{\Omega^{p}}^{-1}(U^{(p)}_{a})$
becomes
$\hat{\varphi}_{a}^{(p)\ast}R_{p}(A_p)=\frac{1}{2}\|(d_{p}^{M})^{\dag}A_{p}\|^{2}+\frac{1}{2}\sum_{i,j=1}^{b_p(M)}(h_{M}^{(p)})_{ij}\varepsilon_{a_i}(A_p)\varepsilon_{a_j}(A_p)$.
The first term is the conventional gauge fixing term in the Lorentz gauge. The second term damps the harmonic degrees of freedom (i.e. zero modes) and due to the Gribov
ambiguity is defined only locally.}. As intended we rewrite \eqref{functional-integral-1} in the form

\begin{equation}
\mathcal{Z}_{M}^{(p)}(\mathcal{O} ) =\sum_{\frak c_{M}^{(p+1)}\in H^{p+1}(M;\mathbb{Z})} \frac{1}{vol(\widehat{H}^{p-1}(M))}\int_{\Omega^{p}(M)}\mathcal{D}A_{p}\ \Xi(A_{p})\
\pi_{\Omega^{p}}^{\ast} \mathcal{O}^{\hat{\frak c}_M^{(p+1)}}(A_{p})\ e^{-\pi_{\Omega^{p}}^{\ast} S^{\hat{\frak c}_M^{(p+1)}}(A_{p})},\label{VEV-1}
\end{equation}
where we have factored out the yet to be defined (regularized) volume of the total gauge group $\widehat{H}^{p-1}(M)$ to account for all redundant gauge degrees of freedom. In
order to compute the functional integral in \eqref{VEV-1} we notice that the subgroup of topologically trivial differential characters $\Omega ^p(M)/\Omega _{\mathbb{Z}}^p(M)$
has the structure of a trivializable vector bundle over the harmonic torus $H^{p}(M;\mathbb{R})/j_{\ast}(H^{p}(M;\mathbb{Z}))$ with typical fiber $im(d_{p+1}^{M})^{\dag}$. Since
$H^{p}(M;\mathbb{R})\cong\mathcal{H}^{p}(M)$ and $j_{\ast}(H^{p}(M;\mathbb{Z}))\cong H_{free}^{p}(M;\mathbb{Z})\cong\mathcal{H}_{\mathbb{Z}}^{p}(M)$ one has

\begin{equation}
H^{p}(M;\mathbb{R})/j_{\ast}(H^{p}(M;\mathbb{Z}))\cong\mathcal{H}^{p}(M)/\mathcal{H}_{\mathbb{Z}}^{p}(M)\cong\mathbb{T}^{b_{p}(M)}.\label{harmonic torus}
\end{equation}
Correspondingly, a global bundle trivialization is given by

\begin{equation}
\underline\varphi_{a}^{(p)}(\vec{z}_{p},\tau_{p})=\left[\sum _{i=1}^{b_p(M)} s_{a_i}(z_i)(\rho _M^{(p)})_i+\tau_{p} \right],\quad \vec{z}_{p}=(z_1,\ldots
,z_{b_{p}(M)})\in\mathbb{T}^{b_{p}(M)},\ \tau_{p}\in im(d_{p+1}^{M})^{\dag}.\label{triv_2}
\end{equation}
With respect to the local diffeomorphisms $\phi_{a}^{(p)}:=\varphi_{a}^{(p)}\circ (\underline\varphi _{a}^{(p)}\times id_{\Omega _{\mathbb{Z}}^p(M)})$ the volume form transforms
as

\begin{equation}
\phi_{a}^{(p)\ast}\mathcal{D}A_{p} =\left(\det h_{M}^{(p)}\right)^{\frac{1}{2}}\left(\det \Delta_{p-1}^{M}|_{im(d_{p}^{M})^{\dag}}\right)^{\frac{1}{2}}dt_{1}\wedge\ldots\wedge
dt_{b_p(M)}\wedge\mathcal{D}\tau_{p}\wedge\mathcal{D}\tau_{p-1} ,\label{jacobian}
\end{equation}
where $\vec{t}=(t_{1},\ldots ,t_{b_p(M)})$ are local coordinates for $\mathbb{T}^{b_p(M)}$. Notice that $\phi_{a}^{(p)\ast}\mathcal{D}A_{p}$ is a globally defined volume form on
$\mathbb{T}^{b_p(M)}\times im(d_{p+1}^{M})^{\dag}\times\Omega^{p}_{\mathbb{Z}}(M)$.\par

Since $\underline\varphi_{a}^{(p)\ast}\mathcal{O}^{\hat{\frak c}_M^{(p+1)}}=\underline\varphi_{b}^{(p)\ast}\mathcal{O}^{\hat{\frak c}_M^{(p+1)}}$ holds on each overlap
$V_a^{(p)}\cap V_b^{(p)}\times im(d_{p+1}^{M})^{\dag}$, the local contributions can be extended to a globally defined function on $\mathbb{T}^{b_p(M)}\times
im(d_{p+1}^{M})^{\dag}$, denoted by $\underline{\mathcal{O}}^{\hat{\frak c}_M^{(p+1)}}$. For fixed $\tau_{p}\in im(d_{p+1}^{M})^{\dag}$,
$\vec{z}_{p}\mapsto\underline{\mathcal{O}}^{\hat{\frak c}_M^{(p+1)}}(\vec{z}_{p},\tau_{p})$ is a function on $\mathbb{T}^{b_p(M)}$, so it can be expanded in a Fourier series
with coefficients

\begin{multline}
\left[\underline{\mathcal{O}}^{\hat{\frak c}_M^{(p+1)}}(\tau_{p})\right]_{(m_1,\ldots,m_{b_{p}(M)})}:=\\ =\int_{0}^{1}dt_{1}\ldots\int_{0}^{1}dt_{b_{p}(M)}\
\underline{\mathcal{O}}^{\hat{\frak c}_M^{(p+1)}}(e^{2\pi it_{1}},\ldots ,e^{2\pi it_{b_{p}(M)}},\tau_{p})e^{-2\pi i\sum_{j=1}^{b_{p}(M)}m_jt_{j}}. \label{fourier}
\end{multline}
We transform the integral in \eqref{VEV-1} into an integral over $\mathbb{T}^{b_p(M)}\times im(d_{p+1}^{M})^{\dag}\times\Omega^{p}_{\mathbb{Z}}(M)$. By using
\eqref{regularized-group}, \eqref{jacobian} and the Fourier series expansion \eqref{fourier} one obtains

\begin{equation}
\begin{split}
\mathcal{Z}_{M}^{(p)}(\mathcal{O})= & (2\pi )^{-\frac{b_p(M)}{2}} (\det h_{M}^{(p)})^{\frac{1}{2}} (\det\Delta_{p-1}^{M}|_{im(d_{p}^{M})^{\dag}})^{\frac{1}{2}} \frac{vol(\Omega
_{\mathbb{Z}}^p(M))}{vol(\widehat{H}^{p-1}(M))}\\ &\times\sum_{\frak c_{M}^{(p+1)}\in H^{p+1}(M;\mathbb{Z})}\int_{im(d_{p+1}^{M})^{\dag}}\mathcal{D}\tau_{p}\
\left[e^{-\underline{S}^{\hat{\frak c}_M^{(p+1)}}(\tau_{p})}\underline{\mathcal{O}}^{\hat{\frak c}_M^{(p+1)}}(\tau_{p})\right]_{(0,\ldots,0)}.\label{VEV-4}
\end{split}
\end{equation}
Notice that $\mathcal{Z}_{M}^{(p)}(\mathcal{O})$ is independent of the partition of unity and the local trivialization. Moreover, it does not depend on the specific choice of
background differential characters either\footnote{Let us consider a different family $\hat{\frak c}_M^{\prime (p+1)}$ of background differential characters related to $\frak
c_{M}^{(p+1)}\in H^{p+1}(M;\mathbb{Z})$. Then there exists a family of harmonic forms $B^{\frak c_{M}^{(p+1)}}\in\mathcal{H}^{p}(M)$, parametrized by $\frak c_{M}^{(p+1)}$, such
that $\hat{\frak c}_M^{\prime (p+1)}=\hat{\frak c}_M^{(p+1)}\cdot j_2([B^{\frak c_{M}^{(p+1)}}])$. By \eqref{triv_2} there exists for each $\frak c_{M}^{(p+1)}$ a
$\vec{w}\in\mathbb{T}^{b_p(M)}$ such that $[B^{\frak c_{M}^{(p+1)}}]=\underline\varphi_{a}^{(p)}(\vec{w},0)$. This gives a homomorphism $\frak c_{M}^{(p+1)}\mapsto [B^{\frak
c_{M}^{(p+1)}}]$ from $H^{p+1}(M;\mathbb{Z})$ to $\mathbb{T}^{b_{p}(M)}$. But then $\underline{\mathcal{O}}^{\hat{\frak c}_M^{\prime
(p+1)}}=l_{\vec{w}}^{\ast}\underline{\mathcal{O}}^{\hat{\frak c}_M^{(p+1)}}$, where $l_{\vec{w}}$ denotes the (left) multiplication by $\vec{w}$. This gives
$[\underline{\mathcal{O}}^{\hat{\frak c}_M^{\prime (p+1)}}(\tau_{p})]_{(0,\ldots ,0)}=[\underline{\mathcal{O}}^{\hat{\frak c}_M^{(p+1)}}(\tau_{p})]_{(0,\ldots ,0)}$.}.\par

Now it remains to determine the quotient of the gauge group volumes in \eqref{VEV-4}. Let $R_{p}$ be an arbitrary but fixed regularizing functional for $\Omega
_{\mathbb{Z}}^p(M)$ and introduce the homomorphism $\hat{\delta}_{1}\colon\widehat{H}^{p-1}(M)\rightarrow im(d_{p}^{M})^{\dag}\times\mathcal{H}_{\mathbb{Z}}^{p}(M)$ by
$\hat{u}_{p-1}\mapsto \left((d_{p}^{M})^{\dag}G_{p}^{M}\delta_{1}(\hat{u}_{p-1}),\pi ^{\mathcal{H}^{p}}(\delta_{1}(\hat{u}_{p-1}))\right)$. Then we define the regularized volume
of the total gauge group $\widehat{H}^{p-1}(M)$ in a natural way by the formal functional integral

\begin{equation}
vol(\widehat{H}^{p-1}(M)):=\int_{\widehat{H}^{p-1}(M)}\mathcal{D}\hat{u}_{p-1}\ e^{-\hat{R}_{p-1}(\hat{u}_{p-1})},\label{regularized-group-1}
\end{equation}
where $\hat{R}_{p-1}:=\hat{\delta}_{1}^{\ast}R_{p}$. In order to compute \eqref{regularized-group-1} and verify that it yields a finite volume we are now in exactly the same
situation as we were before when we calculated \eqref{functional-integral-original}, which finally led to \eqref{VEV-4}. Thus we will use formula \eqref{VEV-4}, but now in
degree $p-1$ and applied to the integrand $e^{-\hat{R}_{p-1}}$ instead of $e^{-S}\mathcal{O}$ as in \eqref{functional-integral-original}. Let us begin with the computation of
the corresponding integral in \eqref{VEV-4}. Since $\underline{\hat{R}}_{p-1}^{\hat{\frak c}_M^{(p)}}(\vec{z}_{p-1},\tau_{p-1})=R_p(\tau_{p-1},\delta_{1}(\hat{\frak
c}_M^{(p)}))$ and $\delta_{1}(\hat{\frak c}_M^{(p)})=(\frak c_M^{(p)})_{f}$, we obtain

\begin{equation}
\sum_{\frak c_M^{(p)}\in H^{p}(M;\mathbb{Z})}\int_{im(d_{p}^{M})^{\dag}}\mathcal{D}\tau_{p-1}\ [e^{-\underline{\hat{R}}_{p-1}^{\hat{\frak
c}_M^{(p)}}(\tau_{p-1})}]_{(0,\ldots,0)}=vol(\Omega _{\mathbb{Z}}^p(M))|H_{tor}^{p}(M;\mathbb{Z})|, \label{volume characters-1}
\end{equation}
where \eqref{regularized-group} was used. The group of global gauge transformations $H^{p-1}(M;U(1))$ can be assigned a volume in a canonical way by noting from
\eqref{sequence-global} that its connected component is the harmonic torus $H^{p-1}(M;\mathbb{R})/j_{\ast}(H^{p-1}(M;\mathbb{Z}))$. With respect to the induced metric on
$\mathcal{H}^{p-1}(M)$ the volume of the harmonic torus reads

\begin{equation}
vol(H^{p-1}(M;\mathbb{R})/j_{\ast}(H^{p-1}(M;\mathbb{Z})))=(\det h_{M}^{(p-1)})^{\frac{1}{2}}.\label{volume harmonic torus}
\end{equation}
According to \eqref{sequence-global} this leads to

\begin{equation}
vol(H^{p-1}(M;U(1)))=\left(\det h_{M}^{(p-1)}\right)^{\frac{1}{2}}|H_{tor}^{p}(M,\mathbb{Z})|.\label{volume global gauge transformations}
\end{equation}
From formula \eqref{VEV-4} we obtain for the volume of the total gauge group

\begin{equation}
\begin{split}
&vol(\widehat{H}^{p-1}(M))=\\ &=(2\pi )^{-\frac{b_{p-1}(M)}{2}}\left(\det \Delta_{p-2}^{M}|_{im(d_{p-1}^{M})^{\dag}}\right)^{\frac{1}{2}} vol(H^{p-1}(M;U(1)))vol(\Omega
_{\mathbb{Z}}^p(M))\frac{vol(\Omega _{\mathbb{Z}}^{p-1}(M))}{vol(\widehat{H}^{p-2}(M))}. \label{volume characters-2}
\end{split}
\end{equation}
By induction this leads to

\begin{equation}
\begin{split}
\frac{vol(\Omega _{\mathbb{Z}}^p(M))}{vol(\widehat{H}^{p-1}(M))}=&\prod _{r=0}^{p-1}(2\pi)^{\frac{b_r(M)}{2}(-1)^{p+1-r}}\ \prod _{r=0}^{p-2}\left(\det{\Delta
_r^{M}|_{im(d_{r+1}^{M})^{\dag}}}\right)^{\frac{1}{2}(-1)^{p+1-r}}\\ &\times\prod _{r=0}^{p-1}vol(H^{r}(M;U(1)))^{(-1)^{p-r}}.
\end{split}\label{ghost-for-ghost-1}
\end{equation}
We interpret this quotient as the ghost-for-ghost contribution for higher abelian gauge theories\footnote{Let us remark that in \cite{kelnhofer-1} the groups
$\Omega^{p-1}(M)\times\mathcal{H}_{\mathbb{Z}}^{p}(M)$ and  $\Omega_{cl}^{p-1}(M)$ were considered as total gauge group and group of global gauge transformations, respectively.
Here we derived $\widehat{H}^{p-1}(M)$ and $H^{p-1}(M;U(1))$ as the corresponding gauge groups. The reduced gauge groups are in both cases $\Omega_{\mathbb{Z}}^{p}(M)$, the
ghost-for-ghost terms differ by the order of the torsion subgroups of $M$, however.}. In the topologically trivial case the quotient \eqref{ghost-for-ghost-1} reduces to the
alternating product of Laplace operators, hence recovering the well-known ghost-for-ghost contribution in the functional integral measure \cite{CT, Townsend, Siegel, Obukhov}.
Substituting \eqref{ghost-for-ghost-1} into \eqref{VEV-4} finally yields

\begin{equation}
\begin{split}
\mathcal{Z}_{M}^{(p)}(\mathcal{O} )= &\prod_{r=0}^{p}\left(\frac{\left(\det{(\frac{1}{2\pi}h_{M}^{(r)})}\right)^{\frac{1}{2}}}{|H_{tor}^{r}(M;\mathbb{Z})|}\right)^{(-1)^{p-r}}
\prod_{r=0}^{p-1} \left(\det\Delta_{r}^{M}|_{im(d_{r+1}^{M})^{\dag}}\right)^{\frac{1}{2}(-1)^{p+1-r}}
\\ &\times\sum_{\frak c_{M}^{(p+1)}\in H^{p+1}(M;\mathbb{Z})}\int_{im(d_{p+1}^{M})^{\dag}}\mathcal{D}\tau_{p}\
\left[e^{-\underline{S}^{\hat{\frak c}_M^{(p+1)}}(\tau_{p})}\underline{\mathcal{O}}^{\hat{\frak c}_M^{(p+1)}}(\tau_{p})\right]_{(0,\ldots,0)}.\label{VEV-4-VEV}
\end{split}
\end{equation}
Since the Laplace operator $\Delta_r^M$ on a compact manifold $M$ is an unbounded positive operator with discrete spectrum
$0=\nu_{0}(\Delta_r^M)<\nu_{1}(\Delta_r^M)\leq\ldots\rightarrow\infty$ and finite dimensional eigenspaces, we will use zeta-function regularization to give the determinants
appearing in \eqref{VEV-4-VEV} a mathematical meaning. In this formulation the determinant of $\Delta_r^M$ is given by

\begin{equation}
\det \Delta_r^M =\exp{\left( -\zeta ^{\prime} (0;\Delta_r^M)\right)}\equiv\exp{\left( -\frac{d}{ds}|_{s=0}\zeta (s;\Delta_r^M)\right)},\label{determinant-zeta}
\end{equation}
where the zeta function of $\Delta_{r}^{M}$ is defined by \cite{Rosenberg, Gilkey}

\begin{equation}
\zeta (s;\Delta_r^M)=\sum _{\nu _{\alpha}(\Delta_r^M)\neq 0}\nu _{\alpha}(\Delta_r^M)^{-s}=\frac{1}{\Gamma (s)}\int _0^{\infty}dt\ t^{s-1} Tr(e^{-t\Delta_r^M}-\pi
^{\mathcal{H}^{r}}),\quad s\in\mathbb{C}.\label{zeta}
\end{equation}
Here the sum runs over the non vanishing eigenvalues of $\Delta_r^M$ only and each eigenvalue appears the same number of times as its multiplicity. The second expression in
\eqref{zeta} is the Mellin transform of the zeta-function. Notice that by construction $\zeta (s;\Delta_r^M)=\zeta (s;\Delta_r^M|_{\mathcal{H}^r(M)^{\perp}})$. The zeta function
is holomorphic for $\Re(s)>\frac{\dim M}{2}$ and has a meromorphic continuation to $\mathbb{C}$ with simple poles at $s_k=\frac{\dim M-k}{2}$ for $k\in\mathbb{N}_0$ and residue
$Res_{s=s_k}[\zeta (s;\Delta_r^M)]=\frac{a_k(\Delta_r^M)}{\Gamma (\frac{\dim M-k}{2})}$ at $s=s_k$. Here $a_{k}(\Delta_r^M)$ denotes the $k$-th Seeley coefficient in the
asymptotic expansion of the $L^2$-trace of the heat kernel $Tr(e^{-t\Delta_r^M})$ \cite{Gilkey}, which is given by

\begin{equation}
Tr(e^{-t\Delta_r^M})\simeq\ \sum _{k=0}^{\infty}a_k(\Delta_r^M)\ t^{\frac{k-\dim M}{2}},\quad\textrm{for}\ t\downarrow 0.\label{seeley1}
\end{equation}
The Seeley coefficients $a_k(\Delta_r^M)$ are integrals over $M$ of polynomials which depend on the metric and its derivatives. However, $a_k(\Delta_r^M)=0$ whenever $k$ is odd.
The zeta function is regular at $s=0$, yielding

\begin{equation}
\zeta (0;\Delta_r^M)=
\begin{cases} -b_{r}(M), &\text{if $\dim M=n+1$ is odd},
\\
a_{n+1}(\Delta_r^M)-b_r(M), &\text{if $\dim M=n+1$ is even}.\end{cases}\label{zeta-0}
\end{equation}
In terms of the Seeley coefficients the Euler characteristics $\chi (M):=\sum _{r=0}^{n+1}(-1)^{r}b_r(M)$ of $M$ can be equivalently written as \cite{Rosenberg, Gilkey}

\begin{equation}
\chi (M)=\sum _{r=0}^{n+1}(-1)^{r}a_{n+1}(\Delta_r^M).\label{Euler}
\end{equation}
From the definition of the determinant \eqref{determinant-zeta} it follows that for any parameter $\lambda\in\mathbb{R}$

\begin{equation}
\det{(\lambda\Delta_r^M)}=\lambda^{\zeta (0;\Delta_r^M)}\det{\Delta_r^M}.\label{determinant-parameter}
\end{equation}
In analogy with the finite dimensional case, $\zeta (0;\Delta_r^M)$ can thus be interpreted as regularized dimension.\par

From a physical perspective, however, the zeta function regularization seems to have a drawback because the eigenvalues of the Laplace operator have a mass dimension leading to
a partition function which is not dimensionless. By including a scale factor $\mu$ of mass-dimension $[\mu ]=1$ this can be restored. Thus instead of \eqref{determinant-zeta} we
take the dimensionless but scale dependent zeta function $\zeta _{\mu} (s;\Delta_r^M)):=\sum _{\nu _{\alpha}(\Delta_r^M))\neq 0}(\mu ^{-2}\nu _{\alpha}(\Delta_r^M)))^{-s}$.
Hence $\zeta _{\mu} (s;\Delta_r^M))=\mu^{2s}\zeta (s;\Delta_r^M))$. Consequently, this leads to the scale dependent but dimensionless determinant

\begin{equation}
\det _{\mu}{\Delta _r^{M}}:=\exp{\left[-\frac{d}{ds}|_{s=0}\ \zeta _{\mu} (s;\Delta _r^{M})\right]},\label{zeta-modified}
\end{equation}
which is related to \eqref{determinant-zeta} by $\det _{\mu}{\Delta _r^{M}}=\mu^{-2\zeta (0;\Delta_r^M)}\det{\Delta _r^{M}}$. Despite this replacement the partition function is
not just yet dimensionless. This traces back to the fact that the gauge fields have dimension and thus the metric $h_{M}^{(r)}$ appearing in \eqref{VEV-4-VEV} admits a dimension
as well which must be corrected accordingly. Following the usual convention we fix the mass dimension of the $p$-form gauge field $A$ in $n+1$ dimensions by $[A]=\frac{n-1}{2}$,
so that the corresponding field strength $F_A=dA$ admits mass dimension $[F_{A}]=\frac{n+1}{2}$. Since higher-abelian gauge fields generalize $p$-form gauge fields, we assign
mass dimension $[\delta_{1}(\hat{u})]=\frac{n+1}{2}$ to the field strength $\delta_{1}(\hat{u})$. The harmonic forms are regarded as specific $p$-form gauge fields so that they
are assigned the same mass dimension as $A$. Hence $[(\rho_{M}^{(p)})_k]=\frac{n-1}{2}$. Consequently, $[(\rho_{M}^{(p+1)})_l]=\frac{n+1}{2}$ for all $k,l$. In fact we choose
$[(\rho_{M}^{(r)})_k]=\frac{n-1}{2}+r-p$. In order to obtain a dimensionless partition function we have to replace

\begin{equation}
\det\Delta _r^M\mapsto\det _{\mu}\Delta _r^{M}\quad h_M^{(r)}\mapsto h_{M;\mu}^{(r)}=\mu^{2(p+1-r)}h_M^{(r)}.\label{dimensionless-jacobian}
\end{equation}
It follows from the heat kernel expansion \eqref{seeley1} that the Seeley coefficients of the dimensionless Laplace operator $\mu^{-2}\Delta _r^{M}$ are related to those of
$\Delta _r^{M}$ by (see also \cite{BVW})

\begin{equation}
a_{k}(\mu^{-2}\Delta_{r}^{M})=\mu^{n+1-k}\ a_{k}(\Delta_{r}^{M}).\label{dimensionless-Seeley}
\end{equation}
Due to the Hodge decomposition and the nilpotency of $d_{r}^{M}$ and $(d_{r}^{M})^{\dag}$ the spectrum of $\Delta_{r}^{M}|_{\mathcal{H}^{r}(M)^{\bot}}$ is the union of the
eigenvalues of $\Delta _{r}^{M}|_{im(d_{r+1}^{M})^{\dag}}$ and $\Delta _{r}^{M}|_{im(d_{r-1}^{M})}$. On the other hand, if $\psi_{M}^{(r-1)}\in im(d_{r}^{M})^{\dag}$ is an
eigenform of $\Delta _{r-1}^{M}|_{im(d_{r}^{M})^{\dag}}$ with eigenvalue $\nu$, then $d_{r-1}^{M}\psi_{M}^{(r-1)}$ is an eigenform of $\Delta _{r}^{M}|_{im(d_{r-1}^{M})}$ with
the same eigenvalue. This correspondence gives a bijection between the sets of non-vanishing eigenvalues and their related eigenforms of the two operators $\Delta
_{r-1}^{M}|_{im(d_{r}^{M})^{\dag}}$ and $\Delta _{r}^{M}|_{im(d_{r-1}^{M})}$. Hence the spectrum of $\Delta_{r}^{M}|_{\mathcal{H}^{r}(M)^{\bot}}$ is the union of eigenvalues of
$\Delta _{r}^{M}|_{im(d_{r+1}^{M})^{\dag}}$ and $\Delta _{r-1}^{M}|_{im(d_{r}^{M})^{\dag}}$. This leads to the following factorization

\begin{equation}
\zeta (s;\Delta_{r}^{M}|_{\mathcal{H}^{r}(M)^{\perp}})=\zeta (s;\Delta_{r}^{M}|_{im(d_{r+1}^{M})^{\dag}})+\zeta
(s;\Delta_{r-1}^{M}|_{im(d_{r}^{M})^{\dag}}),\label{spectrum-split}
\end{equation}
yielding

\begin{equation}
\zeta (s;\Delta_{r}^{M}|_{im(d_{r+1}^{M})^{\dag}})=\sum_{r^{\prime}=0}^{r}(-1)^{r-r^{\prime}}\zeta
(s;\Delta_{r^{\prime}}^{M}|_{\mathcal{H}^{r^{\prime}}(M)^{\perp}}),\label{zeta-relation}
\end{equation}
and finally

\begin{equation}
\sum_{r=0}^{p-1} (-1)^{p-r}\zeta (s;\Delta_{r}^{M}|_{im(d_{r+1}^{M})^{\dag}})=\sum_{r=0}^{p} (-1)^{p-r}(p-r)\zeta
(s;\Delta_{r}^{M}|_{\mathcal{H}^{r}(M)^{\perp}}).\label{relation-determinant}
\end{equation}
Applying the dimensional replacement to the ghost-for-ghost term and using \eqref{relation-determinant}, we obtain from \eqref{VEV-4-VEV} the following general formula for the
dimensionless and regularized partition function for a higher-abelian gauge theory governed by the classical action $S(\hat u_{p})$,

\begin{equation}
\begin{split}
\mathcal{Z}_{M}^{(p)}= & \prod_{r=0}^{p}\left(\frac{\left(\det_{\mu}\Delta_{r}^{M}|_{\mathcal{H}^{r}(M)^{\perp}}\right)^{\frac{1}{2}(p-r)}\
|H_{tor}^{r}(M;\mathbb{Z})|}{\left(\det{(\frac{1}{2\pi}h_{M;\mu}^{(r)})}\right)^{\frac{1}{2}}}\right)^{(-1)^{p+1-r}}
\\ &\times \sum_{\frak c_{M}^{(p+1)}\in H^{p+1}(M;\mathbb{Z})}\int_{im(d_{p+1}^{M})^{\dag}}\mathcal{D}\tau_{p}
\ \left[ e^{-\underline{S}^{\hat{\frak c}_M^{(p+1)}}(\tau_{p})}\right] _{(0,\ldots ,0)}.\label{partition-function-general}
\end{split}
\end{equation}
According to \eqref{dimensionless-jacobian} the scale factor which renders the alternating product in \eqref{partition-function-general} dimensionless is given by

\begin{equation}
\mu^{\sum_{r=0}^{p}(-1)^{p-r}(p-r)a_{n+1}(\Delta_{r}^{M})+\sum_{r=0}^{p}(-1)^{p-r}b_{r}(M)}.
\end{equation}
Finally, the formula for the VEV of an observable $\mathcal{O}$ can be stated in the following general form

\begin{equation}
<\mathcal{O} >=\frac{\sum_{\frak c_{M}^{(p+1)}\in H^{p+1}(M;\mathbb{Z})}\int_{im(d_{p+1}^{M})^{\dag}}\mathcal{D}\tau_{p}\ \left[e^{-\underline{S}^{\hat{\frak
c}_M^{(p+1)}}(\tau_{p} )}\ \underline{\mathcal{O}}^{\hat{\frak c}_M^{(p+1)}}(\tau_{p})\right]_{(0,\ldots ,0)}}{\sum_{\frak c_{M}^{(p+1)}\in
H^{p+1}(M;\mathbb{Z})}\int_{im(d_{p+1}^{M})^{\dag}} \mathcal{D}\tau_{p} \ \left[ e^{-\underline{S}^{\hat{\frak c}_M^{(p+1)}}(\tau_{p})}\right] _{(0,\ldots ,0)}}.
\label{VEV-general}
\end{equation}
Once the concrete action and the observable are selected, the dimensional replacement \eqref{dimensionless-jacobian} must be taken into account when performing the
$\tau_{p}$-integration in \eqref{partition-function-general} or \eqref{VEV-general} in order to obtain the correct dimension. This might lead to additional $\mu$ dependent
terms.\par

\subsection{The partition function of extended higher-abelian Maxwell theory}

In this section we introduce a concrete field theoretical model by extending the Maxwell theory of electromagnetism to higher degrees. Let us begin with the so-called
\textit{higher-abelian (generalized) Maxwell theory} \cite{Szabo, Moore, kelnhofer-1, FMS1, FMS2} of degree $p$ with action

\begin{equation}
S_{0}(\hat u_{p})=\frac{q_p}{2}\|\delta _1(\hat u_{p})\|^2=\frac{q_p}{2}\int _{M}\delta _1(\hat u_{p})\wedge\star\delta _1(\hat u_{p}),\qquad\hat u_{p}\in\widehat{H}^{p}(M).
\label{generalized p-form}
\end{equation}
The coupling constant $q_p$ is taken to be dimensionless. A Hamiltonian analysis of this model was given in \cite{FMS1, FMS2} and the functional integral quantization was
studied in \cite{kelnhofer-1}.\par

Now we extend \eqref{generalized p-form} by including two arbitrary harmonic forms $\gamma_{p}\in\mathcal{H}^{p+1}(M)$ and $\theta_{p}\in\mathcal{H}^{n-p}(M)$ representing two
de Rham cohomology classes on $M$ and define

\begin{equation}
S_{(\gamma_{p},\theta_{p})}(\hat u_{p}):=\frac{q_p}{2}\|\delta _1(\hat u_{p})\ - \gamma_{p} \|^2+2\pi i<\delta_1(\hat u_{p}) -\gamma_{p},\star\ \theta_{p}>.\label{topological
action}
\end{equation}
We call this model \textit{extended higher-abelian Maxwell theory} of degree $p$. These two harmonic forms - named topological fields - are regarded as non-dynamical (external)
fields. Remark that the classical action is invariant under the joint transformation $\hat u_{p}\mapsto\hat u_{p}\cdot\hat v_{p}$ and
$\gamma_{p}\mapsto\gamma_{p}+\delta_{1}(\hat v_{p})$ for all $\hat v_{p}\in\widehat{H}^{p}(M)$\footnote{A more detailed discussion of this model, its symmetries respectively
further extensions will be given elsewhere.}.\par

It follows from \eqref{cheeger-simons} that any tangent vector in $T_{\hat{u}_{p}}\widehat{H}^{p}(M)$ has the form $\hat{u}_{p}\cdot j_2([B])$ with $[B]\in\Omega
^p(M)/d\Omega^{p-1}(M)$. Since

\begin{equation}
\frac{\delta S_{0}}{\delta\hat{u}_{p}}[B]=\frac{d}{dt}|_{t=0}\ S_{0}(\hat u_{p}\cdot j_2[tB])=\frac{d}{dt}|_{t=0}\ S_{(\gamma_{p},\theta_{p})}(\hat u_{p}\cdot j_2[tB])=
\frac{\delta S_{(\gamma_{p},\theta_{p})}}{\delta\hat{u}_{p}}[B]
\end{equation}
both actions $S_0$ and $S_{(\gamma_{p},\theta_{p})}$ lead to the same classical equation of motion, namely

\begin{equation}
(d_{p+1}^{M})^{\dag}\delta _1(\hat u_{p})=0.
\end{equation}
Although classically equivalent, they give rise to different quantum theories. In the next step we will use the general formula \eqref{partition-function-general} to determine
the corresponding partition function. At first

\begin{equation}
\begin{split}
\underline{S}_{(\gamma_{p},\theta_{p})}^{\hat{\frak c}_M^{(p+1)}}(\vec{z}_{p},\tau_{p})=&\frac{q_p}{2}<\tau_{p} ,\Delta_{p}^{M}|_{im(d_{p+1}^{M})^{\dag}}\tau_{p}>+2\pi
i(-1)^{(p+1)(n-p)}\sum_{i=1}^{b_{p+1}(M)}(m_{p+1}^{i}-\gamma_{p} ^{i})\theta_{p} ^{i}\\
&+\frac{q_p}{2}\sum _{i,j=1}^{b_{p+1}(M)}(h_M^{(p+1)})_{ij}(m_{p+1}^{i}-\gamma_{p} ^{i})(m_{p+1}^{j}-\gamma_{p} ^{j}),\label{maxwell-action-adapted}
\end{split}
\end{equation}
where the components $\gamma_{p}^{j}$ and $\theta_{p}^{j}$ are defined with respect to the fixed basis of $\mathcal{H}^{\bullet}(M)$ by

\begin{equation}
\gamma_{p}=\sum_{i=1}^{b_{p+1}(M)}\gamma _{p}^{i}\ (\rho _M^{(p+1)})_i,\qquad\theta _{p}=\sum_{j=1}^{b_{n-p}(M)}\theta _{p}^{j}\ (\rho _M^{(n-p)})_j.
\end{equation}
Let us write $\vec{\gamma} _{p}=(\gamma _{p}^{1},\ldots,\gamma _{p}^{b_{p+1}(M)})$ and $\vec{\theta} _{p}=(\theta _{p}^{1},\ldots,\theta _{p}^{b_{n-p}(M)})$. The $\tau_{p}$
integration in \eqref{partition-function-general} is Gaussian and using the rule \eqref{dimensionless-jacobian} and \eqref{zeta-relation} one gets

\begin{multline} \int_{im(d_{p+1}^{M})^{\dag}}\mathcal{D}\tau_{p}\ e^{-\frac{q_p}{2}<\tau_{p},\Delta
_{p}^{M}|_{im(d_{p+1}^{M})^{\dag}}\tau_{p}>}=\\=\left(\frac{\mu}{\sqrt{q_p}}\right)^{\sum_{r=0}^{p}(-1)^{p-r}\zeta(0;\Delta_{r}^{M})} \prod_{r=0}^{p}
(\det\Delta_{r}^{M}|_{\mathcal{H}^{r}(M)^{\perp}})^{-\frac{1}{2}(-1)^{(p-r)}}.\label{tau-integral}
\end{multline}
The next step is to sum over the topological sectors. The sum over the free part results in a $b_{p+1}(M)$ dimensional Riemann Theta function \eqref{theta-original} and the sum
over the torsion classes yields the order of the $(p+1)$-th torsion subgroup. Substituting \eqref{tau-integral} into \eqref{partition-function-general} finally gives the
following dimensionless partition function for the extended higher-abelian Maxwell theory

\begin{equation}
\begin{split}
\mathcal{Z}_{M}^{(p)}&(q_p;\gamma_{p},\theta_{p})= \\ &=\left[\frac{2\pi}{q_p}\right]^{\frac{1}{2}\sum _{r=0}^{p}(-1)^{p+1-r}b_r(M)}\ \prod_{r=0}^{p}\left[\frac{(\det{\Delta
_{r}^{M}|_{\mathcal{H}^{r}(M)^{\perp}}})^{\frac{(p+1-r)}{2}}\ |H_{tor}^{r}(M;\mathbb{Z})|}{(\det{h_M^{(r)}})^{\frac{1}{2}}}\right]^{(-1)^{p+1-r}}
\\ &\times\Theta _{b_{p+1}(M)}\begin{bmatrix} \vec{\gamma}_{p} \\0\end{bmatrix}\left((-1)^{(p+1)(n-p)}\vec{\theta}_{p}|-\frac{q_p}{2\pi i}\ h_M^{(p+1)}\right)\
|H_{tor}^{p+1}(M;\mathbb{Z})|\\
&\times\mu^{\sum _{r=0}^{p}(-1)^{p-r}(p+1-r)a_{n+1}(\Delta_r^M)}\ q_p^{-\frac{1}{2}\sum _{r=0}^{p}(-1)^{p-r}a_{n+1}(\Delta_r^M)}.\label{partition function Maxwell}
\end{split}
\end{equation}
Due to the modular properties of the Riemann Theta function \eqref{modular-property}, the partition function exhibits the following apparent symmetry properties

\begin{equation}
\begin{split}
&\mathcal{Z}_{M}^{(p)}(q_p;\gamma_{p},\theta_{p})=\mathcal{Z}_{M}^{(p)}(q_p;-\gamma_{p},-\theta_{p})\\
&\mathcal{Z}_{M}^{(p)}(q_p;\gamma_{p}+\omega,\theta_{p})=\mathcal{Z}_{M}^{(p)}(q_p;\gamma_{p},\theta_{p}),
\quad\textrm{for $\omega\in\mathcal{H}_{\mathbb{Z}}^{p+1}(M)$}\\
&\mathcal{Z}_{M}^{(p)}(q_p;\gamma_{p},\theta_{p}+\omega^{\prime})=\mathcal{Z}_{M}^{(p)}(q_p;0,\theta_{p}),
\quad\textrm{for $\gamma_{p}\in\mathcal{H}_{\mathbb{Z}}^{p+1}(M)$, $\omega^{\prime}\in\mathcal{H}_{\mathbb{Z}}^{n-p}(M)$}\\
&\mathcal{Z}_{M}^{(p)}(q_p;\gamma_{p},\theta_{p})=\mathcal{Z}_{M}^{(p)}(q_p;0,0)=:\mathcal{Z}_{M}^{(p)}(q_p),\quad\textrm{for $\gamma_{p}\in\mathcal{H}_{\mathbb{Z}}^{p+1}(M)$,
$\theta_{p}\in\mathcal{H}_{\mathbb{Z}}^{n-p}(M)$}.\label{symmetry-Maxwell}
\end{split}
\end{equation}
In odd dimensions one has $a_{n+1}(\Delta _{r}^{M})=0$, so that the partition function becomes independent of the renormalization scale. Since $a_{n+1}(\Delta _{r}^{M})$ is an
integral over a local polynomial in the metric and its derivatives in even dimensions, the last factor in \eqref{partition function Maxwell} can be absorbed into the action of
the theory. Within a local quantum field theory this amounts to add appropriate gravitational counter-terms. So instead of $\mathcal{Z}_{M}^{(p)}(q_p;\gamma_{p},\theta_{p})$ we
regard

\begin{equation}
\widehat{\mathcal{Z}}_{M}^{(p)}(q_p;\gamma_{p} ,\theta_{p}):= q_p^{\frac{1}{2}\sum _{r=0}^{p}(-1)^{p-r}a_{n+1}(\Delta_r^M)}\ \mathcal{Z}_{M}^{(p)}(q_p;\gamma_{p}
,\theta_{p})\label{reduced partition function}
\end{equation}
as the effective partition function for extended higher-abelian Maxwell theory. Due to \eqref{dimensionless-Seeley} $\widehat{\mathcal{Z}}_{M}^{(p)}$ is dimensionless. The
explicit dependence on $\mu$ on the other hand indicates that there is nevertheless an ambiguity which has to be fixed to obtain a physically reasonable result.\par

In the case $\gamma_{p}=\theta_{p}=0$, our result for the partition function \eqref{reduced partition function} agrees with the dimensionless partition function for
higher-abelian Maxwell theory obtained recently in \cite{DMW}. \par

\subsection{Duality in extended higher-abelian Maxwell theory}

Duality between quantized antisymmetric tensor field theories of different degrees and its relation with the Ray-Singer analytic torsion (see below for the definition) were
analyzed long time ago by Schwarz in his seminal papers \cite{Schwarz, ST}. That there exists a duality in higher-abelian Maxwell theory with action \eqref{generalized p-form}
was already sketched in \cite{Moore} by means of a specific master equation. The case of acyclic manifolds was discussed in \cite{kelnhofer-1}.\par

Does there exist a duality between extended higher-abelian Maxwell theories of different degrees as well? Let $(q_p,\gamma_{p},\theta_{p})$ be the parameters of the theory of
degree $p$ and define the corresponding dual parameters $(q_{n-p-1}^{dual},\gamma_{n-p-1}^{dual},\theta_{n-p-1}^{dual})$ by

\begin{enumerate}
    \item $q_{n-p-1}^{dual}\cdot q_{p}=(2\pi)^{2}$
    \item $\vec{\gamma}_{n-p-1}^{dual}=(-1)^{(p+1)(n-p)}\vec{\theta}_{p}$
    \item $\vec{\theta}_{n-p-1}^{dual}=-\vec{\gamma}_{p}$.
\end{enumerate}
In the following we will compute the quotient of partition functions

\begin{equation}
\frac{\widehat{\mathcal{Z}}_{M}^{(p)}(q_p;\gamma_{p},\theta_{p})}{\widehat{\mathcal{Z}}_{M}^{(n-p-1)}(q_{n-p-1}^{dual},\gamma_{n-p-1}^{dual},\theta_{n-p-1}^{dual})}
\label{duality_general}
\end{equation}
factor per factor using the expression \eqref{reduced partition function}. We begin with the alternating product in the partition function \eqref{partition function Maxwell},
abbreviated by

\begin{equation}
\mathcal{G}^{(p)}:=\prod_{r=0}^{p}\left[\frac{(\det{\Delta _{r}^{M}|_{\mathcal{H}^{r}(M)^{\perp}}})^{\frac{(p+1-r)}{2}}\
|H_{tor}^{r}(M;\mathbb{Z})|}{(\det{h_M^{(r)}})^{\frac{1}{2}}}\right]^{(-1)^{p+1-r}}.
\end{equation}
Since $b_{p+1}(M)=b_{n-p}(M)$ and $\star (\rho_{M}^{(p)})_i=\sum _{j=1}^{b_p(M)}(h_{M}^{(p)})_{ij}(\rho_{M}^{(n+1-p)})_j$, the dual metrics are related by
$(h_M^{(p+1)})^{-1}=h_M^{(n-p)}$. Moreover, we use $H_{tor}^{r+1}(M;\mathbb{Z})\cong H_{tor}^{n+1-r}(M;\mathbb{Z})$ which follows from the universal coefficient theorem and
Poincare duality (see e.g. \cite{Bredon}). Together with the fact that the Hodge operator commutes with the Laplace operator, one finds after a lengthy calculation

\begin{equation}
\frac{\mathcal{G}^{(p)}}{\mathcal{G}^{(n-p-1)}} =\frac{\prod_{r=0}^{n+1}\left[(\det{h_M^{(r)}})^{\frac{1}{2}}\
|H_{tor}^{r}(M;\mathbb{Z})|^{-1}\right]^{(-1)^{r+p}}}{\prod_{r=0}^{n+1}(\det{\Delta _{r}^{M}|_{\mathcal{H}^{r}(M)^{\perp}}})^{\frac{r}{2}(-1)^{r+1+p}}} \left(\det
h_M^{(p+1)}\right)^{\frac{1}{2}}\frac{|H_{tor}^{n-p}(M;\mathbb{Z})|}{|H_{tor}^{p+1}(M;\mathbb{Z})|}. \label{ghost-for-ghost}
\end{equation}
Let us consider the alternating products in \eqref{ghost-for-ghost} in more detail: The product in the denominator is nothing but the Ray-Singer analytic torsion \cite{RS} of
$M$

\begin{equation}
\mathcal{T}_{RS}[M]=\prod_{r=0}^{n+1}(\det{\Delta _{r}^{M}|_{\mathcal{H}^{r}(M)^{\perp}}})^{\frac{r}{2}(-1)^{r+1}}=
\exp{\left(\frac{1}{2}\sum_{r=0}^{n+1}(-1)^{r}r\frac{d}{ds}|_{s=0}\zeta (s;\Delta _{r}^{M}|_{\mathcal{H}^{r}(M)^{\perp}})\right)}.\label{Ray_Singer_torsion}
\end{equation}
Ray and Singer introduced $\mathcal{T}_{RS}[M]$ as an analytic analogue of the Reidemeister-Franz torsion (or $R$-torsion), denoted by $\frak T_{R}[M]$, which is a topological
invariant of $M$. In fact, $\frak T_{R}[M]$ is defined in terms of the combinatorial structure determined by smooth triangulations of the manifold. According to \cite{Cheeger},
Theorem 8.35, (see e.g. \cite{BV} for a review of analytic and Reidemeister torsion) the $R$-torsion of a compact and closed manifold equals the alternating product of the order
of the torsion subgroups of integer cohomology times a so-called regulator. The regulator itself is an alternating product of the volumes of the harmonic tori
$H^{r}(M;\mathbb{R})/j_{\ast}(H^{r}(M;\mathbb{Z}))$, $r=0,\ldots,\dim M$ with respect to the metric induced by identifying real cohomology with harmonic forms. In our notation
this volume is given by \eqref{volume harmonic torus} so that the $R$-torsion admits the form

\begin{equation}
\frak T_{R}[M]=\prod_{r=0}^{n+1}(\det{h_M^{(r)}})^{\frac{1}{2}(-1)^{r}}\ \prod_{r=0}^{n+1}|H_{tor}^{r}(M;\mathbb{Z})|^{(-1)^{r+1}}.\label{Reidemeister torsion}
\end{equation}
Hence the $R$-torsion appears in the numerator of the first factor in \eqref{ghost-for-ghost}. It was conjectured by Ray and Singer and independently proved by Cheeger and
M\"{u}ller \cite{Cheeger, Muller} that the Ray Singer analytic torsion and the $R$-torsion are equal

\begin{equation}
\mathcal{T}_{RS}[M]=\frak T_{R}[M].\label{Cheeger-Muller}
\end{equation}
Thus the first fraction on the right hand side of \eqref{ghost-for-ghost} cancels. The quotient of the contributions from the topological sectors yields

\begin{multline}
\frac{\Theta _{b_{p+1}(M)}\begin{bmatrix} \vec{\gamma}_{p} \\0\end{bmatrix}\left((-1)^{(p+1)(n-p)}\vec{\theta}_{p}|-\frac{q_p}{2\pi i}\
h_M^{(p+1)}\right)|H_{tor}^{p+1}(M;\mathbb{Z})|}{\Theta _{b_{n-p}(M)}\begin{bmatrix} \vec{\gamma}_{n-p-1}^{dual}
\\0\end{bmatrix}\left(\vec{\theta}_{n-p-1}^{dual}|-\frac{q_{n-p-1}^{dual}}{2\pi i}\ h_M^{(n-p)}\right)|H_{tor}^{n-p}(M;\mathbb{Z})|}=
\\= \left[\frac{2\pi}{q_p}\right]^{\frac{1}{2}b_{p+1}(M)}(\det
h_{M}^{(p+1)})^{-\frac{1}{2}}\frac{|H_{tor}^{p+1}(M;\mathbb{Z})|}{|H_{tor}^{n-p}(M;\mathbb{Z})|}\ e^{2\pi i (-1)^{(p+1)(n-p)}\vec{\gamma}_{p}^{T}\vec{\theta}_{p}},
\label{Theta_duality-1}
\end{multline}
where we have used \eqref{decomposition} and \eqref{duality-1}. Evidently, \eqref{duality_general} becomes independent of $\mu$, if the dimension of $M$ is odd. In even
dimensions, one has $a_{n+1}(\Delta_{r}^{M})=a_{n+1}(\Delta_{n+1-r}^{M})$ which gives

\begin{equation}
\sum_{r=0}^{n+1}(-1)^{p+1-r}ra_{n+1}(\Delta_{r}^{M})=(-1)^{p+1}\frac{n+1}{2}\chi(M).\label{duality renormalization}
\end{equation}
From \eqref{ghost-for-ghost}, \eqref{Cheeger-Muller}, \eqref{Theta_duality-1} and \eqref{duality renormalization} we finally get the following duality relation

\begin{equation}
\begin{split}
&\frac{\widehat{\mathcal{Z}}_{M}^{(p)}(q_p;\gamma_{p},\theta_{p})}{\widehat{\mathcal{Z}}_{M}^{(n-p-1)}(q_{n-p-1}^{dual},\gamma_{n-p-1}^{dual},\theta_{n-p-1}^{dual})} =\\
\\&=\begin{cases} \left[\frac{q_p}{2\pi}\right]^{\frac{1}{2}(-1)^{p}\chi (M)}\ e^{2\pi i \int_{M}\theta_{p}\wedge\gamma_{p}}\ \mu^{(-1)^{p}\left(p+1-\frac{n+1}{2}\right)\chi
(M)}, &\text{if $\dim M=n+1$ is even},
\\ \\
e^{2\pi i \int_{M}\theta_{p}\wedge\gamma_{p}} &\text{if $\dim M=n+1$ is odd}.\end{cases}\label{duality_special_1}
\end{split}
\end{equation}
The quotient is independent of the metric of $M$. In odd dimensions, the two extended higher-abelian Maxwell theories are exactly dual whenever the topological fields have
integer periods. In the case of vanishing topological fields (i.e. $\gamma_{p}=\theta_{p}=0$) we recover once again the result previously derived in \cite{DMW}.\par

\section{Thermodynamics of higher-abelian gauge theories}

\subsection{The free energy - general case}

In the remainder of this paper we want to apply the general setting introduced in the previous sections in order to study the effect of the topology on the vacuum structure and
the thermodynamical behavior of higher-abelian gauge theories. Geometrically, a quantum field theory at finite temperature $1/\beta$ at equilibrium is usually realized as
quantum field theory on the product manifold $X_{\beta}:=\mathbb{T}_{\beta}^1\times X$. The temperature is included by equipping a $1$-torus, denoted by $\mathbb{T}_{\beta}^1$,
with temperature dependent metric $g_{\beta}= \beta ^2dt\otimes dt$. Here $t$ is the local coordinate. The $n$-dimensional compact, connected, oriented and closed Riemannian
manifold $X$ with fixed metric $g_X$ is the spatial background and $\widehat{H}^{p}(X_{\beta})$ represents the space of equivalence classes of higher-abelian thermal gauge
fields.\par

We begin with the derivation of a general formula for the free energy of a higher-abelian gauge theory with Euclidean action $S=S(\hat{u}_{p})$, where
$\hat{u}_{p}\in\widehat{H}^{p}(X_{\beta})$. In the second step we will apply this general formula to extended higher-abelian Maxwell theory.\par

According to section 2.1, the corresponding configuration space $\Omega^{p}(X_{\beta})$ of topologically trivial thermal gauge fields is a non-trivializable principal
$\Omega_{\mathbb{Z}}^{p}(X_{\beta})$-bundle over the gauge orbit space $\Omega^{p}(X_{\beta})/\Omega_{\mathbb{Z}}^{p}(X_{\beta})$, which itself is a trivializable vector bundle
over $\mathbb{T}^{b_p(X_{\beta})}$ with typical fiber $im(d_{p+1}^{X_{\beta}})^{\dag}$. The corresponding family of local trivializations is provided by the diffeomorphisms
$\varphi_{a}^{(p)}\circ\underline{\varphi}_{a}^{(p)}$ (see \eqref{triv_0} and \eqref{triv_2}), but now applied to the "thermal" manifold $M=X_{\beta}$.\par

The free energy of the higher-abelian gauge theory of degree $p$ is defined by

\begin{equation}
\mathcal{F}_{X}^{(p)}(\beta):=-\frac{1}{\beta}\ \ln\mathcal{Z}_{X_{\beta}}^{(p)},\label{free-energy-definition}
\end{equation}
where $\mathcal{Z}_{X_{\beta}}^{(p)}$ is the partition function \eqref{partition-function-general} related to the manifold $X_{\beta}$. In the following we will compute term by
term of \eqref{partition-function-general} for the thermal setting.\par

Let us equip $X_{\beta}$ with the product metric $g =g_{\beta}\oplus g_X$. The corresponding volume form $vol_{X_{\beta}}$ splits into the product

\begin{equation}
vol_{X_{\beta}}=pr_{1}^{\ast}vol_{\mathbb{T}_{\beta}^1}\wedge pr_{2}^{\ast}vol_{X},
\end{equation}
where $pr_{1}\colon X_{\beta}\rightarrow \mathbb{T}_{\beta}^1$ and $pr_{2}\colon X_{\beta}\rightarrow X$ are the canonical projections and $vol_{\mathbb{T}_{\beta}^1}$,
$vol_{X}$ are the volume forms on $\mathbb{T}_{\beta}^1$ and $X$, respectively. According to the K\"{u}nneth theorem there exists the decomposition

\begin{equation}
H^{r}(X_{\beta};\mathbb{Z})\cong H^{r}(X;\mathbb{Z})\oplus H^{r-1}(X;\mathbb{Z}).\label{kunneth}
\end{equation}
This implies a split of both the free part $H_{free}^{r}(X_{\beta};\mathbb{Z})\cong H_{free}^{r}(X;\mathbb{Z})\oplus H_{free}^{r-1}(X;\mathbb{Z})$ and the torsion part
$H_{tor}^{r}(X_{\beta};\mathbb{Z})\cong H_{tor}^{r}(X;\mathbb{Z})\oplus H_{tor}^{r-1}(X;\mathbb{Z})$. Hence $b_r(X_{\beta})=b_r(X)+b_{r-1}(X)$. A basis for
$\mathcal{H}_{\mathbb{Z}}^r(X_{\beta})$, denoted by $\{(\rho_{X_{\beta}}^{(r)})_I|I=(i,j)\}$, is generated by

\begin{equation}
\begin{split}
& (\rho_{X_{\beta}}^{(r)})_i :=pr_{2}^{\ast}(\rho_{X}^{(r)})_{i},\quad i=1,\ldots ,b_r(X)\\
& (\rho_{X_{\beta}}^{(r)})_j :=pr_{1}^{\ast}\rho_{\mathbb{T}_{\beta}^1}^{(1)}\wedge pr_{2}^{\ast}(\rho_{X}^{(r-1)})_{j},\quad j=1,\ldots ,b_{r-1}(X),\label{harmonic-basis-1}
\end{split}
\end{equation}
where $\rho_{\mathbb{T}_{\beta}^1}^{(1)}=\frac{1}{\beta}\ vol_{\mathbb{T}_{\beta}^1}$ is the generator for $\mathcal{H}_{\mathbb{Z}}^1(\mathbb{T}_{\beta}^1)$ and
$(\rho_{X}^{(r)})_{k}$ denotes a basis for $\mathcal{H}_{\mathbb{Z}}^{r}(X)$. The dual basis $\rho_{X_{\beta}}^{(n+1-r)}$ satisfying

\begin{equation}
\int_{X_{\beta}}\ (\rho_{X_{\beta}}^{(r)})_{I}\wedge (\rho_{X_{\beta}}^{(n+1-r)})_{J}=\delta_{IJ}
\end{equation}
is then given by

\begin{equation}
\begin{split}
& (\rho_{X_{\beta}}^{(n+1-r)})_j :=pr_{2}^{\ast}(\rho_{X}^{(n+1-r)})_{j},\quad j=1,\ldots ,b_{n+1-r}(X)\\
& (\rho_{X_{\beta}}^{(n+1-r)})_i :=(-1)^{r}pr_{1}^{\ast}\rho_{\mathbb{T}_{\beta}^1}^{(1)}\wedge pr_{2}^{\ast}(\rho_{X}^{(n-r)})_{i},\quad i=1,\ldots ,b_{n-r}(X).
\end{split}
\end{equation}
Let us denote the Hodge star operator on $X$ associated with metric $g_X$ by $\underline{\star}$. A direct calculation gives

\begin{equation}
\begin{split}
& \star pr_{1}^{\ast}\rho_{\mathbb{T}_{\beta}^1}^{(1)}=\frac{1}{\beta}\ pr_{2}^{\ast} vol_X\\
& \star pr_{2}^{\ast}(\rho_{X}^{(r)})_{i}=(-1)^r \beta\ pr_{1}^{\ast}\rho_{\mathbb{T}_{\beta}^1}^{(1)}\wedge pr_{2}^{\ast}\underline{\star} (\rho_{X}^{(r)})
_{i}\\
& \star \left(pr_{1}^{\ast}\rho_{\mathbb{T}_{\beta}^1}^{(1)}\wedge pr_{2}^{\ast}(\rho_{X}^{(r-1)})_{j}\right)=\frac{1}{\beta}\ pr_{2}^{\ast}\ \underline{\star}
(\rho_{X}^{(r-1)})_{j}.
\end{split}
\end{equation}
Therefore the induced metric $h_{X_{\beta}}^{(r)}$ on $\mathcal{H}^r(X_{\beta})$ admits the following matrix representation

\begin{equation}
h_{X_{\beta}}^{(r)}=\begin{pmatrix}
  \beta^{-1}h_{X}^{(r-1)} & 0 \\
  0 & \beta h_{X}^{(r)}
\end{pmatrix},\label{metric-temperature}
\end{equation}
of rank $b_{r-1}(X)+b_r(X)$, where $h_{X}^{(r-1)}$, $h_{X}^{(r)}$ are the induced metrics on $\mathcal{H}^{r-1}(X)$ and $\mathcal{H}^{r}(X)$, respectively. Due to
\eqref{kunneth} every class $\frak c_{X_{\beta}}^{(p+1)}\in H^{p+1}(X_{\beta};\mathbb{Z})$ has the following (non-canonical) decomposition with integer components

\begin{equation}
\frak c_{X_{\beta}}^{(p+1)}= \left(\sum _{i=1}^{b_{p+1}(X)}m_{p+1}^{i}(\frak f_X^{(p+1)})_i+\sum _{j=1}^{r_{p+1}(X)}\ \tilde{m}_{p+1}^{j}(\frak t_X^{(p+1)})_{j},\sum
_{k=1}^{b_{p}(X)}m_{p}^{k}(\frak f_X^{(p)})_k+\sum _{l=1}^{r_{p}(X)}\ \tilde{m}_{p}^{l}(\frak t_X^{(p)})_{l}\right),\label{cohomology-class-thermal}
\end{equation}
where the classes $(\frak f_X^{(p+1)})_i$ and $(\frak f_X^{(p)})_k$ provide a Betti basis for $H_{free}^{p+1}(X_{\beta};\mathbb{Z})$. The generators for
$H_{tor}^{p+1}(X_{\beta};\mathbb{Z})$ are denoted by $(\frak t_X^{(p+1)})_{j}$ and $(\frak t_X^{(p)})_{l}$, respectively. Finally, we choose a family of harmonic background
differential characters $\hat{\frak c}_{X_{\beta}}^{(p+1)}\in\widehat{\mathcal{H}}^{p}(X_{\beta})$ with characteristic classes $\delta _2(\hat{\frak
c}_{X_{\beta}}^{(p+1)})=\frak c_{X_{\beta}}^{(p+1)}$ and field strengths

\begin{equation}
\delta_{1}(\hat{\frak c}_{X_{\beta}}^{(p+1)})=\sum_{j=1}^{b_p(X)}m_{p}^{j}\ pr_{1}^{\ast}\rho_{\mathbb{T}_{\beta}^1}^{(1)}\wedge
pr_{2}^{\ast}(\rho_{X}^{(p)})_{j}+\sum_{k=1}^{b_{p+1}(X)}m_{p+1}^{k}\ pr_{2}^{\ast}(\rho_{X}^{(p+1)})_{k}.\label{cohomology-thermal}
\end{equation}
By using \eqref{zeta-modified}, \eqref{metric-temperature} and the K\"{u}nneth theorem for the torsion subgroup of $X_{\beta}$, one obtains for the free energy the following
formula

\begin{equation}
\begin{split}
\mathcal{F}_{X}^{(p)}(\beta)= & \frac{1}{2\beta}\sum_{r=0}^{p}(-1)^{p+1-r}(p-r)\zeta^{\prime}(0;\Delta_{r}^{X_{\beta}}|_{\mathcal{H}^{r}(X_{\beta})^{\perp}})-
\frac{1}{2\beta}\ln{\det h_{X}^{(p)}}+\frac{b_p(X)}{2\beta}\ln 2\pi\\
& -\frac{1}{2\beta}\left(\sum_{r=0}^{p}(-1)^{p-r}(p-r)a_{n+1}(\Delta_{r}^{X_{\beta}})+b_p(X)\right) \ln\mu^{2}\\
&+\frac{1}{\beta}\left(\frac{b_p(X)}{2}+\sum_{r=0}^{p}(-1)^{(p+1-r)}b_{r}(X)\right)\ln\beta +\frac{1}{\beta}\ln |H_{tor}^{p}(X;\mathbb{Z})|\\
&-\frac{1}{\beta}\ln \left(\sum_{\frak c_{X_{\beta}}^{(p+1)}\in H^{p+1}(X_{\beta};\mathbb{Z})}\int_{im(d_{p+1}^{X_{\beta}})^{\dag}}\mathcal{D}\tau_{p}\
\left[e^{-\underline{S}^{\hat{\frak c}_{X_{\beta}}^{(p+1)}}(\tau_{p} )}\right]_{(0,\ldots ,0)}\right).\label{free-energy-general-zeta}
\end{split}
\end{equation}
As before the sum over the topological sectors is understood as sum over the integer components of $\frak c_{X_{\beta}}^{(p+1)}$. This general formula is the starting point to
derive expressions which are adequate for the low- and the high-temperature regimes, respectively.\par

\subsubsection{Low-temperature regime}

In the next step the Laplace operators appearing in \eqref{free-energy-general-zeta} are rewritten in terms of the geometry of $X$. The subspace $\bigoplus_{\substack{r+s=p}}\
\left( pr_1^{\ast}\Omega^{r}(\mathbb{T}_{\beta}^{1})\otimes pr_2^{\ast}\Omega^{s}(X) \right)$ of $\Omega^{p}(X_{\beta})$ is dense. With respect to the product metric $g$, the
Laplace operator $\Delta _p^{X_{\beta}}$ splits into

\begin{equation}
\Delta _{p}^{X_{\beta}}=\bigoplus_{\substack{r+s=p}}\left( \Delta _r^{\mathbb{T}_{\beta}^1}\otimes 1+1\otimes\Delta _{s}^{X}\right).
\end{equation}
Let $\nu _{\alpha}(\Delta _s^{X})$ denote the eigenvalues of $\Delta _s^{X}$. The eigenvalues of $\Delta _r^{\mathbb{T}_{\beta}^1}$ for $r=0,1$ are $\nu _{k}(\Delta
_r^{\mathbb{T}_{\beta}^1})=(\frac{2\pi k}{\beta})^2$  with $k\in\mathbb{Z}$. Hence the spectrum of $\Delta _p^{X_{\beta}}|_{\mathcal{H}^p(X_{\beta})^{\perp}}$ is given by the
following set of non-vanishing real numbers

\begin{equation}
\begin{split}
& Spec(\Delta _p^{X_{\beta}}|_{\mathcal{H}^p(X_{\beta})^{\perp}})= \\
&=\{\nu _{(k,\alpha)}^{(r,s)}:=\nu _{k}(\Delta _r^{\mathbb{T}_{\beta}^1})+\nu _{\alpha}(\Delta _s^{X})\neq 0|r+s=p,\ (k,\alpha)\in \mathbb{Z}\times I\}.
\end{split}\label{spectrum}
\end{equation}
The zeta function of $\Delta _r^{X_{\beta}}$ admits the following representation in terms of these eigenvalues

\begin{equation}
\zeta (s;\Delta _r^{X_{\beta}}|_{\mathcal{H}^r(X_{\beta})^{\perp}})=
\begin{cases} \sum _{(k,\alpha)\in \mathbb{Z}\times I}\left[\nu
_{(k,\alpha)}^{(0,0)}\right]^{-s}, &\text{if $r=0$},\\ \sum _{(k,\alpha)\in \mathbb{Z}\times I}\left[\nu _{(k,\alpha)}^{(0,r)}\right]^{-s}+\sum _{(k,\alpha)\in \mathbb{Z}\times
I}\left[\nu _{(k,\alpha)}^{(1,r-1)}\right]^{-s}, &\text{if $r \geq1$},\end{cases}\label{zeta-general}
\end{equation}
where each eigenvalue appears as often as its multiplicity. The series converges for $\Re(s)>\frac{n+1}{2}$. Let us introduce the following auxiliary quantity

\begin{equation}
\mathcal{I}(s;\Delta_r^X):=\sum _{(k,\alpha )\in (\mathbb{Z}\times I)^{\prime}}\left[ (\frac{2\pi k}{\beta})^2+ \nu _{\alpha}(\Delta_r^X)\right]^{-s},\label{zeta-quantity-1}
\end{equation}
where the prime indicates that the sum runs over all those pairs $(k,\alpha )$ such that the sum $(\frac{2\pi k}{\beta})^2+\nu _{\alpha}(\Delta_r^X)$ does not vanish. Thus

\begin{equation}
\zeta (s;\Delta _r^{X_{\beta}}|_{\mathcal{H}^r(X_{\beta})^{\perp}})=
\begin{cases} \mathcal{I}(s;\Delta _0^{X}), &\text{if
$r=0$},\\\mathcal{I}(s;\Delta _r^{X})+\mathcal{I}(s;\Delta _{r-1}^{X}) \, &\text{if $r\geq1$}.\end{cases}\label{zeta-general-2}
\end{equation}
Let us now introduce an additional auxiliary quantity to separate the zero-modes

\begin{equation}
\hat{\mathcal{I}}(s;\Delta_r^X):=\sum _{k\in\mathbb{Z}}\sum _{\alpha\in I}\left[ (\frac{2\pi k}{\beta})^2+\nu _{\alpha}(\Delta
_r^{X}|_{\mathcal{H}^r(X)^{\perp}})\right]^{-s}.\label{zeta-quantity-2}
\end{equation}
Hence

\begin{equation}
\mathcal{I}(s;\Delta_r^X)=\hat{\mathcal{I}}(s;\Delta_r^X)+2b_r(X)\ \left(\frac{\beta}{2\pi}\right)^{2s}\zeta _R(2s),\label{zeta-reduction-1-2}
\end{equation}
where $\zeta _R(s)$ is the Riemann zeta function. Applying the Mellin transform to $\hat{\mathcal{I}}(s;\Delta_r^X)$, one gets

\begin{equation}
\begin{split}
\hat{\mathcal{I}}(s;\Delta_r^X) &=\frac{1}{\Gamma (s)}\int _0^{\infty}dt\ t^{s-1}\sum _{k\in\mathbb{Z}}e^{-(\frac{2\pi}{\beta})^2k^2t}\sum _{\alpha\in I}e^{-\nu _{\alpha}(\Delta
_r^{X}|_{\mathcal{H}^r(X)^{\perp}})t}\\ &= \frac{\beta}{2\sqrt{\pi}\Gamma (s)}\int
_0^{\infty}dt\ t^{s-\frac{3}{2}}\sum _{\alpha\in I}e^{-\nu _{\alpha}(\Delta _r^{X}|_{\mathcal{H}^r(X)^{\perp}})t}\\
&\ \ \ +\frac{\beta}{\sqrt{\pi}\Gamma (s)}\int _0^{\infty}dt\ t^{s-\frac{3}{2}}\sum _{k=1}^{\infty}e^{-(\frac{k^2\beta ^2}{4t})}\sum _{\alpha\in I}e^{-\nu _{\alpha}(\Delta
_r^{X}|_{\mathcal{H}^r(X)^{\perp}})t}\\ &= \frac{\beta}{2\sqrt{\pi}}\frac{\Gamma (s-{\frac{1}{2}})}{\Gamma (s)}\zeta (s-{\frac{1}{2}};\Delta_r^X)\\ &\ \ \
+\frac{2^{-s-\frac{3}{2}}\beta ^{s+\frac{1}{2}}}{\sqrt{\pi}\Gamma (s)}\sum _{k\in\mathbb{Z}}\sum _{\alpha\in I}\left[\frac{k}{\sqrt{\nu _{\alpha}(\Delta
_r^{X}|_{\mathcal{H}^r(X)^{\perp}})}}\right]^{s-\frac{1}{2}}K_{s-\frac{1}{2}}(k\beta\sqrt{\nu _{\alpha}(\Delta
_r^{X}|_{\mathcal{H}^r(X)^{\perp}})}),\label{expansion-zeta-function}
\end{split}
\end{equation}
where $K_{\nu}(s)$ is the modified Bessel function of the second kind \cite{gradshteyn}. Let us expand $\zeta (s;\Delta_r^X)$ in a Laurent series at $s=-\frac{1}{2}$

\begin{equation}
\zeta (s-\frac{1}{2};\Delta_r^X)=\frac{Res_{s=-\frac{1}{2}}\left[ \zeta (s;\Delta_r^X)\right]}{s}+FP_{s=-\frac{1}{2}}\left[\zeta (s;\Delta_r^X)\right]+\sum
_{k=1}^{\infty}\tilde{\sigma} _ks^k,\label{zeta-expansion}
\end{equation}
where the finite part $FP$ of the zeta function at $s=-\frac{1}{2}$ is given by

\begin{equation}
FP_{s=-\frac{1}{2}}[\zeta (s;\Delta_r^X)]=\lim _{\epsilon\rightarrow 0}\ \frac{1}{2}\left[\zeta (-\frac{1}{2}-\epsilon;\Delta_r^X)+\zeta
(-\frac{1}{2}+\epsilon;\Delta_r^X)\right].\label{finite-part}
\end{equation}
To determine the residue of the zeta function, we recall from section 2.2 (but here for the $n$-dimensional manifold $X$) that $\zeta (s;\Delta_r^X)$ has simple poles at
$s_{k}=\frac{n-k}{2}$. Hence the residue at $s=-\frac{1}{2}$ is given by

\begin{equation}
Res_{s=-\frac{1}{2}}\left[\zeta (s;\Delta_r^X)\right]=-\frac{a_{n+1}(\Delta_r^X)}{2\sqrt{\pi}}.\label{res-def}
\end{equation}
In order to compute the derivative $\mathcal{I}^{\prime}(s;\Delta_r^X)$ at $s=0$, we make use of the expansions $\frac{1}{\Gamma (s)}=s+\gamma s^2+\mathcal{O}(s^3)$ and $\Gamma
(s-\frac{1}{2})=\Gamma (-\frac{1}{2})(1+\Psi (-\frac{1}{2})s+\mathcal{O}(s^2))$ for small $s$, where $\gamma$ is the Euler-Mascheroni number and $\Psi$ is the Digamma function
\cite{gradshteyn}. Since $\lim_{s\rightarrow 0}\frac{d}{ds}(\frac{h(s)}{\Gamma (s)})=h(0)$ holds for any regular function $h$\footnote{This follows from
$\frac{\Gamma^{\prime}(s)}{\Gamma(s)}=-\gamma-\frac{1}{s}-\sum_{m=1}^{\infty}\left[\frac{1}{s+m}-\frac{1}{m}\right]$ and $\lim_{s\rightarrow
0}\frac{\Gamma^{\prime}(s)}{\Gamma(s)^{2}}=-1$.}, one obtains

\begin{equation}
\begin{split}
\mathcal{I}^{\prime}(0;\Delta_r^X)= & -2b_r(X)\ln\beta -2\sum _{\alpha\in I}\ln \left[ 1-e^{-\beta\sqrt{\nu _{\alpha}(\Delta _r^{X}|_{\mathcal{H}^r(X)^{\perp}})}}\right]\\
& -\beta \left( FP_{s=-\frac{1}{2}}\left[ \zeta (s;\Delta_r^X)\right]+2(1-\ln 2)Res_{s=-\frac{1}{2}}\left[ \zeta (s;\Delta_r^X)\right]\right).\label{I}
\end{split}
\end{equation}
The sum of the second and third term of \eqref{I} is a function $f$ of the eigenvalues of $\Delta_r^X|_{\mathcal{H}^r(X)^{\perp}}$. If written schematically as
$f=f(\Delta_r^X|_{\mathcal{H}^r(X)^{\perp}})$, this function has the property

\begin{equation}
f(\Delta_r^X|_{\mathcal{H}^r(X)^{\perp}})=f(\Delta_r^X|_{im(d_{r+1}^{X})^{\dag}})+f(\Delta_{r-1}^X|_{im(d_{r}^{X})^{\dag}}).\label{I2}
\end{equation}
Hence we find for the first term on the right hand side of \eqref{free-energy-general-zeta}

\begin{equation}
\begin{split}
\sum_{r=0}^{p}(-1)^{p+1-r}(p-r)\zeta^{\prime}(0;\Delta_{r}^{X_{\beta}}|_{\mathcal{H}^{r}(X_{\beta})^{\perp}})
&=\sum_{r=0}^{p-1}(-1)^{p+1-r}\mathcal{I}^{\prime}(0;\Delta_r^X)\\
&=-2\sum_{r=0}^{p-1}(-1)^{p+1-r}b_{r}(X)\ln\beta-f(\Delta_{p-1}^X|_{im(d_{p}^{X})^{\dag}}).\label{I1}
\end{split}
\end{equation}
Now it remains to calculate $a_{n+1}(\Delta_{r}^{X_{\beta}})$ in \eqref{free-energy-general-zeta}. Let us begin by noting that the $L^2$ trace of the heat kernel for $\Delta
_r^{\mathbb{T}_{\beta}^1}$ has the form

\begin{equation}
Tr(e^{-t\Delta _r^{\mathbb{T}_{\beta}^1}})=\Theta _1(0|\sqrt{-1}\ \frac{4\pi t}{\beta ^2})=\frac{\beta}{2\sqrt{\pi t}}\Theta _1(0|\sqrt{-1}\frac{\beta ^2}{4\pi
t})\simeq\frac{\beta}{2\sqrt{\pi t}}+\mathcal{O}(e^{-\frac{1}{t}}),\quad \textrm{for}\ t\downarrow 0,\label{heat-kernel-torus}
\end{equation}
where the duality relation \eqref{duality} was used. This implies $a_k(\Delta _r^{\mathbb{T}_{\beta}^1})=\frac{\beta}{2\sqrt{\pi}}\delta_{k,0}$ for $r\in\{0,1\}$. From the
asymptotic expansions of $\Delta _r^{X_{\beta}}$ and $\Delta _r^{X}$, one derives the following relation between the corresponding Seeley coefficients,

\begin{equation}
a_k(\Delta _r^{X_{\beta}})=
\begin{cases}\frac{\beta}{2\sqrt{\pi}}a_k(\Delta _0^{X}),\quad
&\text{for $r=0$}
\\ \frac{\beta}{2\sqrt{\pi}}\biggl( a_k(\Delta _r^{X})+a_k(\Delta _{r-1}^{X})\biggl).\quad &\text{for $r\geq 1$}\end{cases}\label{Seeley-relation}
\end{equation}
In addition, the $L^2$ trace of the heat kernels for $\Delta_r^X$ and $\Delta_r^X|_{\mathcal{H}^r(X)^{\perp}}$ are related by

\begin{equation}
Tr(e^{-t\Delta_r^X})-\dim\ker\Delta_r^X=Tr(e^{-t\Delta_r^X|_{\mathcal{H}^r(X)^{\perp}}}).\label{heat-kernel-trace}
\end{equation}
From the asymptotic expansions and the spectral properties of the corresponding operators one gets

\begin{equation}
\begin{split}
& a_k(\Delta _r^{X})=a_k(\Delta _r^{X}|_{\mathcal{H}^r(X)^{\perp}})+\delta_{kn}b_r(X),\\ & a_k(\Delta _r^{X}|_{\mathcal{H}^r(X)^{\perp}})=a_k(\Delta _r^{X}|_{im
(d_{r+1}^{X})^{\dag}})+a_k(\Delta _{r-1}^{X}|_{im(d_{r}^{X})^{\dag}}). \label{seeley-relation-1}
\end{split}
\end{equation}
Together with \eqref{res-def} and \eqref{Seeley-relation} the corresponding alternating sum in \eqref{free-energy-general-zeta} can be written as

\begin{equation}
\sum_{r=0}^{p}(-1)^{p-r}(p-r)a_{n+1}(\Delta _r^{X_{\beta}})=-\frac{\beta}{2\sqrt{\pi}}a_{n+1}(\Delta_{r-1}^X|_{im(d_{r}^{X})^{\dag}})=\beta Res_{s=-\frac{1}{2}}\left[\zeta
(s;\Delta_{p-1}^X|_{im(d_{p}^{X})^{\dag}})\right].\label{seeley-thermal-term}
\end{equation}
By substituting \eqref{I}, \eqref{I1} and \eqref{seeley-thermal-term} into \eqref{free-energy-general-zeta}, we obtain the following formula for the free energy of an
higher-abelian gauge theory with action $S$

\begin{equation}
\begin{split}
\mathcal{F}_{X}^{(p)}(\beta)= &-\frac{1}{2}FP_{s=-\frac{1}{2}}\left[\zeta (s;\Delta_{p-1}^X|_{im(d_{p}^{X})^{\dag}})\right]-\frac{1}{2}Res_{s=-\frac{1}{2}}\left[\zeta
(s;\Delta_{p-1}^X|_{im(d_{p}^{X})^{\dag}})\right]\ln{\left(\frac{e\mu}{2}\right)^{2}}\\& -\frac{1}{\beta} \sum _{\alpha\in I}
\ln\left[ 1-e^{-\beta \sqrt{\nu _{\alpha}(\Delta_{p-1}^X|_{im(d_{p}^{X})^{\dag}})}}\right]\\
&+\frac{b_p(X)}{2\beta}\ln{\left[\frac{2\pi}{\beta}\right]}-\frac{1}{2\beta}\ln{\det h_{X}^{(p)}}-\frac{b_p(X)}{2\beta}\ln \mu^{2}+\frac{1}{\beta}\ln
|H_{tor}^{p}(X;\mathbb{Z})|\\ &-\frac{1}{\beta}\ln \left(\sum_{\frak c_{X_{\beta}}^{(p+1)}\in
H^{p+1}(X_{\beta};\mathbb{Z})}\int_{im(d_{p+1}^{X_{\beta}})^{\dag}}\mathcal{D}\tau_{p}\ \left[e^{-\underline{S}^{\hat{\frak c}_{X_{\beta}}^{(p+1)}}(\tau_{p} )}\right]_{(0,\ldots
,0)}\right).\label{free-energy-general-low}
\end{split}
\end{equation}
All terms but the last one are of kinematical origin and stem from the ghost-for-ghost contribution, where the alternating product collapses in the thermal case\footnote{The
"collapse" of determinants of the co-exact Laplace operators has been already pointed out long ago for $p$-form Maxwell theory at finite temperature \cite{D-2002}.}. This
formula - although suited for the low-temperature regime - is valid, however, for all temperatures.

\subsubsection{High-temperature regime}

In the next step we want to derive a formula for the high-temperature limit of the free energy. The aim is to find an analytical continuation of the zeta function which allows
for a separation of the leading terms at high temperatures. The starting point is again \eqref{free-energy-general-zeta}. But instead of \eqref{zeta-reduction-1-2}, the zero
modes are treated differently. Let us introduce the auxiliary quantity

\begin{equation}
\mathcal{K}(s;\Delta _r^{X}|_{\mathcal{H}^r(X)^{\perp}}):=2\sum _{k=1}^{\infty}\sum _{\alpha\in I}\left[\left(\frac{2\pi k}{\beta}\right)^{2}+\nu_{\alpha}(\Delta
_r^{X}|_{\mathcal{H}^r(X)^{\perp}})\right]^{-s}.
\end{equation}
so that $\mathcal{I}(s;\Delta _r^{X})$ admits the decomposition

\begin{equation}
\mathcal{I}(s;\Delta _r^{X})=\zeta (s;\Delta _r^{X}|_{\mathcal{H}^r(X)^{\perp}})+2b_r(X)\ \left(\frac{\beta}{2\pi}\right)^{2s}\zeta _{R}(2s)+\mathcal{K}(s;\Delta
_r^{X}|_{\mathcal{H}^r(X)^{\perp}}).\label{zeta-reduction-2}
\end{equation}
For $\Re (s)>\frac{n}{2}$, the Mellin transformation of $\mathcal{K}$ is given by

\begin{equation}
\mathcal{K}(s;\Delta _r^{X}|_{\mathcal{H}^r(X)^{\perp}})=\frac{2}{\Gamma (s)}\int _0^{\infty}dt\ t^{s-1}\sum _{k=1}^{\infty}e^{-(\frac{2\pi}{\beta})^2k^2t} \sum _{\alpha\in
I}e^{-\nu _{\alpha}(\Delta _r^{X}|_{\mathcal{H}^r(X)^{\perp}})t} \label{k1}
\end{equation}
and can be analytically continued elsewhere \cite{V}. Substituting the heat kernel expansion for $\Delta _r^{X}|_{\mathcal{H}^r(X)^{\perp}}$ and integrating term by term yields
the following asymptotic expansion

\begin{equation}
\mathcal{K}(s;\Delta _r^{X}|_{\mathcal{H}^r(X)^{\perp}}) \simeq 2\left(\frac{2\pi}{\beta}\right)^{-2s}\sum _{m=0}^{\infty}a_m(\Delta
_r^{X}|_{\mathcal{H}^r(X)^{\perp}})\left(\frac{2\pi}{\beta}\right)^{n-m}\frac{\Gamma (s+\frac{m-n}{2})\zeta _{R}(2s+m-n)}{\Gamma (s)}. \label{k2}
\end{equation}
The function $\Gamma (s+\frac{m-n}{2})\zeta _{R}(2s+m-n)$ has simple poles at $m=n$ and $m=n+1$ and can be analytically continued using the well known reflection formula

\begin{equation}
\Gamma(\frac{s}{2})\pi^{-\frac{s}{2}}\zeta_{R}(s)=\Gamma(\frac{1-s}{2})\pi^{-\frac{s-1}{2}}\zeta_{R}(1-s)
\end{equation}
for the Riemann zeta function $\zeta_{R}(s)$. Hence by taking the Laurent expansion of $\zeta _R(s)$ at the pole $s=-1$ and using the series expansion for $\frac{1}{\Gamma (s)}$
once again, a lengthy calculation leads to

\begin{equation}
\begin{split}
\frac{1}{2}\frac{d}{ds}|_{s=0}\mathcal{K}(s;\Delta
_r^{X}|_{\mathcal{H}^j(X)^{\perp}})\simeq &\sum _{\substack{ m=0\\
m\neq n\\ m\neq n+1 }}^{\infty}a_m(\Delta _r^{X}|_{\mathcal{H}^r(X)^{\perp}})\ 2^{n-m}\pi ^{-\frac{1}{2}}\beta ^{m-n}\Gamma
(\frac{n+1-m}{2})\zeta _{R}(n+1-m)\\
& +a_{n+1}(\Delta _r^{X}|_{\mathcal{H}^r(X)^{\perp}})\frac{\beta}{2\pi ^{\frac{1}{2}}}\left(\ln{\left(\frac{\beta}{4\pi}\right)}+\gamma\right)-a_n(\Delta
_r^{X}|_{\mathcal{H}^r(X)^{\perp}})\ln\beta .
\end{split}\label{K2}
\end{equation}
In order to extract the leading terms in \eqref{K2}, we notice that for $m\neq n$, \eqref{seeley-relation-1} gives

\begin{equation}
\sum_{r=0}^{p}(-1)^{p-r}a_m(\Delta _{r}^{X})=a_m(\Delta _p^{X}|_{im(d_{p+1}^{X})^{\dag}}),\label{seeley-relation-3}
\end{equation}
which together with $a_0(\Delta _r^{X})=(4\pi)^{-\frac{n}{2}}\frac{n!}{r!(n-r)!}vol(X)$ \cite{Gilkey} yields

\begin{equation}
a_0(\Delta _p^{X}|_{im(d_{p+1}^{X})^{\dag}})=(4\pi)^{-\frac{n}{2}}\binom{n-1}{p}vol(X).\label{seeley-relation-4}
\end{equation}
Substituting \eqref{zeta-reduction-2} into \eqref{free-energy-general-zeta} and using \eqref{K2} and \eqref{seeley-relation-4} give the final formula for the asymptotic
expansion of the free energy at high temperatures

\begin{equation}
\begin{split}
\mathcal{F}_{X}^{(p)}(\beta)\simeq & \binom{n-1}{p-1}\ \pi^{-\frac{n+1}{2}}\ \Gamma (\frac{n+1}{2})\ \zeta _{R}(n+1)\
\beta ^{-(n+1)}\ vol(X) \\
&+\sum _{\substack{ m=1\\ m\neq n\\ m\neq n+1 }}^{\infty}a_m(\Delta _{p-1}^{X}|_{im(d_{p}^{X})^{\dag}})\ 2^{n-m}\pi
^{-\frac{1}{2}}\ \Gamma (\frac{n+1-m}{2})\ \zeta _{R}(n+1-m)\beta ^{m-n-1}\\
&+\frac{1}{2\beta}\left(\zeta^{\prime}(0;\Delta _{p-1}^{X}|_{im(d_{p}^{X})^{\dag}})-\ln\det h_X^{(p)}\right)-\frac{1}{\beta}\left(\frac{b_p(X)}{2}+a_n(\Delta
_{p-1}^{X}|_{im(d_{p}^{X})^{\dag}})\right) \ln\beta\\ &- Res_{s=-\frac{1}{2}}\left[\zeta (s;\Delta
_{p-1}^{X}|_{im(d_{p}^{X})^{\dag}})\right]\left(\ln{(\frac{\mu\beta}{4\pi})}+\gamma\right)+\frac{b_p(X)}{2\beta}\ln \left(\frac{2\pi}{\mu^{2}}\right)\\
& -\frac{1}{\beta}\ln \left(\sum_{\frak c_{X_{\beta}}^{(p+1)}\in H^{p+1}(X_{\beta};\mathbb{Z})}
\int_{im(d_{p+1}^{X_{\beta}})^{\dag}}\mathcal{D}\tau_{p}\ \left[e^{-\underline{S}^{\hat{\frak c}_{X_{\beta}}^{(p+1)}}(\tau_{p})}\right]_{(0,\ldots ,0)}\right)\\
&+\frac{1}{\beta}\ln |H_{tor}^{p}(X;\mathbb{Z})|.
\end{split}\label{free-energy-general-high}
\end{equation}
Actually, we have only provided the asymptotic expansion for the ghost-for-ghost contribution since the dynamical content is not specified yet. Interestingly, we find that the
leading term is of Stefan-Boltzmann type in $n$ spatial dimensions, however, with the wrong sign.

\subsection{The free energy - extended higher-abelian Maxwell theory}

Now we will focus on the extended higher-abelian Maxwell theory at finite temperature. The starting point is the Euclidean action

\begin{equation}
S_{(\gamma_{p},\theta_{p})}(\hat u_{p})=\frac{q_p}{2}\|\delta _1(\hat u_{p})\ - \gamma_{p} \|^2+2\pi i<\delta_1(\hat u_{p}) -\gamma_{p}, \star
\theta_{p}>,\qquad\hat{u}_{p}\in\widehat{H}^{p}(X_{\beta}). \label{classical thermal action}
\end{equation}
In the thermal context we consider the following topological fields

\begin{equation}
\gamma_{p}=\sum_{j=1}^{b_{p+1}(X)}\gamma_{p}^{j}\ pr_2^{\ast}(\rho _X^{(p+1)})_j\in\mathcal{H}^{p+1}(X_{\beta}),\qquad\theta_{p}=\sum_{k=1}^{b_{n-p}(X)}\theta_{p}^{k}\
pr_2^{\ast}(\rho _X^{(n-p)})_k\in\mathcal{H}^{n-p}(X_{\beta}).\label{thermal-currents}
\end{equation}
In order to determine the corresponding free energy we have to compute just the last term in \eqref{free-energy-general-low}. In terms of local coordinates on
$\mathbb{T}^{b_{p-1}(X)}\times\mathbb{T}^{b_p(X)}\times im(d_{p+1}^{X_{\beta}})^{\dag}$ the action $S$ splits into a topological and a dynamical part,

\begin{equation}
\underline{S}_{(\gamma_{p},\theta_{p})}^{\hat{\frak c}_{X_{\beta}}^{(p+1)}}(\vec{y}_{p-1},\vec{z}_{p},\tau_{p})=S_{(\gamma_{p},\theta_{p})}(\hat{\frak c}_{X_{\beta}}^{(p+1)})+
\frac{q_p}{2}<\tau_{p},\Delta_p^{X_{\beta}}|_{im(d_{p+1}^{X_{\beta}})^{\dag}}\tau_{p}>,
\end{equation}
where $\vec{y}_{p-1}\in\mathbb{T}^{b_{p-1}(X)}$ and $\vec{z}_{p}\in\mathbb{T}^{b_p(X)}$. The sum over the topological sectors gives

\begin{equation}
\begin{split}
\sum_{\frak c_{X_{\beta}}^{(p+1)}\in H^{p+1}(X_{\beta};\mathbb{Z})} e^{-S_{(\gamma_{p},\theta_{p})}(\hat{\frak c}_{X_{\beta}}^{(p+1)})} =
&\Theta_{b_p(X)}\left(\vec{\theta}_{p}|-\frac{q_p}{2\pi i}\beta^{-1}h_{X}^{(p)}\right)|H_{tor}^{p}(X;\mathbb{Z})| \\ &\times\Theta _{b_{p+1}(X)}\begin{bmatrix} \vec{\gamma}_{p}
\\0\end{bmatrix} \left(0|-\frac{q_{p}}{2\pi i}\ \beta h_X^{(p+1)}\right)|H_{tor}^{p+1}(X;\mathbb{Z})|.
\end{split}\label{top-sum-thermal}
\end{equation}
where \eqref{metric-temperature} and \eqref{thermal-currents} were used. Due to the modular property \eqref{modular-property} and \eqref{duality-2} of the Riemann-Theta
function, we can restrict the vectors $\vec{\gamma}_{p}$ and $\vec{\theta}_{p}$ to $\{\vec{\gamma}_{p}\}$ and $\{\vec{\theta}_{p}\}$, respectively, which are defined as follows:
For any $\vec{a}=(a_1,\ldots, a_j,\ldots,a_r)\in \mathbb{R}^{r}$ we introduce $\{\vec{a}\}=(\{a_1\},\ldots, \{a_j\},\ldots,\{a_r\})\in \mathbb{R}^{r}$ such that

\begin{equation}
\{a_{j}\}=
\begin{cases}
\langle a_{j}\rangle, &\text{if $\langle a_{j}\rangle\in [0,\frac{1}{2}]$},\\
1-\langle a_{j}\rangle, &\text{if $\langle a_{j}\rangle\in [\frac{1}{2},1)$}.
\end{cases}\label{reduction}
\end{equation}
Here $\langle a_{j}\rangle =a_{j}-\max \{m\in\mathbb{Z}|m\leq a_{j}\}$ is the fractional part of $a_j$.\par

We use \eqref{tau-integral}, \eqref{zeta-general-2} and \eqref{Seeley-relation} to carry out the $\tau_{p}$-integration in \eqref{free-energy-general-low}. This leads to

\begin{equation}
\begin{split}
&\frac{1}{\beta}\ln\left(\int_{im(d_{p+1}^{X_{\beta}})^{\dag}}\mathcal{D}\tau_{p}\ e^{-\frac{q_p}{2}<\tau_{p},\Delta
_{p}^{X_{\beta}}|_{im(d_{p+1}^{X_{\beta}})^{\dag}}\tau_{p}>}\right)=\\ &=
\frac{1}{2\beta}\sum_{r=0}^{p}(-1)^{p-r}\left[\zeta^{\prime}(0;\Delta_{r}^{X_{\beta}}|_{\mathcal{H}^{r}(X_{\beta})^{\perp}})+
\zeta (0;\Delta_{r}^{X_{\beta}}|_{\mathcal{H}^{r}(X_{\beta})^{\perp}})\ln\left(\frac{\mu^{2}}{q_p}\right)\right]\\
&=\frac{1}{2\beta}\mathcal{I}^{\prime}(0;\Delta _{p}^{X})+\frac{a_{n+1}(\Delta _{p}^{X})}{4\sqrt{\pi}}\ln\left(\frac{\mu^{2}}{q_p}\right)
-\frac{b_{p}(X)}{2\beta}\ln\left(\frac{\mu^{2}}{q_p}\right).
\end{split}\label{tau-integral-thermal}
\end{equation}
Now we substitute \eqref{top-sum-thermal} and \eqref{tau-integral-thermal} together with \eqref{res-def}, \eqref{I} into \eqref{free-energy-general-low}. In addition we use
\eqref{decomposition} and \eqref{duality} to rewrite the Riemann Theta functions in \eqref{top-sum-thermal} in order to separate the temperature independent part. The resulting
expression for the free energy of extended higher-abelian Maxwell theory splits into two parts

\begin{equation}
\mathcal{F}_{X}^{(p)}(q_p;\beta,\gamma_{p},\theta_{p})=\mathcal{F}_{X;dyn}^{(p)}(q_p;\beta,\gamma_{p},\theta_{p})
+\mathcal{F}_{X;top}^{(p)}(q_p;\beta,\gamma_{p},\theta_{p}),\label{free-energy-final-form}
\end{equation}
where

\begin{equation}
\begin{split}
\mathcal{F}_{X;dyn}^{(p)}(q_p;\beta,\gamma_{p},\theta_{p})= &\frac{1}{2}FP_{s=-\frac{1}{2}}\left[\zeta (s;\Delta _p^{X}|_{im
(d_{p+1}^{X})^{\dag}})\right]+\frac{1}{2}Res_{s=-\frac{1}{2}}\left[\zeta (s;\Delta _p^{X}|_{im(d_{p+1}^{X})^{\dag}})\right]\ln{\left(\frac{e\mu}{2}\right)^{2}}\\
&+ \frac{1}{\beta} \sum _{\alpha\in I}\ln\left[ 1-e^{-\beta \sqrt{\nu _{\alpha}(\Delta _p^{X}|_{im (d_{p+1}^{X})^{\dag}})}}\right]-\frac{1}{2}Res_{s=-\frac{1}{2}}\left[\zeta
(s;\Delta _{p}^{X})\right]\ln{q_p}
\end{split}\label{free-energy-final-form-dynamical}
\end{equation}
is related to the dynamical (propagating) modes and

\begin{equation}
\begin{split}
\mathcal{F}_{X;top}^{(p)}(q_p;\beta,\gamma_{p},\theta_{p})= &\frac{2\pi^{2}}{q_p}\sum_{j,k=1}^{b_p(X)}(h_X^{(p)})_{jk}^{-1}\{\theta_{p}^{j}\}\{\theta_{p}^{k}\}
+\frac{q_p}{2}\sum_{j,k=1}^{b_{p+1}(X)}(h_X^{(p+1)})_{jk}\{\gamma_{p}^{j}\}\{\gamma_{p}^{k}\}\\&-\frac{1}{\beta}\ln \Theta _{b_p(X)}\left( \frac{2\pi i}{q_p}\beta
(h_X^{(p)})^{-1}\{\vec{\theta}_{p}\}|\frac{2\pi i}{q_p}\beta (h_X^{(p)})^{-1}\right) \\
&-\frac{1}{\beta}\ln{\Theta _{b_{p+1}(X)} \left( \frac{q_p}{2\pi i}\beta h_X^{(p+1)}\{\vec{\gamma}_{p}\}|-\frac{q_p}{2\pi i}\beta h_X^{(p+1)}\right)}\\
&-\frac{1}{\beta}\ln\left(|H_{tor}^{p+1}(X;\mathbb{Z})|\right)
\end{split}\label{free-energy-final-form-topological}
\end{equation}
is related to the topologically inequivalent configurations. In section 2.3 the partition function was redefined in \eqref{reduced partition function} in order to absorb the
$q_p$ dependent local term into the action. In the thermal case this amounts to go over from $\mathcal{F}_{X}^{(p)}$ to

\begin{equation}
\widehat{\mathcal{F}}_{X}^{(p)}(q_p;\beta,\gamma_{p},\theta_{p}):=\mathcal{F}_{X}^{(p)}(q_p;\beta,\gamma_{p},\theta_{p}) +\frac{1}{2}Res_{s=-\frac{1}{2}}\left[\zeta(s;\Delta
_{p}^{X})\right]\ln{q_p},\label{free-energy-final-low-modified}
\end{equation}
which affects only the vacuum contribution. Instead of $\mathcal{F}_{X}^{(p)}$ we will work in the following with $\widehat{\mathcal{F}}_{X}^{(p)}$ and refer to it as the
effective free energy of extended higher-abelian Maxwell theory. From \eqref{free-energy-final-form-dynamical} and \eqref{free-energy-final-form-topological} one can read off
the corresponding vacuum (Casimir) energy, namely

\begin{equation}
\begin{split}
\widehat{\mathcal{F}}_{X;vac}^{(p)}(q_p;\gamma_{p},\theta_{p})= &\frac{1}{2}FP_{s=-\frac{1}{2}}\left[\zeta (s;\Delta _p^{X}|_{im
(d_{p+1}^{X})^{\dag}})\right]+\frac{1}{2}Res_{s=-\frac{1}{2}}\left[\zeta (s;\Delta _p^{X}|_{im
(d_{p+1}^{X})^{\dag}}))\right]\ln{\left(\frac{e\mu}{2}\right)^{2}}\\
&+\frac{2\pi^{2}}{q_p}\sum_{j,k=1}^{b_p(X)}(h_X^{(p)})_{jk}^{-1}\{\theta_{p}^{j}\}\{\theta_{p}^{k}\}
+\frac{q_p}{2}\sum_{j,k=1}^{b_{p+1}(X)}(h_X^{(p+1)})_{jk}\{\gamma_{p}^{j}\}\{\gamma_{p}^{k}\}.
\end{split}\label{free-energy-vacuum}
\end{equation}
The topological contribution to the vacuum (Casimir) energy is absent whenever the topological fields vanish or have integer components. On the other hand, the thermal
excitations are always affected by a non-trivial topology of $X$.\par

The free energy has the following symmetry properties

\begin{equation}
\begin{split}
&\widehat{\mathcal{F}}_{X}^{(p)}(q_p;\gamma_{p},\theta_{p})=\widehat{\mathcal{F}}_{X}^{(p)}(q_p;-\gamma_{p},-\theta_{p})\\
&\widehat{\mathcal{F}}_{X}^{(p)}(q_p;\gamma_{p}+\omega,\theta_{p}+\omega^{\prime})=\widehat{\mathcal{F}}_{X}^{(p)}(q_p;\gamma_{p}, \theta_{p}), \qquad
\omega\in\mathcal{H}_{\mathbb{Z}}^{p+1}(X),\ \omega^{\prime}\in\mathcal{H}_{\mathbb{Z}}^{n-p}(X).\label{symmetry-Maxwell-free-energy}
\end{split}
\end{equation}
Notice that the expression \eqref{free-energy-final-low-modified} for the free energy is valid for the whole temperature range. However, in order to obtain an expression which
is suited to exhibit the high-temperature structure, we substitute \eqref{zeta-reduction-2} into \eqref{tau-integral-thermal}. To extract the relevant contribution from the
topological sector, we apply the duality relation \eqref{duality-1} to the $b_{p+1}(X)$ dimensional Riemann Theta function in \eqref{top-sum-thermal}. A lengthy calculation
finally yields the following asymptotic expansion for the free energy of extended higher-abelian Maxwell theory in the high-temperature regime

\begin{equation}
\begin{split}
\widehat{\mathcal{F}}_{X}^{(p)}&(q_p;\beta ,\gamma_{p},\theta_{p})\simeq  -\binom{n-1}{p}\ \pi^{-\frac{n+1}{2}}\ \Gamma (\frac{n+1}{2})\ \zeta _{R}(n+1)\
\beta ^{-(n+1)}\ vol(X) \\
&-\sum _{\substack{ m=1\\ m\neq n\\ m\neq n+1 }}^{\infty}a_m(\Delta _p^{X}|_{im (d_{p+1}^{X})^{\dag}})\ 2^{n-m}\pi
^{-\frac{1}{2}}\beta ^{m-n-1}\ \Gamma (\frac{n+1-m}{2})\ \zeta _{R}(n+1-m)\\
&-\frac{1}{2\beta}\left[\zeta^{\prime}(0;\Delta _p^{X}|_{im (d_{p+1}^{X})^{\dag}})+ \ln{\left(\frac{\det{\left( (\frac{q_p}{2\pi})h_X^{(p)}\right)}}{\det{\left(
(\frac{q_p}{2\pi})h_X^{(p+1)}\right)}}\ |H_{tor}^{p+1}(X;\mathbb{Z})|^2\right)}\right]\\
&+\frac{1}{2\beta}\left[ 2a_n(\Delta _p^{X}|_{im (d_{p+1}^{X})^{\dag}})+b_p(X)+b_{p+1}(X)\right]\ln\beta
\\ &+ Res_{s=-\frac{1}{2}}\left[\zeta (s;\Delta _p^{X}|_{im
(d_{p+1}^{X})^{\dag}})\right]\left(\ln{\left(\frac{\mu\beta}{4\pi}\right)}+\gamma\right)\\
&- \frac{1}{\beta}\ln\Theta _{b_p(X)}\left( \vec{\theta}_{p} |-\frac{q_p}{2\pi i}\beta ^{-1} h_X^{(p)}\right) -\frac{1}{\beta}\ln\Theta _{b_{p+1}(X)}\left(
\vec{\gamma}_{p}|\frac{2\pi i}{q_p} \beta ^{-1} (h_X^{(p+1)})^{-1}\right).
\end{split}\label{K3}
\end{equation}
In this limit the contributions from the propagating modes become dominant. The leading term is extensive and exhibits the Stefan-Boltzmann dependency on the temperature in $n$
spatial dimensions. The remaining terms represent modifications due to the topology and geometry of $X$.\par

In summary, we have shown explicitly how the topology and geometry of $X$ affect both the vacuum (Casimir) energy and the finite temperature part related to the occupied states
of the thermal ensemble. The ambiguity in the free energy is expressed by the renormalization scale $\mu$ and appears, whenever $Res_{s=-\frac{1}{2}}\left[\zeta (s;\Delta
_p^{X}|_{im (d_{p+1}^{X})^{\dag}})\right]\neq 0$. In that case renormalization issues have to be taken into account. On the other hand, if $\zeta (s;\Delta _p^{X}|_{im
(d_{p+1}^{X})^{\dag}})$ is finite at $s=-\frac{1}{2}$, the scale dependency disappears and the free energy is uniquely determined. Finally, let us remark that in the special
case $p=1$ and $\gamma_{1}=0$, we recover the expression for the free energy of the photon gas \cite{kelnhofer-2}, which back then were obtained within the framework of
principal $U(1)$-bundles with connections.\par

Let us recall that in its original meaning the Casimir effect is caused by the change of the vacuum energy or, at finite temperature, of the free energy due to constraints
imposed on the quantum fields compared with the free energy of the unconstrained system. The difference between these two free energies is called \textit{Casimir free energy}
(see e.g. \cite{PMG} for a detailed discussion on this topic). In this respect, the formulae which we derived for the free energy represent always the total free energy of the
constrained system. An obvious candidate for a (unconstrained) "reference configuration" would be a gas of higher-abelian gauge fields placed in a large box in flat Euclidean
space. In the infinite volume limit all related topological and finite size effects can be neglected. Since we applied zeta function regularization the vacuum energy in the
reference configuration is implicitly disregarded. Thus only the thermal contribution of the propagating degrees of freedom remains. In the large volume limit the discrete index
$\alpha$ in the third term of \eqref{free-energy-final-form-dynamical} can be replaced by the continuous $n$-dimensional wave vector $\vec{k}\in\mathbb{R}^{n}$. Hence the sum
can be replaced by an integration with respect to the measure $\frac{d^n\vec{k}}{(2\pi)^n}$. Using \cite{gradshteyn} the integration gives for the free energy density

\begin{equation}
\binom{n-1}{p}\frac{1}{\beta}\int_{\mathbb{R}^n}\ \frac{d^n\vec{k}}{(2\pi)^n}\ \ln{\left(1-e^{-\beta |\vec{k}|}\right)}=-\binom{n-1}{p}\pi^{-\frac{n+1}{2}}\ \Gamma
(\frac{n+1}{2})\ \zeta _{R}(n+1)\ \beta ^{-(n+1)},\label{euclidean}
\end{equation}
where the factor $\binom{n-1}{p}$ is the number of independent propagating degrees of freedom. As expected we get the free energy density of black-body radiation of $p$-form
Maxwell theory in $n$ dimensions\footnote{Let us remark that there exists another renormalization approach for the Casimir free energy where also the finite temperature
contributions are appropriately renormalized \cite{GKM, Bezerra-Klimchitskaya-Mostepanenko-Romero 2011, Bezerra-Mostepanenko-Mota-Romero 2011, Bezerra-Mota-Muniz 2014,
Mota-Bezerra 2015}. The underlining concept is based on the proposal that the renormalized Casimir free energy should satisfy the classical limit at high temperatures.}.

\subsubsection{The equation of state}

In this section we want to derive the equation of state for the gas of higher-abelian gauge fields. In terms of the (effective) free energy, the internal energy
$\widehat{\mathcal{U}}_{X}^{(p)}$, entropy $\widehat{\mathcal{S}}_{X}^{(p)}$ and pressure $\widehat{\mathcal{P}}_{X}^{(p)}$ are given by
\begin{equation}
\widehat{\mathcal{U}}_{X}^{(p)} =\frac{\partial}{\partial\beta}\biggl(\beta\widehat{\mathcal{F}}_{X}^{(p)}\biggl),\quad \widehat{\mathcal{S}}_{X}^{(p)} =\beta ^2\
\frac{\partial}{\partial\beta}\ \widehat{\mathcal{F}}_{X}^{(p)},\quad \widehat{\mathcal{P}}_{X}^{(p)} =-\frac{\partial}{\partial V}\ \widehat{\mathcal{F}}_{X}^{(p)}.
\label{thermodynamic-funct}
\end{equation}
Like the free energy, every thermodynamic function splits into a dynamical and into a topological part. Under a constant scale transformation $g\mapsto\lambda ^2g$,
$\lambda\in\mathbb{R}$, of the background metric of $X_{\beta}$, one gets $\beta\mapsto\lambda\beta$ and

\begin{alignat}{2}
& vol_{X}\mapsto\lambda ^nvol_{X}, &\qquad\quad \nu _{k}^{(i)}(\Delta _r^{\mathbb{T}_{\beta}^1})\mapsto \lambda ^{-2}\nu _{k}^{(i)}(\Delta _r^{\mathbb{T}_{\beta}^1})
\nonumber\\
& h_{X}^{(p)}\mapsto\lambda ^{n-2p}h_{X}^{(p)}, &\qquad\quad  \nu _{l}^{(j)}(\Delta _s^{X})\mapsto \lambda ^{-2}\nu _{l}^{(j)}(\Delta _s^{X}).\nonumber
\end{alignat}
The free energy transforms as

\begin{equation}
\begin{split}
&\widehat{\mathcal{F}}_X^{(p)}\mapsto\frac{1}{\lambda}\widehat{\mathcal{F}}_X^{(p)}+\frac{\ln\lambda}{\lambda} Res_{s=-\frac{1}{2}}[\zeta (s;\Delta _p^{X}|_{im
(d_{p+1}^{X})^{\dag}})]\\
&+\frac{(2\pi)^2}{q_p\lambda}(\lambda^{2p+1-n}-1)\sum_{j,k=1}^{b_p(X)}(h_X^{(p)})_{jk}^{-1}\{\theta^j\}\{\theta^k\}+
\frac{q_p}{2\lambda}(\lambda^{n-2p-1} -1)\sum_{j,k=1}^{b_{p+1}(X)}(h_X^{(p+1)})_{jk}\{\gamma^j\}\{\gamma^k\}\\
&+\frac{1}{\lambda\beta}\ln{\left(\frac{\Theta _{b_p(X)}\left( \frac{2\pi i}{q_p}\beta (h_X^{(p)})^{-1}\{\vec{\theta}\} |\frac{2\pi i}{q_p}\beta (h_X^{(p)})^{-1}\right)}{\Theta
_{b_p(X)}\left( \frac{2\pi i}{q_p}\lambda^{2p+1-n}\beta (h_X^{(p)})^{-1}\{\vec{\theta}\} |\frac{2\pi i}{q_p}\lambda^{2p+1-n}\beta
(h_X^{(p)})^{-1}\right)}\right)} \\
&+\frac{1}{\lambda\beta}\ln{\left(\frac{\Theta _{b_{p+1}(X)}\left(\frac{q_p}{2\pi i}\beta h_X^{(p+1)}\{\vec{\gamma}\} |-\frac{q_p}{2\pi i}\beta h_{X}^{(p+1)}\right)}{\Theta
_{b_{p+1}(X)}\left(\frac{q_p}{2\pi i}\lambda^{n-2p-1}\beta (h_X^{(p+1)})\{\vec{\gamma}\} |\frac{q_p}{2\pi i}\lambda^{n-2p-1}\beta h_{X}^{(p+1)}\right)}\right)}.
\label{trafo-free-energy}
\end{split}
\end{equation}
The main observation is that in general the free energy does not transform homogeneously of degree $-1$, meaning that the free energy is no longer an extensive quantity. There
are two obstructions: The first one is caused by the explicit dependence on the renormalization scale $\mu$. The second one is related to non-trivial
$H_{free}^{p}(X;\mathbb{Z})\oplus H_{free}^{p+1}(X;\mathbb{Z})$. The resulting equation of state is obtained from the right hand side of \eqref{trafo-free-energy} by taking the
derivative with respect to $\lambda$ at $\lambda =1$. A lengthy calculation yields

\begin{equation}
\widehat{\mathcal{F}}_{X}^{(p)}=n\widehat{\mathcal{P}}_{X}^{(p)}\ vol(X)-\beta ^{-1}\widehat{\mathcal{S}}_{X}^{(p)}+\Gamma ^{(p)},
\end{equation}
with anomalous term

\begin{equation}
\begin{split}
\Gamma^{(p)} &=Res_{s=-\frac{1}{2}}[\zeta (s;\Delta _p^{X}|_{im
(d_{p+1}^{X})^{\dag}})]\\
&+(2p+1-n)\left(\frac{2\pi^2}{q_p}\sum_{j,k=1}^{b_p(X)}(h_X^{(p)})_{jk}^{-1}\{\theta_{p}^j\}\{\theta_{p}^k\}-
\frac{q_p}{2}\sum_{j,k=1}^{b_{p+1}(X)}(h_X^{(p+1)})_{jk}\{\gamma_{p}^j\}\{\gamma_{p}^k\}\right)\\
&+\frac{2p+1-n}{\beta}\frac{\partial}{\partial \lambda}|_{\lambda =1}\ln{\left(\frac{\Theta _{b_{p+1}(X)}\left(\frac{q_p}{2\pi i}\lambda\beta
h_{X}^{(p+1)}\{\vec{\gamma_{p}}\}|-\frac{q_p}{2\pi i}\lambda\beta h_{X}^{(p+1)}\right)}{\Theta _{b_p(X)}\left(\frac{2\pi i}{q_p}\lambda\beta (h_X^{(p)})^{-1}\{\vec{\theta_{p}}\}
|\frac{2\pi i}{q_p}\lambda \beta (h_X^{(p)})^{-1}\right)}\right)}.\label{trafo-1}
\end{split}
\end{equation}
Since the thermodynamic functions are interrelated by $\widehat{\mathcal{F}}_{X}^{(p)}=\widehat{\mathcal{U}}_{X}^{(p)}-\beta ^{-1}\widehat{\mathcal{S}}_{X}^{(p)}$, we finally
obtain the following equation of state:

\begin{equation}
\widehat{\mathcal{P}}_{X}^{(p)}=\frac{1}{n\ vol(X)}\left[ \widehat{\mathcal{U}}_{X}^{(p)}-\Gamma^{(p)}\right].\label{eq-state}
\end{equation}
The topological contribution to $\Gamma^{(p)}$ vanishes for $n=2p+1$. If in addition the zeta function of $\Delta _p^{X}|_{im (d_{p+1}^{X})^{\dag}}$ is finite at
$s=-\frac{1}{2}$, then the anomalous term $\Gamma^{(p)}$ vanishes identically.\par

\subsubsection{Thermal duality}

Now we want to discuss the relation between thermodynamic functions of extended higher-abelian Maxwell theories of degree $p$ and $n-p-1$. Let $(q_p,\gamma_p,\theta_p)$ be the
corresponding set of parameters in degree $p$. We introduce the dual parameters by

\begin{enumerate}
    \item $q_{n-p-1}^{dual}\cdot q_{p}=(2\pi)^{2}$
    \item $\vec{\gamma}_{n-p-1}^{dual}=\vec{\theta}_{p}$
    \item $\vec{\theta}_{n-p-1}^{dual}=\vec{\gamma}_{p}$.
\end{enumerate}
Since $Spec(\Delta _{n-p}^{X}|_{im(d_{n-p-1}^{X})})=Spec(\Delta _{n-p-1}^{X}|_{im(d_{n-p}^{X})^{\dag}})$ and $\underline{\star}\ \Delta _p^{X}|_{im(d_{p+1}^{X})^{\dag}}=\Delta
_{n-p}^{X}|_{im(d_{n-p-1}^{X})}\ \underline{\star}$, one has $Spec(\Delta _p^{X}|_{im(d_{p+1}^{X})^{\dag}})=Spec(\Delta _{n-p-1}^{X}|_{im(d_{n-p}^{X})^{\dag}})$. Using that
$H_{tor}^{p+1}(X;\mathbb{Z})\cong H_{tor}^{n-p}(X;\mathbb{Z})$ one can verify directly from the explicit expression \eqref{free-energy-final-form} that

\begin{equation}
\widehat{\mathcal{F}}_X^{(p)}\left(q_p;\beta,\gamma_{p},\theta_{p}\right)=\widehat{\mathcal{F}}_{X}^{(n-p-1)}\left(\frac{(2\pi)^{2}}{q_p};
\beta,\theta_{p},\gamma_{p}\right).\label{duality-free-energy-final-form}
\end{equation}
As a consequence, all thermodynamic functions are equal implying that these theories are exactly dual to each other.\par

Alternatively, \eqref{duality-free-energy-final-form} can be obtained directly from the general duality relation \eqref{duality_special_1}: Since the Euler characteristics of a
product splits and $\chi(\mathbb{T}_{\beta}^{1})=0$, one gets $\chi (X_{\beta})=\chi (\mathbb{T}_{\beta}^{1})\chi (X)=0$. Due to the specific choice for $\gamma_{p}$ and
$\theta_{p}$ \eqref{thermal-currents} the pairing term is absent.\par

\subsection{Extended higher-abelian Maxwell theory on the $n$-torus}

In this section we want to present an explicit example by studying the extended higher-abelian Maxwell theory on the $n$-torus $X=\mathbb{T}^n$. The aim is to compute exact
expressions for the corresponding thermodynamic functions in the low- and the high-temperature regimes, respectively.\par

Let $(t_1,\ldots ,t_n)$ denote the local coordinates of $\mathbb{T}^n$. We equip the $n$-torus with the flat metric $g_{\mathbb{T}^{n}}=L^{2}\sum_{i=1}^{n}dt^i\otimes dt^i$, so
that $vol(\mathbb{T}^n)=L^n$. \par

The starting point for the computation of the free energy is the general formula \eqref{free-energy-final-form}. The eigenvalues of $\Delta _{p}^{\mathbb{T}^n}|_{im
(d_{p+1}^{\mathbb{T}^n})^{\dag}}$ are given by the sequence of positive real numbers $\nu _{\vec{m}}(\Delta _{p}^{\mathbb{T}^n}|_{im
(d_{p+1}^{\mathbb{T}^n})^{\dag}})=\left(\frac{2\pi}{L}\right)^{2}\sum_{j=1}^{n}m_{j}^{2}$, where $\vec{m}=(m_1,\ldots,m_n)\in\mathbb{Z}_{0}^{n}:=\mathbb{Z}^{n}\backslash 0$. For
each fixed $\vec{m}\in\mathbb{Z}_{0}^{n}$, let us choose an orthonormal basis $\varepsilon_{\vec{m}}^{r}\in \vec{m}^{\perp}$ with $r=1,\ldots ,n-1$. The components of
$\varepsilon_{\vec{m}}^{r}$ are denoted by $\varepsilon_{\vec{m};j}^{r}$, with $j=1,\ldots ,n$. Then the co-exact $p$-forms

\begin{equation}
\psi_{r_1,\ldots ,r_p;\vec{m}}^{(p)}:=\sum_{1\leq j_1<\ldots <j_p\leq n}\varepsilon_{\vec{m};j_{1}}^{r_1}\cdots\varepsilon_{\vec{m};j_{p}}^{r_p}\ e^{2\pi
i\sum_{k=1}^{n}m_kt^{k}}dt^{j_1}\wedge\ldots\wedge dt^{j_p}\in im (d_{p+1}^{\mathbb{T}^n})^{\dag}\otimes\mathbb{C}
\end{equation}
are eigenforms of $\Delta _{p}^{\mathbb{T}^n}|_{im (d_{p+1}^{\mathbb{T}^n})^{\dag}}$ associated to the eigenvalue $\nu _{\vec{m}}(\Delta _{p}^{\mathbb{T}^n}|_{im
(d_{p+1}^{\mathbb{T}^n})^{\dag}})$. Moreover, for each fixed $\vec{m}\in\mathbb{Z}_{0}^{n}$ the eigenforms $\psi_{r_1,\ldots ,r_p;\vec{m}}^{(p)}$ with $1\leq r_1<\ldots <r_p\leq
n-1$ provide a basis for the $\binom{n-1}{p}$-dimensional eigenspace, whose dimension gives the number of independent polarization states associated to the propagating degrees
of freedom. For the corresponding zeta-function we obtain

\begin{equation}
\zeta (s;\Delta _p^{\mathbb{T}^n}|_{im (d_{p+1}^{\mathbb{T}^n})^{\dag}}) =\binom{n-1}{p}\sum_{\vec{m}\in\mathbb{Z}_{0}^{n}}\left[\sum_{j=1}^{n}\left(\frac{2\pi
m_j}{L}\right)^{2}\right]^{-s} =\left(\frac{2\pi}{L}\right)^{-2s}\binom{n-1}{p}E_n(s;1,\ldots ,1),\label{zeta-function-torus}
\end{equation}
where $E_n$ denotes the Epstein zeta function in $n$ dimensions \eqref{epstein-conventional}. Using the reflection formula \eqref{reflection-formula}, it follows that
\eqref{zeta-function-torus} is regular at $s=-\frac{1}{2}$, so that its finite part \eqref{finite-part} reads

\begin{equation}
FP_{s=-\frac{1}{2}}\left[\zeta (s;\Delta _p^{\mathbb{T}^n}|_{im (d_{p+1}^{\mathbb{T}^n})^{\dag}})\right]=-\frac{1}{L}\binom{n-1}{p}\pi^{-\frac{n+1}{2}}\Gamma
(\frac{n+1}{2})E_n(\frac{n+1}{2};1,\ldots ,1).\label{finite-part-torus}
\end{equation}
As a consequence, the resulting free energy does not depend on the renormalization scale $\mu$. In order to determine the topological contribution, we notice that the $p$-forms

\begin{equation}
(\rho_{\mathbb{T}^n}^{(p)})_{i_1,\ldots ,i_p}=dt^{i_1}\wedge\ldots\wedge dt^{i_p},\qquad 1\leq i_1<\ldots <i_p\leq n\label{basis-torus}
\end{equation}
provide a basis for $\mathcal{H}_{\mathbb{Z}}^{p}(\mathbb{T}^n)$ and induce the metric

\begin{equation}
(h_{\mathbb{T}^n}^{(p)})_{ij}=L^{n-2p}\delta_{ij},\quad i,j=1,\ldots ,\binom{n}{p}. \label{harmonic-metric-torus}
\end{equation}
Substituting \eqref{finite-part-torus} and \eqref{harmonic-metric-torus} into \eqref{free-energy-final-form} leads to the following expression for the free energy

\begin{equation}
\begin{split}
\widehat{\mathcal{F}}_{\mathbb{T}^{n}}^{(p)}(q_p;\beta,\gamma_{p},\theta_{p}) =&-\frac{1}{2L}\binom{n-1}{p}\ \pi ^{-\frac{n+1}{2}}\Gamma(\frac{n+1}{2})
E_n(\frac{n+1}{2};1,\ldots ,1)+\frac{2\pi^{2}}{q_p}L^{2p-n}\sum_{j=1}^{\binom{n}{p}}\{\theta_{p}^j\}^2\\
&+\frac{q_p}{2}L^{n-2p-2}\sum_{j=1}^{\binom{n}{p+1}}\{\gamma_{p}^j\}^2 + \frac{1}{\beta}\binom{n-1}{p} \sum _{\vec{m}\in\mathbb{Z}_{0}^{n}}\ln{\left[ 1-e^{-2\pi\beta
L^{-1}|\vec{m}|}\right]}
\\ & -\frac{1}{\beta}\sum _{j=1}^{\binom{n}{p}}\ln{\Theta _1\left(\frac{2\pi i}{q_p}\beta L^{2p-n} \{\theta_{p}^j\}|\frac{2\pi i}{q_p}\beta L^{2p-n}\right)}\\ &
-\frac{1}{\beta}\sum _{j=1}^{\binom{n}{p+1}}\ln{\Theta _1 \left( \frac{q_p}{2\pi i}\beta L^{n-2p-2} \{\gamma_{p}^j\}|-\frac{q_p}{2\pi i}\beta L^{n-2p-2}\right)}.
\end{split}\label{free-enery-torus}
\end{equation}
The contributions (first and fourth term) of the propagating modes depend on the degree $p$ only via the factor giving the number of independent polarization states. When
considering the product $L\widehat{\mathcal{F}}_{\mathbb{T}^{n}}^{(p)}$ these contributions become a function of the inverse scaled temperature $\frac{\beta}{L}$ and thus they
are scale invariant under the joint transformation $\beta\mapsto\lambda\beta$ and $L\mapsto\lambda L$ for $\lambda\in\mathbb{R}$. By contrast, the topological part depends
intrinsically on degree $p$ and - except in odd dimensions and for degree $p=\frac{n-1}{2}$ - spoils this scale invariance. In that exceptional case, the equation of state
\eqref{eq-state} reduces to the "conventional" form

\begin{equation}
\widehat{\mathcal{P}}_{\mathbb{T}^{2p+1}}^{(p)}\ L^{2p+1}=\frac{1}{2p+1}\ \widehat{\mathcal{U}}_{\mathbb{T}^{2p+1}}^{(p)},\label{eq-state-torus}
\end{equation}
without anomalous term. In the following we will provide low- and high-temperature expansions for the thermodynamic functions.
\bigskip

\subsubsection{Low-temperature regime}

Let us denote the sum of the last two Riemann-Theta functions in \eqref{free-enery-torus} by $f_{low}^{(p)}$. Since $\{\theta_{p}^j\}$ and $\{\gamma_{p}^j\}$ belong to the
interval $[0,\frac{1}{2}]$, condition \eqref{constraint} is satisfied. By applying \eqref{theta-limit} to $f_{low}^{(p)}$, one gets the following series expansion

\begin{equation}
\begin{split}
f_{low}^{(p)} :=&\frac{1}{\beta}\ln 2^{N_{\theta}+N_{\gamma}}+\frac{1}{\beta}\sum_{m=1}^{\infty}\frac{1}{m}\ \frac{\binom{n}{p}+2(-1)^m N_{\theta}}{1-e^{\frac{(2\pi)^2}{q_p}
\beta L^{2p-n}m}}+\frac{1}{\beta}\sum_{m=1}^{\infty}\frac{1}{m}\ \frac{\binom{n}{p+1}+2(-1)^m N_{\gamma}}{1-e^{q_p \beta L^{n-2p-2}m}}\\
& +\frac{1}{\beta}\sum_{\substack{ j=1 \\\{\theta_{p}^j\}\neq\frac{1}{2} }}^{\binom{n}{p}}\sum_{m=1}^{\infty}\frac{(-1)^{m+1}}{m}\ \frac{e^{-\frac{(2\pi)^2}{q_p} \beta
L^{2p-n}m(\frac{1}{2}+\{\theta_{p}^j\})}+e^{-\frac{(2\pi)^2}{q_p} \beta L^{2p-n}m(\frac{1}{2}-\{\theta_{p}^j\})}}{1-e^{-\frac{(2\pi)^2}{q_p} \beta L^{2p-n}m}}
\\
& +\frac{1}{\beta}\sum_{\substack{ j=1 \\\{\gamma_{p}^j\}\neq\frac{1}{2} }}^{\binom{n}{p+1}}\sum_{m=1}^{\infty}\frac{(-1)^{m+1}}{m}\ \frac{e^{-q_p \beta
L^{n-2p-2}m(\frac{1}{2}-\{\gamma_{p}^j\})}+e^{-q_p \beta L^{n-2p-2}m(\frac{1}{2}+\{\gamma_{p}^j\})}}{1-e^{-q_p \beta L^{n-2p-2}m}} ,\label{asym-expansion-2}
\end{split}
\end{equation}
where $N_{\theta}$ and $N_{\gamma}$ are the numbers of components of $\vec{\theta}_{p}$ and $\vec{\gamma}_{p}$, such that $\{\theta_{p}^{i}\}=\frac{1}{2}$ or
$\{\gamma_{p}^{i}\}=\frac{1}{2}$. All but the first term decrease exponentially for $\beta\rightarrow\infty$.\par

The internal energy can be written in the form

\begin{equation}
\widehat{\mathcal{U}}_{\mathbb{T}^{n}}^{(p)}(q_p;\beta,\gamma_{p},\theta_{p}) =\widehat{\mathcal{U}}_{\mathbb{T}^{n};Cas}^{(p)}(q_p;\gamma_{p},\theta_{p}) +
\frac{2\pi}{L}\binom{n-1}{p} \sum _{\vec{k}\in\mathbb{Z}_{0}^{n}}\frac{|\vec{k}|}{e^{2\pi\beta L^{-1}|\vec{k}|}-1}-\frac{\partial}{\partial\beta}(\beta f_{low}^{(p)}).
\label{internal-energy-torus-low}
\end{equation}
where the last two terms decrease exponentially for low temperatures and $\widehat{\mathcal{U}}_{\mathbb{T}^{n};Cas}^{(p)}$ denotes the regularized vacuum (Casimir) energy,
given by

\begin{equation}
\begin{split}
\widehat{\mathcal{U}}_{\mathbb{T}^{n};Cas}^{(p)}(q_p;\gamma_{p},\theta_{p}):=
&\lim_{\beta\rightarrow\infty}\widehat{\mathcal{U}}_{\mathbb{T}^{n}}^{(p)}(q_p;\beta,\gamma_{p},\theta_{p})
=\lim_{\beta\rightarrow\infty}\widehat{\mathcal{F}}_{\mathbb{T}^{n}}^{(p)}(q_p;\beta,\gamma_{p},\theta_{p})=\\
= &-\frac{1}{2L}\binom{n-1}{p}\ \pi
^{-\frac{n+1}{2}}\Gamma(\frac{n+1}{2}) E_n(\frac{n+1}{2};1,\ldots ,1)+\frac{2\pi^{2}}{q_p}L^{2p-n}\sum_{j=1}^{\binom{n}{p}}\{\theta_{p}^j\}^2\\
&+\frac{q_p}{2}L^{n-2p-2}\sum_{j=1}^{\binom{n}{p+1}}\{\gamma_{p}^j\}^2.
\end{split}\label{free-enery-zero}
\end{equation}
Whereas the vacuum energy of the propagating modes is negative, the topological fields give always a positive contribution to the total vacuum energy.\par

The pressure splits into

\begin{equation}
\widehat{\mathcal{P}}_{\mathbb{T}^{n}}^{(p)}(q_p;\beta,\gamma_{p},\theta_{p}) =
\widehat{\mathcal{P}}_{\mathbb{T}^{n};Cas}^{(p)}(q_p;\gamma_{p},\theta_{p})+\frac{2\pi}{n}L^{-(n+1)}\binom{n-1}{p} \sum
_{\vec{k}\in\mathbb{Z}_{0}^{n}}\frac{|\vec{k}|}{e^{2\pi\beta L^{-1}|\vec{k}|}-1}+\frac{\partial}{\partial L} f_{low}^{(p)}, \label{pressure-torus-low}
\end{equation}
where again the last two terms decrease exponentially with increasing temperature. The vacuum (Casimir) pressure is defined as the limit

\begin{equation}
\begin{split}
\widehat{\mathcal{P}}_{\mathbb{T}^{n};Cas}^{(p)}(q_p;\gamma_{p},\theta_{p}):=&\lim_{\beta\rightarrow\infty}\widehat{\mathcal{P}}_{\mathbb{T}^{n}}^{(p)}(q_p;\beta,\gamma_{p},\theta_{p})
\\ = & -\frac{1}{2n}\binom{n-1}{p}\ \pi ^{-\frac{n+1}{2}}\Gamma(\frac{n+1}{2}) E_n(\frac{n+1}{2};1,\ldots ,1)\ L^{-(n+1)}\\ &-\frac{2\pi^{2}}{q_p}L^{2(p-n)}\
\frac{2p-n}{n}\sum_{j=1}^{\binom{n}{p}}\{\theta_{p}^j\}^2 -\frac{q_p}{2}L^{-2(p+1)}\ \frac{n-2p-2}{n}\sum_{j=1}^{\binom{n}{p+1}}\{\gamma_{p}^j\}^2.
\end{split}\label{pressure-torus-zero}
\end{equation}
The vacuum pressure exerted by the propagating degrees of freedom is negative for all $p$. On the other hand, the sign of the pressure induced by the topological fields depends
on $p$. Finally, we find for the entropy

\begin{equation}
\begin{split}
\widehat{\mathcal{S}}_{\mathbb{T}^{n}}^{(p)}(q_p;\beta,\gamma_{p},\theta_{p}) =& -\binom{n-1}{p}\sum _{\vec{k}\in\mathbb{Z}_{0}^{n}}\ln{\left[ 1-e^{-2\pi\beta
L^{-1}|\vec{k}|}\right]}+2\pi\frac{\beta}{L}\binom{n-1}{p} \sum _{\vec{k}\in\mathbb{Z}_{0}^{n}}\frac{|\vec{k}|}{e^{2\pi\beta L^{-1}|\vec{k}|}-1}\\
&-\beta^{2}\frac{\partial}{\partial\beta}( f_{low}^{(p)}).
\end{split}\label{entropy-torus-low}
\end{equation}
In the zero temperature limit the entropy related to the propagating modes vanishes. According to the series expansion \eqref{asym-expansion-2}, the topological modes (i.e. the
last term in \eqref{entropy-torus-low}) contribute, whenever the topological fields have components which are multiples of $\frac{1}{2}$. All other terms decrease exponentially.
In summary, the entropy converges to

\begin{equation}
\widehat{\mathcal{S}}_{\mathbb{T}^{n};Cas}^{(p)}(q_p;\gamma_{p},\theta_{p}):=\lim_{\beta\rightarrow\infty}
\widehat{\mathcal{S}}_{\mathbb{T}^{n}}^{(p)}(q_p;\beta,\gamma_{p},\theta_{p})=\ln
2^{N_{\theta}+N_{\gamma}},
\end{equation}
indicating that the ground state of the system is degenerate of degree $2^{N_{\theta}+N_{\gamma}}$. The maximum entropy at zero temperature is obtained for $N_{\theta
;max}=\binom{n}{p}$ and $N_{\gamma ;max}=\binom{n}{p+1}$, resulting in

\begin{equation}
\widehat{\mathcal{S}}_{\mathbb{T}^{n};Cas}^{(p)}(q_p;\gamma_{p},\theta_{p})|_{N_{\theta ;max},N_{\gamma ;max}}=\binom{n+1}{p+1}\ln2.
\end{equation}
\par

\subsubsection{High-temperature regime}

To study the high-temperature behavior we will not apply the general formula for the asymptotic expansion \eqref{K3} but derive an exact formula for the free energy and
subsequently for the other thermodynamic functions. The starting point is \eqref{free-energy-general-zeta} together with \eqref{top-sum-thermal} and the second equation in
\eqref{tau-integral-thermal}, but now applied to $\mathbb{T}_{\beta}^{n}:=\mathbb{T}_{\beta}^{1}\times\mathbb{T}^{n}$. The eigenvalues of $\Delta
_{r}^{\mathbb{T}_{\beta}^n}|_{im (d_{r+1}^{\mathbb{T}_{\beta}^n})^{\dag}}$ read

\begin{equation}
\nu _{k_{0},\vec{k}}=\left(\frac{2\pi k_0}{\beta}\right)^2+\sum _{i=1}^n\left(\frac{2\pi k_i}{L}\right)^2,\quad (k_0,\vec{k})=(k_0,k_1,\ldots
,k_n)\in\mathbb{Z}_{0}^{n+1}.\label{eigenvalues-torus}
\end{equation}
such that the resulting zeta function becomes

\begin{equation}
\zeta (s;\Delta _{r}^{\mathbb{T}_{\beta}^n}|_{im (d_{r+1}^{\mathbb{T}_{\beta}^n})^{\dag}})=\binom{n}{p}\ E_{n+1}(s;\frac{2\pi}{\beta},\frac{2\pi}{L},\ldots
,\frac{2\pi}{L}).\label{epstein-torus-high}
\end{equation}
In the first step we apply the Chowla-Selberg formula \eqref{epstein-expansion} to \eqref{epstein-torus-high} by setting $r=n+1$, $l=n$, $c_1=\ldots c_{n}=\frac{2\pi}{L}$ and
$c_{n+1}=\frac{2\pi}{\beta}$. In the second step we extract the leading term for the high-temperature regime from the corresponding $b_{p+1}(\mathbb{T}^{n})$-dimensional Riemann
Theta function in \eqref{top-sum-thermal} by using \eqref{duality-1}. A lengthy calculation using \eqref{relation-determinant} gives the following expression for the free energy

\begin{equation}
\begin{split}
\widehat{\mathcal{F}}_{\mathbb{T}^{n}}^{(p)}(q_p;\beta,\gamma_{p},\theta_{p})= &-\binom{n-1}{p}\pi^{-\frac{n+1}{2}} \Gamma (\frac{n+1}{2})\zeta _{R}(n+1)
\beta ^{-(n+1)}L^{n}\\
&-2\binom{n-1}{p}\beta ^{-\frac{n+2}{2}}L^{\frac{n}{2}} \sum _{k_{n+1}=1}^{\infty}\ \sum
_{\vec{k}\in\mathbb{Z}_{0}^{n}}k_{n+1}^{\frac{n}{2}}|\vec{k}|^{-\frac{n}{2}}K_{\frac{n}{2}}\left(2\pi k_{n+1}\frac{L}{\beta}|\vec{k}|\right)\\ &
-\frac{1}{2\beta}\left[\binom{n-1}{p}E_{n}^{\prime}(0;1,\ldots ,1)-\ln \left(
(2\pi)^{\Lambda_{1}}L^{\Lambda_{2}}\beta^{\Lambda_{3}}q_p^{\Lambda_{4}}\right)\right]\\
& -\frac{1}{\beta}\sum _{j=1}^{\binom{n}{p}}\ln{\Theta _1\left(\{\theta_{p}^j\}|-\frac{q_p}{2\pi i}\frac{L^{n-2p}}{\beta}\right)} -\frac{1}{\beta}\sum
_{j=1}^{\binom{n}{p+1}}\ln{\Theta _1 \left(\{\gamma_{p}^j\}|\frac{2\pi i}{q_p} \frac{L^{2p+2-n}}{\beta}\right)},
\end{split}\label{free-energy-high}
\end{equation}
with the constants

\begin{equation}
\begin{split}
&\Lambda_{1}:=\binom{n}{p}-\binom{n}{p+1}-2\binom{n-1}{p}\\ &\Lambda_{2}:=2\binom{n-1}{p}-\binom{n}{p}(n-2p)+\binom{n}{p+1}(n-2p-2)\\
&\Lambda_{3}:=\binom{n}{p+1}+\binom{n}{p}-2\binom{n-1}{p}\\ &\Lambda_{4}:=\binom{n}{p+1}-\binom{n}{p}.
\end{split}\label{free-energy-constants}
\end{equation}
As expected from the general structure, the leading contribution is once again the Stefan-Boltzmann term in $n$ dimensions. Let us denote the last two terms in
\eqref{free-energy-high} by $f_{high}^{(p)}$. Since the condition \eqref{constraint} is satisfied, one gets the following series expansion for $f_{high}^{(p)}$, using
\eqref{asymptotic-expansion},

\begin{multline}
f_{high}^{(p)}=\frac{1}{\beta}\left[\binom{n}{p}\sum_{m=1}^{\infty}\frac{1}{m}\ \frac{1}{1-e^{q_p \frac{L^{n-2p}}{\beta}m}}+
2\sum_{j=1}^{\binom{n}{p}}\sum_{m=1}^{\infty}\frac{(-1)^m}{m}\ \frac{e^{\frac{q_p}{2}\frac{L^{n-2p}}{\beta}m}\cos (2\pi m\{\theta_{p}^{j}\})}
{1-e^{q_p \frac{L^{n-2p}}{\beta}m }}\right]\\
+\frac{1}{\beta}\left[\binom{n}{p+1}\sum_{m=1}^{\infty}\frac{1}{m}\ \frac{1}{1-e^{\frac{(2\pi)^{2}}{q_p} \frac{L^{2p+2-n}}{\beta}m
}}+2\sum_{j=1}^{\binom{n}{p+1}}\sum_{m=1}^{\infty}\frac{(-1)^m}{m}\ \frac{e^{\frac{2\pi^{2}}{q_p}\frac{L^{2p+2-n}}{\beta}m}\cos (2\pi m\{\gamma_{p}^{j}\})}
{1-e^{\frac{(2\pi)^{2}}{q_p}\frac{L^{2p+2-n}}{\beta}m }}\right].\label{asym-expansion-3}
\end{multline}
All terms of $f_{high}^{(p)}$ decrease exponentially for $\beta\rightarrow 0$. Since $K_{\nu}(z)\simeq\sqrt{\frac{\pi}{2z}}e^{-z}\left(1+\mathcal{O}(\frac{1}{z})\right)$ for
$|z|\rightarrow\infty$ \cite{gradshteyn}, the summands in the second term of \eqref{free-energy-high} show an exponential decrease in the limit $\beta\rightarrow 0$ as well. Let
us now briefly display the other thermodynamic functions in the high temperature regime. For the internal energy one gets

\begin{equation}
\begin{split}
\widehat{\mathcal{U}}_{\mathbb{T}^{n}}^{(p)}(q_p;\beta,\gamma_{p},\theta_{p}) =&n\binom{n-1}{p}\pi^{-\frac{n+1}{2}} \Gamma (\frac{n+1}{2})\zeta _{R}(n+1) \frac{L^{n}}{\beta
^{n+1}} +\frac{\Lambda_{3}}{2\beta}-\frac{\partial}{\partial\beta}\left(\beta f_{high}^{(p)}\right)\\ & -4\pi\binom{n-1}{p}L^{\frac{n+2}{2}}\beta^{-\frac{n+4}{2}}\sum
_{k_{n+1}=1}^{\infty}\ \sum _{\vec{k}\in\mathbb{Z}_{0}^{n}}k_{n+1}^{\frac{n+2}{2}}\ |\vec{k}|^{\frac{2-n}{2}}\ K_{\frac{n}{2}-1}\left(2\pi
k_{n+1}\frac{L}{\beta}|\vec{k}|\right).\label{internal-energy-high}
\end{split}
\end{equation}
Here we have used that $\frac{d}{dz}K_{\nu}(z)=-K_{\nu-1}(z)-\frac{\nu}{z}K_{\nu}(z)$ \cite{gradshteyn}. In addition to the leading term of Stefan-Boltzmann type and the
exponentially decreasing last two terms in \eqref{internal-energy-high}, there appears a topological term linear in the temperature. For the thermal pressure we obtain

\begin{equation}
\begin{split}
\widehat{\mathcal{P}}_{\mathbb{T}^{n}}^{(p)}(q_p;\beta,\gamma_{p},\theta_{p}) =&\binom{n-1}{p}\pi^{-\frac{n+1}{2}} \Gamma (\frac{n+1}{2})\zeta _{R}(n+1) \beta
^{-(n+1)}-\frac{\Lambda_{2}}{2n\beta L^{n}}+\frac{1}{nL^{n-1}}\frac{\partial}{\partial L}f_{high}^{(p)}\\ & -\frac{4\pi}{n}
\binom{n-1}{p}L^{\frac{2-n}{2}}\beta^{-\frac{n+4}{2}}\sum _{k_{n+1}=1}^{\infty}\ \sum _{\vec{k}\in\mathbb{Z}_{0}^{n}}k_{n+1}^{\frac{n+2}{2}}\ |\vec{k}|^{\frac{2-n}{2}}\
K_{\frac{n}{2}-1}\left(2\pi k_{n+1}\frac{L}{\beta}|\vec{k}|\right),\label{pressure-high}
\end{split}
\end{equation}
where the last two terms show an exponential decrease for $\beta\rightarrow 0$. Like before the corresponding topological subleading term is linear in the temperature. Finally,
the entropy admits the following form

\begin{equation}
\begin{split}
\widehat{\mathcal{S}}_{\mathbb{T}^{n}}^{(p)}(q_p;\beta,\gamma_{p},\theta_{p}) =& (n+1)\binom{n-1}{p}\pi^{-\frac{n+1}{2}} \Gamma (\frac{n+1}{2})\zeta _{R}(n+1)
\left(\frac{L}{\beta}\right)^{n} -\beta^{2}\frac{\partial}{\partial\beta}f_{high}^{(p)}\\ & +2\binom{n-1}{p}\left(\frac{L}{\beta}\right)^{\frac{n}{2}} \sum
_{k_{n+1}=1}^{\infty}\ \sum _{\vec{k}\in\mathbb{Z}_{0}^{n}}k_{n+1}^{\frac{n}{2}}|\vec{k}|^{-\frac{n}{2}}K_{\frac{n}{2}}\left(2\pi k_{n+1}\frac{L}{\beta}|\vec{k}|\right)\\ &
-4\pi\binom{n-1}{p}\left(\frac{L}{\beta}\right)^{\frac{n+2}{2}} \sum _{k_{n+1}=1}^{\infty}\ \sum _{\vec{k}\in\mathbb{Z}_{0}^{n}}k_{n+1}^{\frac{n+2}{2}}\
|\vec{k}|^{\frac{2-n}{2}}\ K_{\frac{n}{2}-1}\left(2\pi k_{n+1}\frac{L}{\beta}|\vec{k}|\right)\\ &+\frac{1}{2}\left[\binom{n-1}{p} E_{n}^{\prime}(0;1,\ldots,1)+\Lambda_{3}-\ln
\left( (2\pi)^{\Lambda_{1}}L^{\Lambda_{2}}\beta^{\Lambda_{3}}q_p^{\Lambda_{4}}\right)\right].\label{entropy-high}
\end{split}
\end{equation}
In the high temperature limit all terms except the first and the last one in \eqref{entropy-high} are exponentially suppressed.\par

In summary, we have explicitly shown that in the high temperature limit all thermodynamic functions are dominated by terms of Stefan-Boltzmann type which are related to the
propagating degrees of freedom. The topological contributions appear as subleading corrections. Although the formulae for the thermodynamic functions look quite different in the
low- and high temperature regimes, each of them is valid for the whole temperature range.\par

Before closing this section we want to address briefly two topics: the thermodynamic limit of the free energy density and the entropy to energy ratio. Concerning the first one
the starting point is \eqref{free-energy-high}. Depending on the degree $p$ one has to use the series expansions either \eqref{asym-expansion-2} or \eqref{asym-expansion-3} to
compute the large volume limit for the topological contributions. In summary, it can be shown that the topological effects disappear so that all terms but the Stefan-Boltzmann
term vanish and we end up with (see \eqref{euclidean})

\begin{equation}
\lim_{L\rightarrow\infty}\widehat{\mathcal{F}}_{\mathbb{T}^{n}}^{(p)}/L^{n}=-\binom{n-1}{p}\pi^{-\frac{n+1}{2}} \Gamma (\frac{n+1}{2})\zeta _{R}(n+1) \beta ^{-(n+1)}.
\end{equation}
Regarding the second topic we give an estimate for the entropy to internal energy ratio in the high-temperature limit using \eqref{internal-energy-high} and
\eqref{entropy-high}, however, neglecting exponentially decreasing terms. This ratio can be written in the form

\begin{equation}
\frac{\widehat{\mathcal{S}}_{\mathbb{T}^{n}}^{(p)}}{\widehat{\mathcal{U}}_{\mathbb{T}^{n}}^{(p)}}\approx \frac{n+1}{n}\beta +
\mathcal{O}\left(\beta^{n+1}(1+\ln\beta)\right),\quad \beta\rightarrow 0.
\end{equation}
The leading term is linear in $\beta$ and derives from the corresponding Stefan-Boltzmann terms. This term is in some sense universal since it depends only on the dimension of
the torus and is independent of $p$. The topological and geometrical properties contribute through further subleading terms in the expansion. For instance, in dimension $n=3$
and for $p=1$ the first term agrees exactly with the classical result for the entropy to energy ratio of the black body photon gas enclosed in a large box. For general $p$, our
result compares further to \cite{BMT}, where the same leading term was obtained for the $p$-form Maxwell theory at finite temperature on hyperbolic spaces. Once again, this
result highlights that in the infinite temperature limit the system of higher-abelian gauge fields is controlled by the propagating degrees of freedom.

\section{Polyakov loop operator in higher-abelian gauge theories}

\subsection{The static brane antibrane free energy}

So far we have discussed the impact of the topology of the spatial background on the thermodynamic functions of higher-abelian gauge fields. In the next step we want to probe
the topological effect on the two-point correlation function of a higher-abelian generalization of the Polyakov loop operator.

In ordinary gauge theory at finite temperature the Polyakov loop operator is a variant of the Wilson loop operator and measures the holonomy of the gauge potential 1-form along
the periodic (thermal) time direction $\mathbb{T}_{\beta}^{1}$. The corresponding correlation functions are interpreted as free energy in the presence of static charges relative
to the free energy of the pure gauge field background. In particular, the two-point correlation function gives the effective potential between a pair of oppositely charged
static particles. Moreover, the Polyakov loop operator defines an order parameter for the confinement-deconfinement transition, even in the abelian case (see e.g.
\cite{Sachs-Wipf, GSST-2, GSST-1}).\par

In order to generalize the Polyakov loop operator to higher-abelian gauge fields, we replace the static charge located at a point $x\in X$ (which is a singular 0-cycle) by a
static, closed brane, which is represented by a smooth singular $(p-1)$-cycle $\Sigma\in Z_{p-1}(X;\mathbb{Z})$. Correspondingly, the world-line
$\mathbb{T}_{\beta}^{1}\times\{x\}\in Z_{1}(X_{\beta};\mathbb{Z})$ of the static particle placed at $x\in X$ along periodic (thermal) time is replaced by the world-volume
$\Sigma_{\beta}:=\mathbb{T}_{\beta}^{1}\times\Sigma\in Z_{p}(X_{\beta};\mathbb{Z})$ of the static brane $\Sigma$. \par

For a smooth singular $(p-1)$-cycle $\Sigma\in Z_{p-1}(X;\mathbb{Z})$ we define the \textit{higher-abelian Polyakov loop operator of degree $p$} by

\begin{equation}
\Sigma\mapsto\frak P_{\Sigma}^{(p)}(\hat{u}_{p}):=\hat{u}_{p}(\Sigma_{\beta}),\qquad\Sigma_{\beta}:=\mathbb{T}_{\beta}^{1}\times\Sigma\in Z_{p}(X_{\beta};\mathbb{Z})
\end{equation}
For topologically trivial differential characters which are of the form $j_2([A_{p}])\in\widehat H^p(X_{\beta})$, for $A_{p}\in\Omega^{p}(X_{\beta})$, the higher-abelian
Polyakov loop operator reduces to

\begin{equation}
\frak P_{\Sigma}^{(p)}(j_2([A_{p}]))=\exp{2\pi i\left(\int_{\Sigma_{\beta}}A_{p}\right)},
\end{equation}
which is nothing but the minimal coupling of the $p$-form gauge field $A_{p}$ to the world-volume $\Sigma_{\beta}$.\par

In the following we will assign charges $\pm 1$ to the brane and antibrane, respectively\footnote{In general we could, however, assign a charge $\tilde{q}\in\mathbb{Z}$ to the
brane by replacing $\Sigma$ by $\tilde{q}\Sigma$. Notice that the total brane charge is identically zero on a compact and closed manifold.}. In analogy to the case of point
particles, we introduce the \textit{static brane antibrane free energy} $\frak{F}_{\Sigma^{(1)},-\Sigma^{(2)}}^{(p)}$ by the two-point correlation function of the higher-abelian
Polyakov loop operator

\begin{equation}
e^{-\beta\frak{F}_{\Sigma^{(1)},-\Sigma^{(2)}}^{(p)}}:=<\frak{P}_{\Sigma^{(1)}}^{(p)}\ \overline{\frak{P}_{\Sigma^{(2)}}^{(p)}}>.\label{static free energy}
\end{equation}
We interpret $\frak{F}_{\Sigma^{(1)},-\Sigma^{(2)}}^{(p)}$ as change in the free energy of the system in the presence of a pair $\Sigma^{(1)}, \Sigma^{(2)}\in
Z_{p-1}(X;\mathbb{Z})$ of static, closed branes with opposite charges with respect to the background at the same temperature in the absence of branes.\par

Let us now compute the static brane antibrane free energy for extended higher-abelian Maxwell theory using the general formula \eqref{VEV-general}, yet specified to $X_{\beta}$.
For sake of simplicity we consider the torsion free case, i.e. $H_{p-1;tor}(X;\mathbb{Z})=0$. From

\begin{equation}
\left(\frak P_{\Sigma_{1}}^{(p)}\ \overline{\frak P_{\Sigma^{(2)}}^{(p)}}\right)(\hat{u}_{p})= \frak P_{\Sigma^{(1)}}^{(p)}(\hat{u}_{p})\ \overline{\frak
P_{\Sigma^{(2)}}^{(p)}}(\hat{u}_{p})= \frak P_{\Sigma^{(1)}-\Sigma^{(2)}}^{(p)}(\hat{u}_{p})=\hat{u}_{p}\left(\Sigma_{\beta}^{(1)}-\Sigma_{\beta}^{(2)}\right),
\end{equation}
where $\Sigma_{\beta}^{(1)}-\Sigma_{\beta}^{(2)}=\mathbb{T}_{\beta}^{1}\times (\Sigma^{(1)}-\Sigma^{(2)})$ is understood as difference in $C_{p}(X_{\beta};\mathbb{Z})$, we get
in terms of the local coordinates $\vec{y}_{p-1}=(y_1,\ldots,y_{b_{p-1}(X)})\in\mathbb{T}^{b_{p-1}(X)}$, $\vec{z}_{p}=(z_{1},\ldots,z_{b_{p}(X)})\in\mathbb{T}^{b_{p}(X)}$ and
$\tau_{p}\in im(d_{p+1}^{X_{\beta}})^{\dag}$,

\begin{multline}
\left(\underline{\frak P_{\Sigma^{(1)}-\Sigma^{(2)}}^{(p)}}\right)^{\hat{\frak c}_{X_{\beta}}^{(p+1)}}(\vec{y}_{p-1},\vec{z}_{p},\tau_{p} )= \\
=\hat{\frak c}_{X_{\beta}}^{(p+1)}\left(\Sigma_{\beta}^{(1)}-\Sigma_{\beta}^{(2)}\right)\ \exp{2\pi i\left(\int_{\Sigma_{\beta}^{(1)}-\Sigma_{\beta}^{(2)}}\tau_{p}\right)}\
\prod_{j=1}^{b_{p-1}(X)}y_{j}^{(\int_{\Sigma^{(1)}-\Sigma^{(2)}}(\rho_{X}^{(p-1)})_{j})}.
\end{multline}
Then the relevant Fourier coefficient \eqref{fourier} becomes

\begin{multline}
\left[\left(\underline{\frak P_{\Sigma^{(1)}-\Sigma^{(2)}}^{(p)}}\right)^{\hat{\frak c}_{X_{\beta}}^{(p+1)}}(\tau_{p})\right]_{(0,\dots,0)}=\\=
\begin{cases}
0, &\text{if $\int_{\Sigma^{(1)}-\Sigma^{(2)}}(\rho_{X}^{(p-1)})_{j}\neq 0$},\\
\hat{\frak c}_{X_{\beta}}^{(p+1)}\left(\Sigma_{\beta}^{(1)}-\Sigma_{\beta}^{(2)}\right)\ \exp{2\pi i\left(\int_{\Sigma_{\beta}^{(1)}-\Sigma_{\beta}^{(2)}}\tau_{p}\right)},
&\text{if $\int_{\Sigma^{(1)}-\Sigma^{(2)}}(\rho_{X}^{(p-1)})_{j}= 0$}.
\end{cases}\label{VEV-Polyakov-1}
\end{multline}
Since the basis of harmonic forms with integer periods $(\rho_{X}^{(p-1)})_{j}$ generates $H_{free}^{p-1}(X;\mathbb{Z})$, the two-point correlator vanishes unless
$\Sigma^{(1)}-\Sigma^{(2)}$ is a boundary\footnote{This statement on the two-point correlation function is valid even if $H_{p-1}(X;\mathbb{Z})$ has torsion. In fact, let $\frak
W_{\sigma}^{(p)}$ denote the (higher-abelian) Wilson operator assigned to a $p$-cycle $\tilde\Sigma\in Z_{p}(M;\mathbb{Z})$ which is defined by $\frak
W_{\tilde\Sigma}^{(p)}(\hat{u}_{p}):=\hat{u}_{p}(\tilde\Sigma)$. It was shown in \cite{kelnhofer-1} that the VEV of the Wilson loop operator vanishes on nontrivial cycles. Since
$\frak P_{\Sigma}^{(p)}=\frak W_{\mathbb{T}_{\beta}^{1}\times\Sigma}^{(p)}$, one gets $ <\frak P_{\Sigma^{(1)}}^{(p)}\ \overline{\frak P_{\Sigma^{(2)}}^{(p)}}>=0$ unless
$[\Sigma^{(1)}-\Sigma^{(2)}]= 0$ in $H_{p-1}(X;\mathbb{Z})$.}. Let $\sigma\in C_{p}(X;\mathbb{Z})$ be the $p$-chain such that $\Sigma^{(1)}-\Sigma^{(2)}=\partial\sigma$. By
substituting \eqref{cohomology-thermal} into \eqref{VEV-Polyakov-1} one gets

\begin{equation}
\left[\left(\underline{\frak P_{\partial\sigma}}\right)^{\hat{\frak c}_{X_{\beta}}^{(p+1)}}(\tau_{p})\right]_{(0,\dots,0)}= \exp{2\pi i\left(
\sum_{j=1}^{b_p(X)}m_{p}^{j}\int_{\sigma}(\rho_{X}^{(p)})_{j}+\int_{(\partial\sigma)_{\beta}}\tau_{p}\right)}.
\end{equation}
where $(\partial\sigma)_{\beta}:=\mathbb{T}_{\beta}^{1}\times\partial\sigma$. According to \eqref{VEV-general} we have to take the sum over all topological sectors, leading to

\begin{equation}
\begin{split}
&\sum_{\frak c_{X_{\beta}}^{(p+1)}\in H^{p+1}(X_{\beta})}e^{-S_{(\gamma_{p},\theta_{p})}(\hat{\frak c}_{X_{\beta}}^{(p+1)})+2\pi i\sum_{j=1}^{b_p(X)}m_{p}^{j}
\int_{\sigma}(\rho_{X}^{(p)})_{j}}=\\&=\Theta _{b_{p+1}(X)}\begin{bmatrix} \vec{\gamma}_{p} \\0\end{bmatrix} \left(0|-\frac{q_{p}}{2\pi i}\beta h_X^{(p+1)}\right)\Theta
_{b_{p}(X)}\left((-1)^{(p+1)(n-p)}\vec{\theta}_{p}+\int_{\sigma}\vec{\rho}_{X}^{\ (p)}\bigg|-\frac{q_{p}}{2\pi i}\beta^{-1} h_X^{(p)}\right).\label{brane top sector}
\end{split}
\end{equation}
Here $\int_{\sigma}\vec{\rho}_{X}^{\ (p)}$ stands for the vector $(\int_{\sigma}(\rho_{X}^{(p)})_{1},\ldots ,\int_{\sigma}(\rho_{X}^{(p)})_{b_p(X)})$. In order to perform the
integration over $im(d_{p+1}^{X_{\beta}})^{\dag}$ we introduce two $p$-currents $J_{\Sigma_{\beta}^{(1)}}$, $J_{\Sigma_{\beta}^{(2)}}$, which are associated with the two cycles
$\Sigma_{\beta}^{(1)}$ and $\Sigma_{\beta}^{(2)}$ by

\begin{equation}
\int_{\Sigma_{\beta}^{(j)}}B = \int_{X_{\beta}}B\wedge \star J_{\Sigma_{\beta}^{(j)}}=<B,J_{\Sigma_{\beta}^{(j)}}>, \quad \forall B\in\Omega^{p}(X_{\beta}), \quad j=1,2.
\end{equation}
In general, the two $p$-currents belong to the dual of the space of differential forms, which can be given a precise meaning in the context of de Rham currents\footnote{Here we
regard the de Rham currents as differential forms with distributional coefficients.} \cite{deRham}. Since the Hodge decomposition theorem also holds for de Rham currents, one
has

\begin{equation} \int_{\Sigma_{\beta}^{(j)}}\tau_{p} = <\tau_{p} ,\Pi ^{im(d_{p+1}^{X_{\beta}})^{\dag}}(J_{\Sigma_{\beta}^{(j)}})>,
\quad\forall\tau_{p}\in im(d_{p+1}^{X_{\beta}})^{\dag},\quad j=1,2,\label{current-brane}
\end{equation}
where $\Pi ^{im(d_{p+1}^{X_{\beta}})^{\dag}}$ denotes the projector onto the co-exact de Rham currents. Together with \eqref{brane top sector} the Gaussian integration over
$\tau_{p}$ leads to

\begin{equation}
\begin{split}
<\frak P_{\Sigma_{1}}^{(p)}\ \overline{\frak{P}_{\Sigma^{(2)}}^{(p)}}>=&\frac{\Theta _{b_{p}(X)}\left((-1)^{(p+1)(n-p)}\vec{\theta}_{p}+\int_{\sigma}\vec{\rho}_{X}^{\
(p)}\big|-\frac{q_{p}}{2\pi i} \beta^{-1}h_X^{(p)}\right)}{\Theta _{b_{p}(X)}\left(\vec{\theta}_{p}|-\frac{q_{p}}{2\pi i}\beta^{-1} h_X^{(p)}\right)}\\ &\times
\exp\left(-\frac{2\pi^{2}}{q_p}<\Pi ^{im(d_{p+1}^{X_{\beta}})^{\dag}}(J_{\Sigma_{\beta}^{(1)}}-J_{\Sigma_{\beta}^{(2)}}),G_{p}^{X_{\beta}}\Pi
^{im(d_{p+1}^{X_{\beta}})^{\dag}}(J_{\Sigma_{\beta}^{(1)}}-J_{\Sigma_{\beta}^{(2)}})>\right).\label{polyakov_wilson_loop}
\end{split}
\end{equation}
Notice that the two-point function is independent of the external field $\gamma_{p}$. In the next step we want to determine the temperature dependence of the current-current
term in \eqref{polyakov_wilson_loop}. For this let us introduce the normalized eigenforms $\chi_{m}^{(0)}(t):=\beta^{-\frac{1}{2}}e^{2\pi imt}$ and
$\chi_{m}^{(1)}(t):=\beta^{\frac{1}{2}}e^{2\pi imt}dt$ of $\Delta_{r}^{\mathbb{T}_{\beta}^1}$ (for $r=0,1$). Furthermore, let
$\{\nu_{\alpha}(\Delta_{r}^{X}|_{im(d_{r+1}^{X})^{\dag}})|\alpha\in I\footnote{For sake of simplifying notation we denote the index sets for labeling the eigenvalues and
eigenforms of the Laplace operators collectively by the same symbol $I$.}\}$ be the set of eigenvalues of $\Delta_{r}^{X}|_{im(d_{r+1}^{X})^{\dag}}$ and let
$\{\psi_{X;\alpha}^{(r)}\in im(d_{r+1}^{X})^{\dag}|\alpha\in I\}$ denote an orthonormal $L^2$-basis of associated co-exact eigenforms

\begin{equation}
\Delta_{r}^{X}|_{im(d_{r+1}^{X})^{\dag}}\ \psi_{X;\alpha}^{(r)}=\nu_{\alpha}(\Delta_{r}^{X}|_{im(d_{r+1}^{X})^{\dag}})\ \psi_{X;\alpha}^{(r)}.\label{ONB-XX}
\end{equation}
An orthonormal $L^2$-basis of co-exact eigenforms of $\Delta_{p}^{X_{\beta}}|_{im(d_{p+1}^{X_{\beta}})^{\dag}}$ can be constructed by

\begin{equation}
\begin{split}
&(\Psi_{1}^{(p)})_{m\alpha}=pr_{1}^{\ast}\chi_{m}^{(0)}\wedge pr_2^{\ast}\psi_{X;\alpha}^{(p)},\quad m\in\mathbb{Z}, \alpha\in I \\
&(\Psi_{2}^{(p)})_{mj}=pr_{1}^{\ast}\chi_{m}^{(0)}\wedge pr_2^{\ast}(\tilde\rho_{X}^{(p)})_{j},\quad m\in\mathbb{Z}_{0},\\
&(\Psi_{3}^{(p)})_{\alpha}=pr_{1}^{\ast}\chi_{0}^{(1)}\wedge pr_2^{\ast}\psi_{X;\alpha}^{(p-1)},\quad \alpha\in I,
\end{split}\label{ONB-coexact}
\end{equation}
where $(\tilde\rho_{X}^{(p)})_{j}$ is an orthonormal basis of $\mathcal{H}^{p}(X)$. With respect
to this basis one derives from \eqref{current-brane}

\begin{equation}
\Pi ^{im(d_{p+1}^{X_{\beta}})^{\dag}}(J_{\Sigma_{\beta}^{(j)}})=\sqrt{\beta}\ \sum_{\alpha\in I} \left(\int_{\Sigma^{(j)}}\psi_{X;\alpha}^{(p-1)}\right)
(\Psi_{3}^{(p)})_{\alpha},\label{current-result}
\end{equation}
which gives for the current-current term

\begin{equation}
<\Pi ^{im(d_{p+1}^{X_{\beta}})^{\dag}}(J_{\Sigma_{\beta}^{(1)}}-J_{\Sigma_{\beta}^{(2)}}),G_{p}^{X_{\beta}}\Pi
^{im(d_{p+1}^{X_{\beta}})^{\dag}}(J_{\Sigma_{\beta}^{(1)}}-J_{\Sigma_{\beta}^{(2)}})>=\beta \sum_{\alpha\in I}\frac{|\int_{\sigma} d\psi_{X;\alpha}^{(p-1)}|^{2}}{\nu
_{\alpha}(\Delta _{p-1}^{X}|_{im(d_{p}^{X})^{\dag}})\ }.\label{formula_current}
\end{equation}
Here the eigenforms and eigenvalues appear as often as their multiplicity in the sum over $I$. Due to the distributional structure of the currents, this infinite sum is singular
and requires regularization\footnote{An easy example can be given in the case of electrodynamics ($p=1$) on a circle $X=\mathbb{T}^{1}$: Let two charges be located at the points
$z_{1,2}=e^{2\pi i t_{1,2}}$. The points can be joined by a path $s\mapsto\sigma (s)=e^{2\pi i(t_{1}+s(t_{2}-t_{1}))}$. Together with the eigenfunctions $\psi
_{m}^{(0)}(\vec{t})=L^{-\frac{1}{2}}e^{2\pi imt}$ of $\Delta _{0}^{\mathbb{T}^1}|_{im(d_{1}^{\mathbb{T}^1})^{\dag}}$ associated to the eigenvalues $\nu _{m}(\Delta
_{0}^{\mathbb{T}^1}|_{im(d_{1}^{\mathbb{T}^1})^{\dag}})=\left(\frac{2\pi m}{L}\right)^{2}$, $m\in\mathbb{Z}_{0}$,  the sum in \eqref{current-result} becomes

\begin{equation}
\begin{split}
\frac{L}{q_1}\sum_{m\in\mathbb{Z}_{0}}\frac{1-e^{2\pi im(t_{2}-t_{1})}}{m^{2}} &=\frac{L}{q_1}\left(E_1\begin{bmatrix} 0
\\0\end{bmatrix} (1;\mathbf{1})-E_1\begin{bmatrix} 0
\\t_{2}-t_{1}\end{bmatrix} (1;\mathbf{1})\right)\\ &=\frac{L}{q_1}\Big( 2\zeta_{R}(2)-2\pi^{2}B_2(\langle t_{2}-t_{1}\rangle)\Big)
=\frac{2\pi^2}{q_1}\Big( L\langle t_{2}-t_{1}\rangle\big(1- \frac{L\langle t_{2}-t_{1}\rangle}{L}\big)\Big),\nonumber
\end{split}
\end{equation}
where we have regularized the infinite sum on the left hand side by the Epstein zeta function \eqref{epstein-original}. $B_2$ denotes the periodic second Bernoulli polynomial
and $\langle t_{2}-t_{1}\rangle$ is the fractional part of $t_{2}-t_{1}$. Notice that $L\langle t_{2}-t_{1}\rangle$ is the distance between the two charges. Thus the regularized
sum is exactly the Coulomb energy between two charges on a circle which is linear and quadratic in the distance between the two charges (see e.g. \cite{RZ}).}. There exists an
alternative (formal) expression for \eqref{formula_current}: We choose an orthonormal $L^2$-basis

\begin{equation}
\tilde{\psi}_{X;\alpha}^{(p)}:=\nu _{\alpha}(\Delta _{p-1}^{X}|_{im(d_{p}^{X})^{\dag}})^{-\frac{1}{2}}d\psi _{X;\alpha}^{(p-1)},\label{ONB-X}
\end{equation}
of eigenforms of $\Delta_p^{X}|_{im(d_{p-1}^{X})}$ with eigenvalues $\nu _{\alpha}(\Delta _{p-1}^{X}|_{im(d_{p}^{X})^{\dag}})$. Let us define the de Rham current $J_{\sigma}$
associated to $\sigma\in C_p(X;\mathbb{Z})$ by

\begin{equation}
\int_{\sigma}C = \int_{X}C\wedge \underline{\star}\ J_{\sigma}=<C,J_{\sigma}>,\quad \forall C\in\Omega^{p}(X).
\end{equation}
By expanding $\Pi ^{im(d_{p-1}^{X})}(J_{\sigma})$ in terms of the basis \eqref{ONB-X}, one finds

\begin{equation}
\sum _{\alpha\in I}\frac{|\int_{\sigma} d\psi_{X;\alpha}^{(p-1)}|^{2}}{\nu _{\alpha}(\Delta _{p-1}^{X}|_{im(d_{p}^{X})^{\dag}})\ }=\sum _{\alpha\in
I}|<\tilde{\psi}_{X;\alpha}^{(p)},\Pi ^{im(d_{p-1}^{X})}(J_{\sigma})>|^{2}=\|\Pi ^{im(d_{p-1}^{X})}(J_{\sigma})\|^{2}.\label{vacuum-contribution-dynamical-original}
\end{equation}
In order to separate the temperature independent part in the contribution of the topological sector, we apply \eqref{duality} to the $b_p(X)$-dimensional Riemann Theta function
in \eqref{brane top sector}. Substituting \eqref{formula_current} into \eqref{polyakov_wilson_loop} leads to the final result for the static brane antibrane free energy

\begin{equation}
\begin{split}
\frak F_{\Sigma^{(1)},-\Sigma^{(2)}}^{(p)}& (\beta;\theta_{p})=\\=&\frac{2\pi^2}{q_p} \sum _{\alpha\in I}\frac{|\int_{\Sigma^{(1)}} \psi_{X;\alpha}^{(p-1)}-\int_{\Sigma^{(2)}}
\psi_{X;\alpha}^{(p-1)}|^{2}}{\nu _{\alpha}(\Delta _{p-1}^{X}|_{im(d_{p}^{X})^{\dag}})\ }- \frac{2\pi^2}{q_p}\sum_{j,k=1}^{b_p(X)}(h_X^{(p)})_{jk}^{-1}
\{\theta_{p}^{j}\}\{\theta_{p}^{k}\}\\ &+ \frac{2\pi^2}{q_p}\sum_{j,k=1}^{b_p(X)}(h_X^{(p)})_{jk}^{-1}\left\{(-1)^{(p+1)(n-p)}\theta_{p}^{j}+
\int_{\sigma}(\rho_{X}^{(p)})_{j}\right\}\left\{(-1)^{(p+1)(n-p)}\theta_{p}^{k}+\int_{\sigma}(\rho_{X}^{(p)})_{k}\right\}\\ &-\frac{1}{\beta}\ln\left[ \frac{\Theta
_{b_{p}(X)}\left( \frac{2\pi i}{q_{p}}\beta (h_X^{(p)})^{-1}\{(-1)^{(p+1)(n-p)}\vec{\theta}_{p}+\int_{\sigma}\vec{\rho}_{X}^{\ (p)}\}\big|\frac{2\pi i}{q_{p}}\beta
(h_X^{(p)})^{-1}\right)}{\Theta _{b_{p}(X)}\left( \frac{2\pi i}{q_{p}}\beta (h_X^{(p)})^{-1}\{\vec{\theta}_{p}\}|\frac{2\pi i}{q_{p}}\beta
(h_X^{(p)})^{-1}\right)}\right].\label{static-potential-final-split}
\end{split}
\end{equation}
The static brane antibrane free energy does not depend on the choice for the $p$-chain $\sigma$ connecting $\Sigma^{(1)}$ and $\Sigma^{(2)}$. In addition to the temperature
independent interaction term of Coulomb type, the topologically inequivalent field configurations give rise to both a temperature independent and a temperature dependent
contribution which are present even if $\theta_{p}\in \mathcal{H}_{\mathbb{Z}}^{n-p}(X_{\beta})$. In ordinary $p$-form Maxwell theory, however, only the Coulomb term is
present.\par

The static brane antibrane free energy is symmetric under interchanging the charges only if the components of $\vec{\theta}_{p}$ are either integers or half-integers, namely

\begin{equation}
\frak F_{\Sigma^{(1)},-\Sigma^{(2)}}^{(p)}(\beta;\theta_{p})=\frak F_{-\Sigma^{(1)},\Sigma^{(2)}}^{(p)}(\beta;\theta_{p}),\qquad\vec{\theta}_{p}\in
\frac{1}{2}\mathbb{Z}^{b_{p}(X)}. \label{static-potential-final-symmetry-1}
\end{equation}
For all other values of $\vec{\theta}_{p}$ the charge conjugation symmetry is broken. However, if in addition $\vec{\theta}_{p}$ is replaced by $-\vec{\theta}_{p}$ the symmetry
can be restored such that

\begin{equation}
\frak F_{\Sigma^{(1)},-\Sigma^{(2)}}^{(p)}(\beta;\theta_{p})=\frak F_{-\Sigma^{(1)},\Sigma^{(2)}}^{(p)}(\beta;-\theta_{p}).\label{static-potential-final-symmetry}
\end{equation}

\subsection{An example - the case $p=n$}

In the case $p=n$, the extended higher-abelian Maxwell theory does not have any propagating degrees of freedom. Let $\Sigma^{(1)}$ and $\Sigma^{(2)}$ be two smoothly embedded
$(n-1)$-dimensional submanifolds of $X$ representing the two static, oppositely charged branes and let $\sigma\subset X$ be a $n$-dimensional submanifold of $X$ such that
$\partial\sigma =\Sigma^{(1)}-\Sigma^{(2)}$. Then the de Rham current associated to $\sigma$ is given by

\begin{equation}
J_{\sigma}=\delta_{\sigma}vol_X,
\end{equation}
where $\delta_{\sigma}$ is the characteristic function on $\sigma$. This current is square summable \cite{deRham}. Since
$\Pi^{\mathcal{H}^{n}}(J_{\sigma})=\frac{vol(\sigma)}{vol(X)}vol_{X}$, one finds for the current-current term \eqref{vacuum-contribution-dynamical-original}

\begin{equation}
\|\Pi ^{imd_{n-1}}(J_{\sigma})\|^{2}=\|(id_{\Omega^{n}(X)}-\Pi^{\mathcal{H}^{n}})J_{\sigma}\|^{2}=vol(\sigma)\left(
1-\frac{vol(\sigma)}{vol(X)}\right),\label{vacuum-contribution-dynamical}
\end{equation}
where $vol(\sigma):=\int_{\sigma}vol_X$ denotes the volume of the submanifold $\sigma$. Compared to the example of electrodynamics on a circle discussed before, the distance
between two charged particles is now replaced by the volume of the boundary submanifold $\sigma$ connecting the two static branes. In addition to the term linear in the volume
of $\sigma$, there is also the quadratic term due to the compactness of $X$. From \eqref{static-potential-final-split} we get the following two expressions

\begin{equation}
\begin{split}
\frak F_{\Sigma^{(1)},-\Sigma^{(2)}}^{(n)}(\beta;\theta_{n})= & \frac{2\pi^2}{q_n}vol(\sigma)\left(1-\frac{vol(\sigma)}{vol(X)}\right)
+\frac{2\pi^2}{q_n}vol(X)\left(\{\theta_{n}+\frac{vol(\sigma)}{vol(X)}\}^{2}-\{\theta_{n}\}^{2}\right)\\ &-\frac{1}{\beta}\ln\left[ \frac{\Theta _{1}\left( \frac{2\pi i}{q_{n}}
\beta vol(X)\{\theta_{n}+\frac{vol(\sigma)}{vol(X)}\} |\frac{2\pi i}{q_{n}}\beta vol(X)\right)}{\Theta _{1}\left( \frac{2\pi i}{q_{n}} \beta vol(X)\{\theta_{n}\} |\frac{2\pi
i}{q_{n}}\beta vol(X)\right)}\right]\\ = &\frac{2\pi^2}{q_n}vol(\sigma)\left(1-\frac{vol(\sigma)}{vol(X)}\right)-\frac{1}{\beta}\ln\left[ \frac{\Theta _{1}\left(
\{\theta_{n}+\frac{vol(\sigma)}{vol(X)}\} |-\frac{q_{n}}{2\pi i}\beta^{-1} vol(X)^{-1}\right)}{\Theta _{1}\left( \{\theta_{n}\} |-\frac{q_{n}}{2\pi i}\beta^{-1}
vol(X)^{-1}\right)}\right].\label{static-free-energy-p=n}
\end{split}
\end{equation}
The Coulomb term is always invariant under the replacement $vol(\sigma)\mapsto vol(X)-vol(\sigma)$. However, the other terms break this symmetry unless $\theta_{n}$ is either an
integer or an half-integer.\par

Let us consider the zero-temperature limit $\frak f_{\Sigma^{(1)},-\Sigma^{(2)}}^{(n)}(\theta_{n}):=\lim_{\beta\rightarrow\infty}\frak
F_{\Sigma^{(1)},-\Sigma^{(2)}}^{(n)}(\beta;\theta_{n})$. According to \eqref{theta-limit} the thermal contribution in the first equation of \eqref{static-free-energy-p=n} shows
an exponential decrease. However, two cases must be distinguished:
\bigskip

1. $\langle\theta_{n}\rangle\in [0,\frac{1}{2}]$

\begin{equation}
\begin{split}
\frak f_{\Sigma^{(1)},-\Sigma^{(2)}}^{(n)}(\theta_{n}) =
\begin{cases} \frac{2\pi^2}{q_n}vol(\sigma)\big( 1+2\langle\theta_{n}\rangle\big),
&\text{if $0\leq \frac{vol(\sigma)}{vol(X)}\leq\frac{1}{2}-\langle\theta_{n}\rangle$},\\
\frac{2\pi^2}{q_n}vol(X)\left( 1-\frac{vol(\sigma)}{vol(X)}\right)\big( 1-2\langle\theta_{n}\rangle\big), &\text{if $\frac{1}{2}-\langle\theta_{n}\rangle\leq
\frac{vol(\sigma)}{vol(X)} \leq 1$}.
\end{cases}\label{charge-anticharge-potential-1}
\end{split}
\end{equation}

2. $\langle\theta_{n}\rangle\in [\frac{1}{2},1)$

\begin{equation}
\begin{split}
\frak f_{\Sigma^{(1)},-\Sigma^{(2)}}^{(n)}(\theta_{n})=
\begin{cases} \frac{2\pi^2}{q_n}vol(\sigma)\big( 2\langle\theta_{n}\rangle -1\big),
&\text{if $0\leq \frac{vol(\sigma)}{vol(X)}\leq\frac{3}{2}-\langle\theta_{n}\rangle$},\\
\frac{2\pi^2}{q_n}vol(X)\left( 1-\frac{vol(\sigma)}{vol(X)}\right) \big( 3-2\langle\theta_{n}\rangle\big), &\text{if $\frac{3}{2}-\langle\theta_{n}\rangle\leq
\frac{vol(\sigma)}{vol(X)} \leq 1$}.
\end{cases}\label{charge-anticharge-potential-2}
\end{split}
\end{equation}
Depending on the value of $\theta_{n}$, there is a cusp either at $\frac{1}{2}-\langle\theta_{n}\rangle$ or at $\frac{3}{2}-\langle\theta_{n}\rangle$. For
$\langle\theta_{n}\rangle =\frac{1}{2}$, the interaction potential between the two branes is compensated, since

\begin{equation}
\frak f_{\Sigma^{(1)},-\Sigma^{(2)}}^{(n)}(\theta_{n})=0. \label{high-temp-contribution-dynamical}
\end{equation}
Using \eqref{asymptotic-expansion} it follows from the second equation in \eqref{static-free-energy-p=n} that the contributions from the topological degrees of freedom are
exponentially suppressed in the high temperature limit. Hence

\begin{equation}
\lim_{\beta\rightarrow 0}\frak F_{\Sigma^{(1)},-\Sigma^{(2)}}^{(n)}(\beta;\theta_{n})= \frac{2\pi^2}{q_n} vol(\sigma)\left(
1-\frac{vol(\sigma)}{vol(X)}\right).\label{high-temp-contribution-dynamical2}
\end{equation}
In both temperature regimes we find in the large volume limit a "confining volume law" for the static brane antibrane free energy as it increases linearly with the volume of the
submanifold connecting the two branes. This confining law is the $n$-dimensional generalization of the well-known confining property of electrodynamics in $1+1$ dimensions.

\section{Acknowledgements}
I would like to express my gratitude to H. H\"{u}ffel for his encouragement and his valuable comments. I also thank A. Cap for valuable discussions.
\bigskip\bigskip

\begin{appendix}


\section{Appendix: Riemann Theta function and Epstein Zeta function}
In this appendix we want to summarize the main facts regarding the Riemann Theta function and the Epstein zeta function which are used in the present paper. For more detailed
information refer to \cite{Mumford, Siegel-1, Erdelyi}. \par

\subsection{Riemann Theta function}

The $r$-dimensional Riemann Theta function with characteristics $(\vec{a},\vec{b})\in\mathbb{R}^{r}\times\mathbb{R}^{r}$ is defined by

\begin{equation}
\Theta _r\begin{bmatrix} \vec{a} \\\vec{b}\end{bmatrix} (\vec{u}|B)=\sum\limits _{\vec{m}\in\mathbb{Z}^r}e^{\pi i(\vec{m}+\vec{a})^{T}B (\vec{m}+\vec{a})+2\pi
i(\vec{m}+\vec{a})^{T}(\vec{u}+\vec{b})},\label{theta-original}
\end{equation}
where $B$ is a symmetric complex $r\times r$-dimensional matrix whose imaginary part is positive definite and $\vec{u}\in\mathbb{C}^r$. It has the following modular properties

\begin{equation}
\begin{split}
&\Theta _r\begin{bmatrix} \vec{a}+\vec{k} \\\vec{b}+\vec{l}\end{bmatrix} (\vec{u}|B)=e^{2\pi i\vec{a}^{T}\vec{l}}\ \Theta _r\begin{bmatrix} \vec{a}\\\vec{b}\end{bmatrix}
(\vec{u}|B),\qquad \vec{k},\vec{l}\in\mathbb{Z}^r,\\
&\Theta _r\begin{bmatrix} \vec{a} \\\vec{b}\end{bmatrix} (\vec{u}+\vec{m}|B)=e^{2\pi i\vec{a}^{T}\vec{m}}\ \Theta _r\begin{bmatrix} \vec{a}\\\vec{b}\end{bmatrix}
(\vec{u}|B),\qquad \vec{m}\in\mathbb{Z}^r,\\
&\Theta _r\begin{bmatrix} \vec{a}\\\vec{b}\end{bmatrix} (\vec{u}+B\vec{m}|B)=e^{-2\pi i\vec{b}^{T}\vec{m}-\pi i\vec{m}^{T}B\vec{m}-2\pi i\vec{m}^{T}\vec{u}}\ \Theta
_r\begin{bmatrix} \vec{a}\\\vec{b}\end{bmatrix} (\vec{u}|B),\qquad \vec{m}\in\mathbb{Z}^r.\label{modular-property}
\end{split}
\end{equation}
We write $\Theta _{r}(\vec{u}|B):=\Theta _r\begin{bmatrix} \vec{0} \\\vec{0}\end{bmatrix} (\vec{u}|B)$ for the Riemann Theta function with characteristics $(\vec{0},\vec{0})$.
By construction

\begin{equation}
\Theta _r\begin{bmatrix} \vec{a} \\\vec{b}\end{bmatrix} (\vec{u}|B)=e^{\pi i\vec{a}^{T}B\vec{a}+2\pi i\vec{a}^{T}(\vec{u}+\vec{b})}\ \Theta
_{r}(\vec{u}+B\vec{a}+\vec{b}|B)\qquad \vec{a},\vec{b}\in\mathbb{R}^r.\label{decomposition}
\end{equation}
The Riemann Theta function satisfies the following important "duality" formula

\begin{equation}
\Theta _r (\vec{u}|B)=\det{(-iB)}^{-\frac{1}{2}}\ e^{-\pi i\vec{u}^{T}B^{-1}\vec{u}}\ \Theta _r (B^{-1}\vec{u}|-B^{-1}),\label{duality}
\end{equation}
which implies

\begin{equation}
\Theta _r\begin{bmatrix} \vec{a} \\\vec{0}\end{bmatrix} (\vec{u}|B)= \det (-iB)^{-\frac{1}{2}}e^{-\pi i \vec{u}^{T}B^{-1}\vec{u}}\ \Theta _r
(\vec{a}+B^{-1}\vec{u}|-B^{-1}).\label{duality-1}
\end{equation}
In particular, let us consider the Riemann Theta function with $\vec{b}=0$. Then we have

\begin{equation}
\Theta _r(-\vec{u}|B)=\Theta _r(\vec{u}|B),\qquad\Theta _r\begin{bmatrix} -\vec{a}\\\vec{0}\end{bmatrix} (\vec{0}|B)=\Theta _r\begin{bmatrix} \vec{a}\\\vec{0}\end{bmatrix}
(\vec{0}|B).\label{duality-2}
\end{equation}
Due to \eqref{modular-property} and \eqref{duality-2} one can replace any vector $\vec{a}\in\mathbb{R}^{r}$ by the translated vector $\{\vec{a}\}:=(\{a_{1}\},\ldots,\{a_{r}\})$,
defined in \eqref{reduction}. Hence $\{-a_{j}\}=\{a_{j}\}$ and $\{a_j+m_j\}=\{a_j\}$ for $m_j\in \mathbb{Z}$. For $\vec{u}\in\mathbb{R}^{r}$ one gets

\begin{equation}
\Theta _r (\vec{u}|B)=\Theta _r (\{\vec{u}\}|B),\qquad \Theta _r\begin{bmatrix} \vec{a}\\\vec{0}\end{bmatrix} (\vec{0}|B)=\Theta _r\begin{bmatrix}
\{\vec{a}\}\\\vec{0}\end{bmatrix} (\vec{0}|B).\label{theta-reduced-1}
\end{equation}
For the one dimensional Riemann Theta function we will give a series expansion by using the infinite product representation (Jacobi-triple-product)

\begin{equation}
\Theta _1(u|B)=\prod_{k=1}^{\infty}(1-q^{2k})(1+q^{2k-1}e^{2\pi iu})(1+q^{2k-1}e^{-2\pi iu}),\label{jacobi}
\end{equation}
where $q=e^{\pi iB}$. Consider now $B=i\upsilon$, with $\upsilon\in\mathbb{R}$ and $\upsilon >0$. Let $\upsilon$ and $u=\Re u+i\Im u\in\mathbb{C}$ satisfy the relation

\begin{equation}
\upsilon - 2|\Im u|\geq 0.\label{constraint}
\end{equation}
Using the series expansion for the logarithm and the geometric series formula, one derives from \eqref{jacobi}

\begin{equation}
\ln{\Theta _1(u|i\upsilon)}= \sum_{m=1}^{\infty}\frac{1}{m}\ \frac{1}{1-e^{2\pi\upsilon m}}+2\sum_{m=1}^{\infty}\frac{(-1)^{m+1}}{m}\ \frac{e^{-\pi\upsilon m}\ \cos(2\pi
mu)}{1-e^{-2\pi\upsilon m}}.\label{asymptotic-expansion}
\end{equation}
Of particular interest are two cases: In the first case let $u\in\mathbb{R}$. According to \eqref{theta-reduced-1} $u$ can be replaced by $\{u\}$ and one finds that the two
terms in $\ln{\Theta _1(\{u\}|i\upsilon)}$ decrease exponentially for $\upsilon\rightarrow\infty$. In the second case let $u=i\upsilon\{\varepsilon\}$ where
$\varepsilon\in\mathbb{R}$. Extracting from \eqref{asymptotic-expansion} the contribution, where $\{\varepsilon\}=\frac{1}{2}$, gives

\begin{equation}
\begin{split}
\ln{\Theta _1(i\upsilon\{\varepsilon\}|i\upsilon)}=&\delta_{\{\varepsilon\},\frac{1}{2}}\ln 2+ \sum_{m=1}^{\infty}\frac{1}{m}\
\frac{1+2(-1)^{m}\delta_{\{\varepsilon\},\frac{1}{2}}}{1-e^{2\pi\upsilon m}}\\
&+(1-\delta_{\{\varepsilon\},\frac{1}{2}})\sum_{m=1}^{\infty}\frac{(-1)^{m}}{m}\ \frac{ e^{-2\pi\upsilon m(\{\varepsilon\}-\frac{1}{2})}+e^{2\pi\upsilon
m(\{\varepsilon\}+\frac{1}{2})}}{1-e^{2\pi\upsilon m}}.\label{theta-limit}
\end{split}
\end{equation}
In the limit $\upsilon\rightarrow\infty$, all terms but the first one decrease exponentially.

\subsection{Epstein Zeta function}

Let us now introduce briefly the Epstein Zeta function and display some properties \cite{Siegel-1, Erdelyi}, which are used in this paper. Let
$(\vec{a},\vec{b})\in\mathbb{R}^r\times\mathbb{R}^r$ and let $\mathbf{Q}$ be the quadratic form associated to a $r\times r$-dimensional positive-definite matrix $Q=(Q_{jk})$.
Then $r$-dimensional Epstein zeta function with characteristics $(\vec{a},\vec{b})$ is defined by

\begin{equation}
E _r\begin{bmatrix} \vec{a} \\\vec{b}\end{bmatrix} (s;\mathbf{Q})=\sum\limits _{\substack{ \vec{m}+\vec{a}\neq 0\\
\vec{m}\in \mathbb{Z}^{r}}}\frac{e^{2\pi i\vec{m}^{T}\vec{b}}}{[\mathbf{Q}(\vec{m}+\vec{a})]^{s}},\qquad s\in\mathbb{C}.\label{epstein-original}
\end{equation}
This series is absolutely convergent for $\Re(s)>\frac{r}{2}$ and satisfies the so-called reflection formula

\begin{equation}
\pi^{-s}\Gamma(s)E _r\begin{bmatrix} \vec{a} \\\vec{b}\end{bmatrix} (s;\mathbf{Q})=(\det Q)^{-\frac{1}{2}}\pi^{(s-\frac{r}{2})}\ \Gamma\left(\frac{r}{2}-s\right)e^{-2\pi
i\vec{a}^{T}\vec{b}}\ E _r\begin{bmatrix} \vec{b} \\-\vec{a}\end{bmatrix} \left(\frac{r}{2}-s;\mathbf{Q}^{-1}\right),\label{reflection-formula}
\end{equation}
where $\mathbf{Q}^{-1}$ is the quadratic form associated by the inverse matrix $Q^{-1}$. The Epstein zeta function has an analytic continuation to all $s\in\mathbb{C}$ which is
an entire function whenever $\vec{b}\not\in\mathbb{Z}^{r}$. If $\vec{b}\in\mathbb{Z}^{r}$, then it is meromorphic on $\mathbb{C}$ with the only pole at $s=\frac{r}{2}$ and
residue $(\det Q)^{-\frac{1}{2}}\pi^{\frac{r}{2}}\Gamma(\frac{r}{2})^{-1}$. \par

If $\vec{a}=\vec{b}=\vec{0}$ and $\textbf{Q}$ is defined by the diagonal matrix $Q=diag(c_{1}^{2},\ldots,c_{r}^{2})$, then we write

\begin{equation}
E_r(s;c_{1},\ldots,c_{r}):= E _r\begin{bmatrix} \vec{0} \\\vec{0}\end{bmatrix} (s;\mathbf{Q})=\sum\limits
_{\vec{m}\in\mathbb{Z}_{0}^{r}}\left(\sum_{j=1}^{r}(c_jm_{j})^{2}\right)^{-s}.\label{epstein-conventional}
\end{equation}
where $\mathbb{Z}_{0}^{r}:=\mathbb{Z}^{r}\setminus\{0\}$. In the special case of $r=1$, the Epstein zeta function reduces to the Riemann zeta function $\zeta_{R}$, namely
$E_1(s;1)=2\zeta_{R}(2s)$.\par

Under a constant scale transformation $c_i\mapsto \lambda c_i$, $\lambda\in\mathbb{R}$, the Epstein function transforms as

\begin{equation}
E_{r}(s;\lambda c_1,\ldots ,\lambda c_r)=\lambda ^{-2s}E_{r}(s;c_1,\ldots ,c_r).\label{epstein-trafo}
\end{equation}
This implies for the derivative of $E_r$ that
\begin{equation}
E_{r}^{\prime}(0;\lambda c_1,\ldots ,\lambda c_r)=2\ln{\lambda}+E_{r}^{\prime}(0;c_1,\ldots ,c_r).
\end{equation}
Epstein zeta functions in different dimensions are related by the Chowla-Selberg formula,

\begin{equation}
\begin{split}
E_{r}(s;c_1,\ldots ,c_r)= &E_{l}(s;c_1,\ldots ,c_l)+\frac{\pi ^{\frac{l}{2}}\Gamma (s-\frac{l}{2})}{\prod _{i=1}^{l}c_i\Gamma (s)}E_{r-l}(s-\frac{l}{2};c_{l+1},\ldots ,c_r)\\
&+\frac{1}{\Gamma (s)}T_{r,l}(s;c_1,\ldots ,c_r),
\end{split}\label{Chowla_Selberg}
\end{equation}
where
\begin{equation}
\begin{split}
T_{r,l}(s;c_1,\ldots ,c_r) = &\frac{2\pi ^{s}}{\prod _{i=1}^{l}c_i}\ \sum _{(k_1,\ldots ,k_l)\in\mathbb{Z}_0^l}\ \sum _{(k_{l+1},\ldots
,k_r))\in\mathbb{Z}_0^{r-l}}\left[ \frac{\sum_{i=1}^{l}(\frac{k_i}{c_i})^2}{\sum_{i=l+1}^{r}(k_ic_i)^2} \right]^{\frac{2s-l}{4}}\\
&\times K_{s-\frac{l}{2}}\left(2\pi\sqrt{\left(\sum_{i=1}^{l}(\frac{k_i}{c_i})^2\right)\left(\sum_{i=l+1}^{r}(k_ic_i)^2\right)}\right).
\end{split}
\end{equation}
Here $K_{\nu}(z)$ denotes the modified Bessel function of the second kind \cite{gradshteyn}. The function $s\mapsto T_{r,l}(s;c_1,\ldots ,c_r)$ is analytic on $\mathbb{C}$. We
get at $s=0$

\begin{equation}
E_{r}(0;c_1,\ldots ,c_r)=E_{1}(0;c_1)=2\zeta _{R}(0)=-1.\label{epstein=0}
\end{equation}
Finally, the derivative of the Chowla-Selberg formula \eqref{Chowla_Selberg} at $s=0$ reads

\begin{equation}
E_{r}^{\prime}(0;c_1,\ldots ,c_r)=E_{l}^{\prime}(0;c_1,\ldots ,c_l)+\frac{\pi ^{-\frac{r}{2}}\Gamma (\frac{r}{2})}{\prod
_{i=1}^{r}c_i}E_{r-l}(\frac{r}{2};\frac{1}{c_{l+1}},\ldots ,\frac{1}{c_r})+T_{r,l}(0;c_1,\ldots ,c_r).\label{epstein-expansion}
\end{equation}

\end{appendix}


\end{document}